\documentclass{aa}  

\usepackage{graphicx}
\usepackage{txfonts}
\usepackage{aas_macros}
\usepackage{hyperref}
\usepackage{multirow}
\hypersetup{
	colorlinks,
	citecolor=blue,
	filecolor=blue,
	linkcolor=blue,
	urlcolor=blue
}

\usepackage{gensymb, stfloats}
\usepackage{natbib}
\begin{document}

   \title{A High Geometric Albedo for LTT9779b Points Towards a Metal-rich Atmosphere and Silicate Clouds}

   \titlerunning{A High Geometric Albedo for LTT9779b}
   \authorrunning{Saha et al.}

   \author{Suman Saha\inst{1,2},
          James S. Jenkins\inst{1,2},
          Vivien Parmentier\inst{3},
          Sergio Hoyer\inst{4},
          Magali Deleuil\inst{4},
          Ian J. M. Crossfield\inst{5},
          Pablo A. Peña R.\inst{1,2},
          Jose I. Vines\inst{6},
          R. Ram\'irez Reyes\inst{7,8},
          Mat\'ias R. D\'iaz\inst{9}}
   \institute{Instituto de Estudios Astrofísicos, Facultad de Ingeniería Ciencias, Universidad Diego Portales, Av. Ejército Libertador 441, Santiago, Chile\\
        \email{suman.saha@mail.udp.cl}
        \and
        Centro de Excelencia en Astrofísica y Tecnologías Afines (CATA), Camino El Observatorio 1515, Las Condes, Santiago, Chile
        \and
        Université Côte d’Azur, Observatoire de la Côte d’Azur, CNRS, Laboratoire Lagrange, France
        \and
        Aix Marseille Univ, CNRS, CNES, LAM, 38 reu Fr\'ed\'erik Joliot-Curie, 13388 Marseille, France
        \and
        Department of Physics and Astronomy, University of Kansas, Lawrence, KS, USA
        \and
        Instituto de Astronom\'ia, Universidad Cat\'olica del Norte, Angamos 0610, 1270709 Antofagasta, Chile
        \and
        Instituto de Astrofisica, Departamento de Fisica y Astronomia, Facultad Ciencias Exactas, Universidad Andres Bello,
        Fernandez Concha 700, Las Condes, Santiago, Chile
        \and
        Departamento de Astronom\'ia, Universidad de Chile,
        Camino el Observatorio 1515, Las Condes, Santiago, Chile
        \and
        Las Campanas Observatory, Carnegie Institution for Science, Raul Bitr\'an 1200, La Serena, Chile.
        }

   \date{}
 
  \abstract
   {} 
   {In this work, we aim to confirm the high albedo of the benchmark ultrahot Neptune LTT9779b using 20 secondary eclipse measurements of the planet observed with CHEOPS. In addition, we perform a search for variability in the reflected light intensity of the planet as a function of time.}
   {First, we used the TESS follow-up data of LTT9779b from three sectors (2, 29 and 69) to remodel the transit signature and estimate an updated set of transit and ephemeris parameters, which were directly used in the modeling of the secondary eclipse lightcurves. This involved a critical noise-treatment algorithm, including sophisticated techniques such as wavelet denoising and Gaussian Process (GP) regression, to constrain noise levels from various sources. In addition to using the officially released reduced aperture photometry data from CHEOPS DRP, we also reduced the raw data using an independent PSF photometry pipeline, known as PIPE, to verify the robustness of our analysis. The extracted secondary eclipse lightcurves were modeled using the PYCHEOPS package, where we have detrended the background noise correlated with the spacecraft roll angle, originating from the inhomogeneous and asymmetric shape of the CHEOPS point spread function, using an N-order glint function.}
   {Our independent lightcurve analyses have resulted in consistent estimations of the eclipse depths, with values of 89.9$\pm$13.7 ppm for the DRP analysis and 85.2$\pm$13.1 ppm from PIPE, indicating a high degree of statistical agreement. Adopting the DRP value yields a highly constrained geometric albedo of 0.73$\pm$0.11. No significant eclipse depth variability is detected down to a level of $\sim$37~ppm.}
   {Our results confirm that LTT9779b exhibits a strikingly high optical albedo, which substantially reduces the internal energy budget of the planet compared to more opaque atmospheres. By modeling our new and precise eclipse measurements, we find that the planet's atmosphere is likely highly metal-rich, with silicate clouds probably present. Our models find it difficult to explain the high geometric albedo, since fitting these high optical bands leads to a decrease in the NIR planetary emission, well below the current observations. However, we discuss additional physical processes that could circumvent this problem, such as the introduction of strong particle backscattering.}


   \maketitle
%

\section{Introduction}\label{sec:sec1}

Observations of optical secondary eclipses of planets can provide a detailed understanding of the reflected light properties of these worlds, and with that comes a more complete picture of the energy budget and atmospheric structure, or surface constituents. Due to the photometric precision requirements necessary to make these occultation measurements possible, the majority of reflected light studies have been performed on hot Jupiters (HJs), since their larger physical size presents a more favorable target for the outgoing photon flux from the host star \citep[e.g.,][]{2022A&A...659L...4B, 2023A&A...672A..24K, 2024A&A...682A.102P}.

Although HJs appear excellent targets for optical reflected light studies, some of the first searches for this signature foreshadowed what was to come later when we had even more precise facilities.  The lack of any confirmed detections for planets like Tau Bootis~b (\citealp{leigh03a,cameron04}), HD75289b (\citealp{leigh03b}), and HD209458b (\citealp{rowe08}), could place upper limits on the geometric albedo (A$_g$) in the range of $\sim$0.15$\--$0.40.  Indeed, subsequent studies were able to show that HJs in general are very dark planets, having A$_g$ values of $\sim$0.1 (\citealp{esteves15}), meaning they don't reflect a large fraction of the incident stellar flux falling on their upper atmospheres.  

The albedo of a planet's atmosphere is set by the competition between the absorption and reflection of stellar light. Given the high temperatures of hot and ultra-hot Jupiters, numerous absorbers can be in the vapor phase: titanium and vanadium oxide for the hottest planets, sodium, and potassium for the cooler ones ~\citep{Fortney2008}. In HJs, Rayleigh scattering by the gas is usually much smaller than the absorption by these optical absorbers (see class IV planets by \citealp{Sudarsky2000}). Therefore, these planets could have a high albedo only if condensed particles reflect back a significant portion of the light. The most likely species to make a thick enough cloud cover to affect the albedo are silicate clouds. As shown by the class V planets of \citet{Sudarsky2000}, the inclusion of such a thick silicate cloud would raise the albedo of planetary atmospheres to high values. 

However, Jupiter-sized planets hotter than 2000K, aka ultra-hot Jupiters, are unlikely to form these clouds on their dayside. Indeed, these planets have a strong thermal inversion (predicted by \citealp{Fortney2008} and recently unambiguously observed by JWST in \citealp{Coulombe2023}), driven by the optical absorption of the light by TiO and VO in the upper atmospheres. Their dayside atmospheres therefore become hot enough to thermally dissociate molecules~\citep{Parmentier2018}, and are clearly too hot to host any type of clouds on the dayside.

\begin{table*}
    \centering
    \caption{Details of the CHEOPS observations}
    \label{tab:tab1}
    $\begin{array}{lccccccc}
    \hline
    \hline
    \text{Visit} & \text{Epoch} & \text{Obs. start [UTC]} & \text{Obs. end [UTC]}& \text{Duration [hr]} & \text{Exposure time [s]}& \text{No. of frames} & \text{Eclipse coverage [$\%$]}\\
    \hline
    1 & 964 & \text{2020-09-25T13:56} & \text{2020-09-25T18:41} & 4.74 & 60 & 212 & 100.0\\
    2 & 965 & \text{2020-09-26T08:38} & \text{2020-09-26T13:20} & 4.7 & 60 & 196 & 61.5\\
    3 & 966 & \text{2020-09-27T03:32} & \text{2020-09-27T08:17} & 4.74 & 60 & 208 & 74.1\\
    4 & 967 & \text{2020-09-27T22:59} & \text{2020-09-28T03:29} & 4.5 & 60 & 221 & 77.9\\
    5 & 968 & \text{2020-09-28T17:43} & \text{2020-09-28T23:17} & 5.55 & 60 & 232 & 75.7\\
    6 & 971 & \text{2020-10-01T02:48} & \text{2020-10-01T07:16} & 4.47 & 60 & 210 & 100.0\\
    7 & 986 & \text{2020-10-12T23:55} & \text{2020-10-13T04:21} & 4.44 & 60 & 212 & 100.0\\
    8 & 991 & \text{2020-10-16T22:47} & \text{2020-10-17T04:20} & 5.54 & 60 & 227 & 39.6\\
    9 & 993 & \text{2020-10-18T13:25} & \text{2020-10-18T17:41} & 4.27 & 60 & 184 & 52.2\\
    10 & 1003 & \text{2020-10-26T11:05} & \text{2020-10-26T15:55} & 4.82 & 60 & 170 & 100.0\\
    11 & 1841 & \text{2022-08-21T02:28} & \text{2022-08-21T10:34} & 8.1 & 60 & 303 & 70.4\\
    12 & 1878 & \text{2022-09-19T10:33} & \text{2022-09-19T18:19} & 7.77 & 60 & 368 & 100.0\\
    13 & 1881 & \text{2022-09-21T20:13} & \text{2022-09-22T04:04} & 7.85 & 60 & 358 & 35.3\\
    14 & 1882 & \text{2022-09-22T14:45} & \text{2022-09-22T22:46} & 8.02 & 60 & 345 & 100.0\\
    15 & 1883 & \text{2022-09-23T10:10} & \text{2022-09-23T17:58} & 7.8 & 60 & 367 & 66.3\\
    16 & 1889 & \text{2022-09-28T02:56} & \text{2022-09-28T11:02} & 8.1 & 60 & 344 & 100.0\\
    17 & 1892 & \text{2022-09-30T12:34} & \text{2022-09-30T23:38} & 11.07 & 60 & 466 & 56.4\\
    18 & 1895 & \text{2022-10-02T21:00} & \text{2022-10-03T05:19} & 8.32 & 60 & 363 & 100.0\\
    19 & 1900 & \text{2022-10-06T20:04} & \text{2022-10-07T04:10} & 8.1 & 60 & 343 & 91.3\\
    20 & 1906 & \text{2022-10-11T14:50} & \text{2022-10-11T22:34} & 7.74 & 60 & 316 & 77.1\\
    \hline
    \end{array}$
\end{table*}

At the other end of the planetary spectrum, small planets that physically present less favorable targets have still been studied in some detail, in particular the super-Earth population.  Although still a relatively small statistical sample, discoveries from the Kepler Mission (\citealp{borucki10}) allowed the first look at the reflective properties of these planets, finding a pattern similar to the HJs, that in general super-Earths and sub-Neptunes can be as opaque as HJs, having A$_g$ values of 0.11$\pm$0.06 and 0.05$\pm$0.04 respectively, whereas the super-Neptunes are a little more reflective, with A$_g$ values of $\sim$0.23$\pm$0.11 (\citealp{sheets17}). As can be seen by the significantly larger uncertainty on the super-Neptune albedo mean value, the sample used was small, only containing 12 systems, since they were validated to have false positive probabilities of less than 1\%, even though they were not fully confirmed as genuine planets at the time of publication. However, these results hint that super-Neptunes may present a unique group of worlds that showcase atmospheric structures more susceptible to reflectivity, possibly due to the presence of a high atmospheric metallicity and so many host structures like high-altitude silicate cloud decks.

In 2023, the first optical secondary eclipse of a confirmed ultrahot Neptune (UHN), which is also a super-Neptune planet with a radius of $\sim$4.5R$_{\oplus}$, was published (\citealp{2023A&A...675A..81H}, hereafter H23). Using the CHaracterising ExOPlanets Satellite (CHEOPS; \citealp{2021ExA....51..109B}), H23 observed 10 optical eclipses of LTT9779b, arriving at a mean eclipse depth of 115$\pm$24~ppm in the CHEOPS bandpass, which gives an A$_g$ of 0.80$\pm$0.17, similar to that of Venus in our solar system.  Such a high albedo points towards the presence of a highly reflective silicate cloud layer in the upper atmosphere.  The team further showed that the presence of these refractory clouds at such a high equilibrium temperature ($\sim$2000~K; \citealp{2020NatAs...4.1148J}) is possible if the atmosphere is extremely metal-rich and saturated in metals. Indeed, observations from Spitzer, JWST, WINERED, and ESPRESSO all point towards a planet with a metallicity greater than 20$\times$ solar (\citealp{crossfield20,radica24, 2024ApJ...962L..19V, ramirez25}), and the CHEOPS data also indicated the lower limit was 400$\times$ solar. On the other hand, while \citet{2023AJ....166..158E} and the recently published \citet{2025arXiv250114016C} reported lower retrieved metallicities from HST transmission and JWST emission spectra, respectively, both studies acknowledge limitations arising from the restricted wavelength coverage, which hinders accurate detection of carbon-bearing molecules.

One of the interesting issues H23 faced when trying to model the spectral energy distribution (SED) of the planet with 1D radiative transfer models, was capturing the strong CHEOPS optical eclipse depth, whilst also fitting well the large discrepancy between the Spitzer 3.6 and 4.5 $\mu$m bands (see \citealp{dragomir20} who suggested the presence of CO, CO2, and/or clouds to explain this observation).  Various atmospheric metallicities and heat redistribution factors were tested in order to best fit the data, finding that the atmosphere is likely super metal-rich and there is only intermediate heat redistribution around the planet.  Indeed, the Spitzer phase curve analysis presented in \citet{crossfield20} also indicated an inefficient heat redistribution around the atmosphere, along with a possible metal-rich atmosphere as mentioned above.

With such a high atmospheric metallicity being suggested for this planet, attempts to constrain the atmospheric metals further using space-based low-to-medium resolution spectrophotometry with JWST, and/or ground-based high-resolution spectroscopy using instruments like ESPRESSO have been made. \citet{radica24} have found $>5\sigma$ detection of suppressed spectral features in transmission using NIRISS/SOSS onboard JWST; however, after exploring various water and methane-dominated atmospheres, they realized that a number of different scenarios can fit the data equally well, particularly dependent on the actual atmospheric metallicity.  \citet{ramirez25} also studied the planet in transmission from the ground using high-resolution spectroscopy with ESPRESSO, finding no clear detections of single-line elements like sodium, nor other elements like TiO using the cross-correlation technique. However, this work again pointed towards a super metal-rich atmosphere for LTT9779b, since increasing the atmospheric metallicity decreases the scale height, rendering absorption signatures too weak to detect with the current data.  Therefore, taking all these previous efforts together, there is a clear line of evidence that suggests LTT9779b hosts a large rocky core surrounded by a compact but super metal-rich atmosphere.  Further constraining the structure and atmospheric chemistry of this atmosphere is a high priority to better understand the formation and evolutionary mechanisms that occur for extreme planets deeply embedded in the Neptune desert.  

This paper is set out to discuss in detail the results from observing a further 10 CHEOPS secondary eclipses of LTT9779b beyond those already examined in the earlier work of H23.  With these 20 occultations, we can provide even stricter constraints on the geometric albedo of this planet, along with providing a more accurate modeling framework to understand the atmospheric structure. In Sect.~\ref{sec:sec2} we discuss the photometric time series observations from both the Transiting Exoplanet Survey Satellite (TESS; \citealp{ricker15}) and CHEOPS. In Sect.~\ref{sec:sec3} we discuss in detail the processing and modeling of these lightcurves.  In Sect.~\ref{sec:sec4} we detail our modeling efforts to fit these new optical eclipse depths, in addition to the already published near-infrared eclipse depths. Finally, in Sect.~\ref{sec:sec5} we contextualize our results within the broader picture of planet formation and evolution, and summarize our findings.

\begin{figure}
	\centering
	\includegraphics[width=\linewidth]{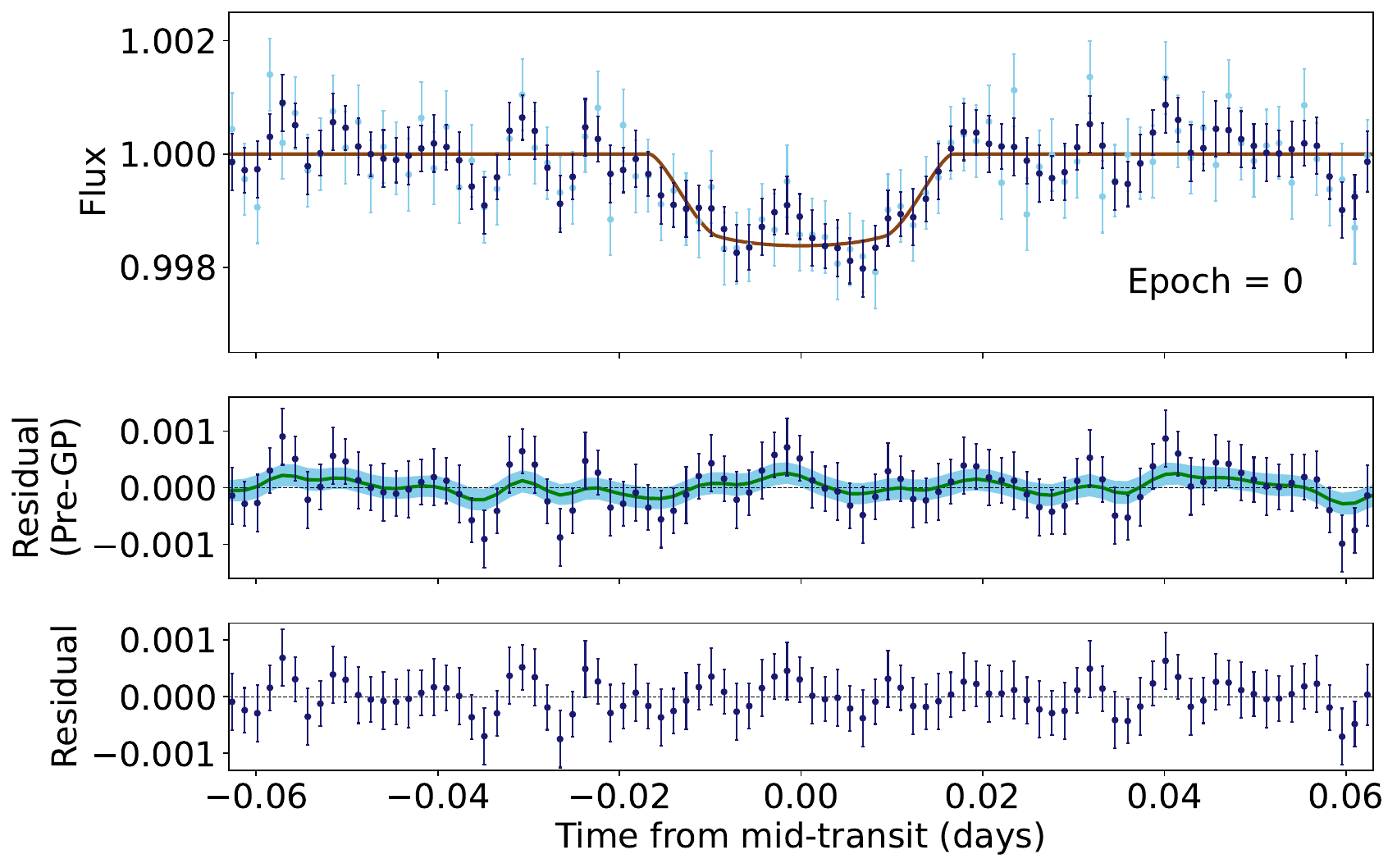}\vspace{0.3cm}
    \includegraphics[width=\linewidth]{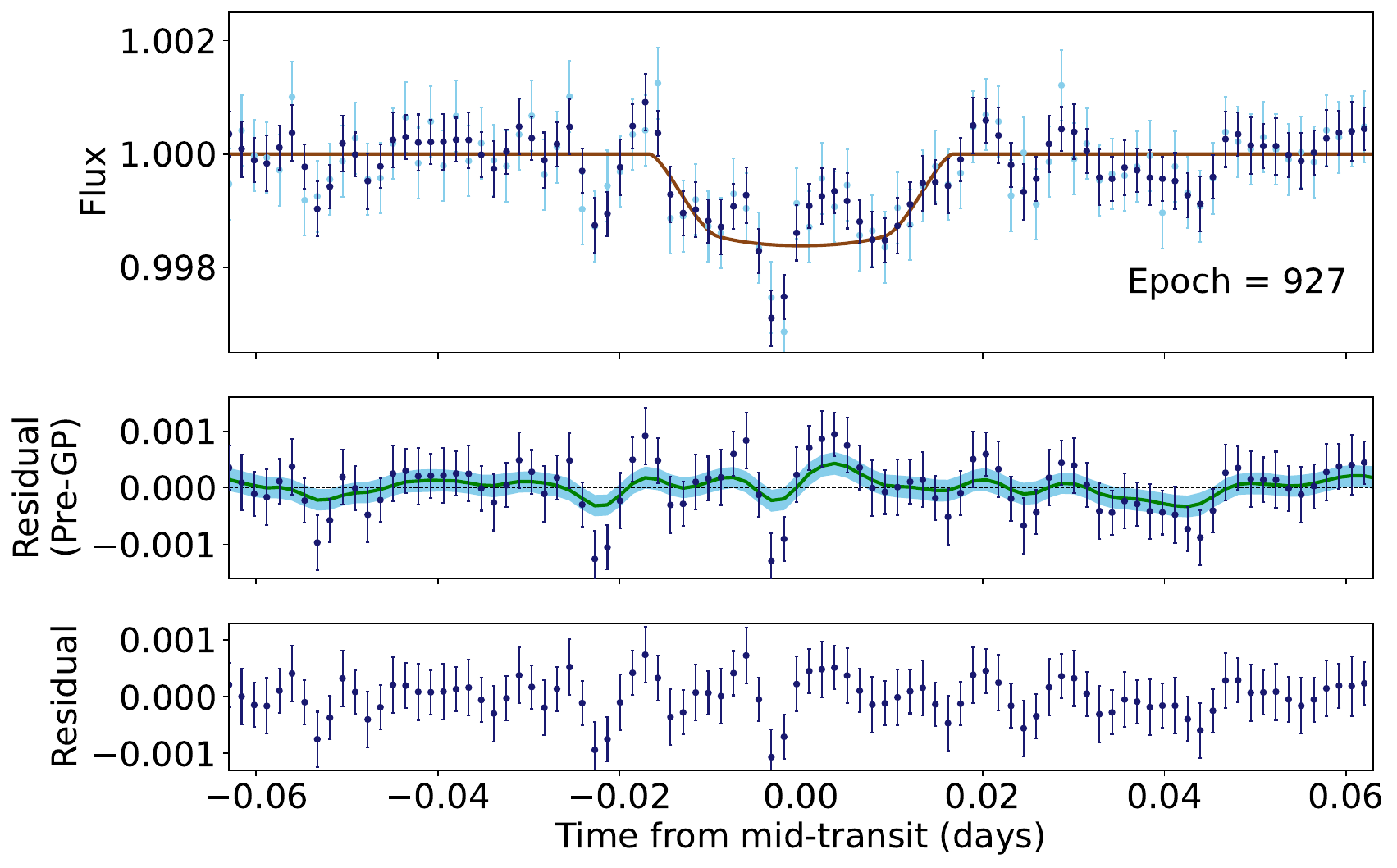}\vspace{0.3cm}
    \includegraphics[width=\linewidth]{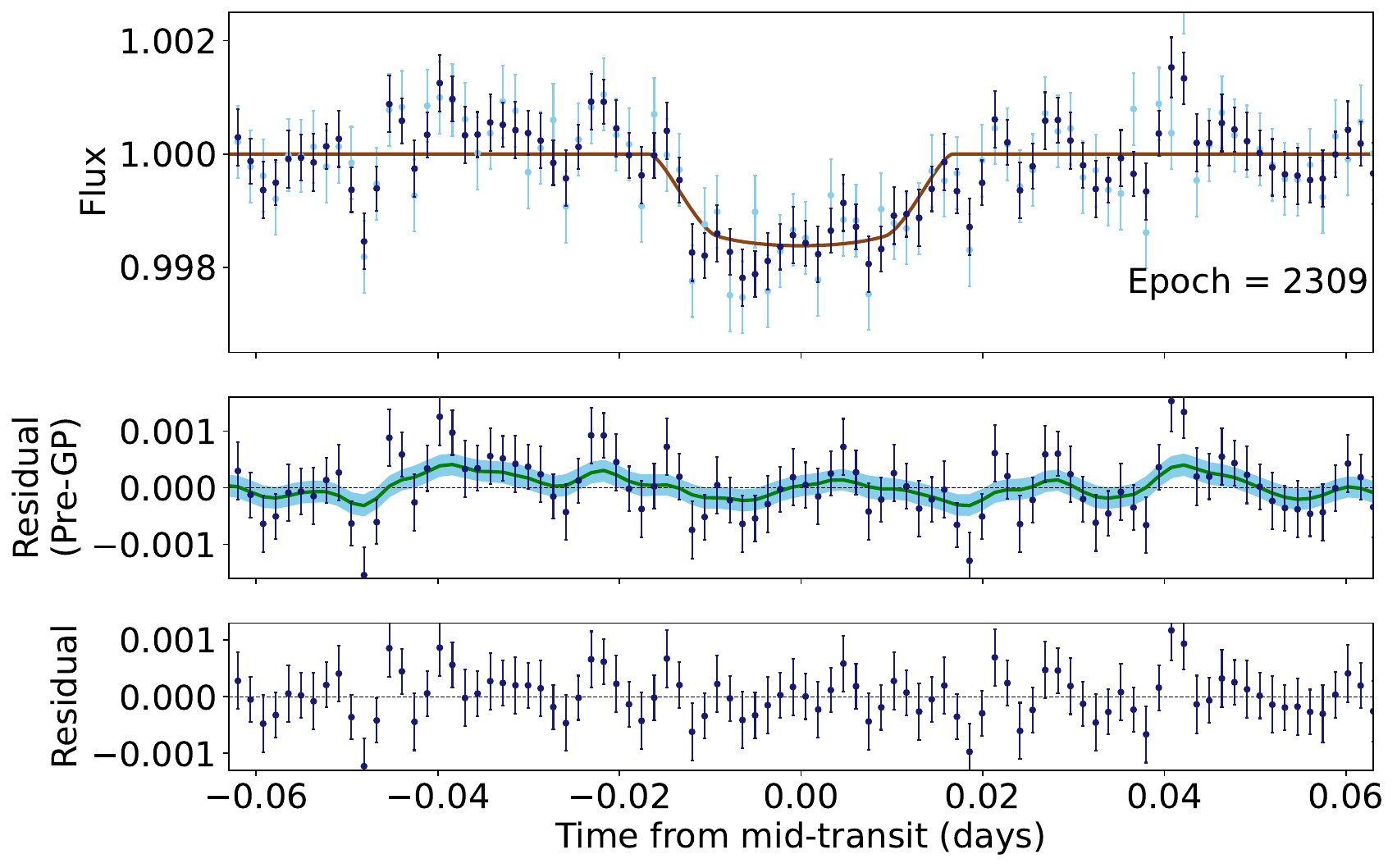}\vspace{0.2cm}
	\caption{Observed and best-fit model light-curves corresponding to the first transit observed from each TESS sector. For each transit, the top panel shows the unprocessed lightcurve (light blue), the lightcurve after wavelet denoising (blue), and the best-fit transit model (brown); the middle panel shows the residual flux before GP regression (blue), and the mean (green) and 1-$\sigma$ interval (light blue) of the best-fit GP regression model; and the bottom panel shows the final mean residual flux (blue). The mean residual flux corresponds to the residual flux considering the mean of the best-fit GP regression model.}
	\label{fig:fig1}
\end{figure}

\section{Observations}\label{sec:sec2}

\subsection{TESS transit observations}

TESS observed LTT9779 during sectors 2, 29, and 69, the corresponding sector mid-points being in September 2018, September 2020, and September 2023, respectively. The previous studies that have reported the transit properties of LTT9779b \citep[e.g.][]{2020NatAs...4.1148J} have used the TESS observations only from sector 2. Since a precise estimation of the transit parameters, including the ephemeris, is extremely important for precise modeling of secondary eclipse observations, we have decided to model all the transit lightcurves observed from all three TESS sectors for an updated estimation of the parameters.

We have used the Barbara A. Mikulski Archive for Space Telescope (MAST)\footnote{https://mast.stsci.edu} to download the publicly available TESS data for LTT9779 corresponding to the mentioned sectors. For all three sectors, the standard 120s cadence PDCSAP lightcurves, processed using the Science Processing Operations Center \cite[SPOC,][]{2016SPIE.9913E..3EJ, 2020RNAAS...4..201C} pipeline, were available, which we have adopted for our analyses. We have identified 87 full transit observations from these lightcurve data. More details on the analyses and modeling of the lightcurves are given in Section \ref{sec:sec31}.

\begin{table}
    \centering
    \caption{Estimated physical properties for LTT9779b from the transit modelling of TESS lightcurves}
    \label{tab:tab2}
    $\begin{array}{lc}
    \hline
    \hline
    \text{Parameter} & \text{Value}\\
    \hline
    \text{Transit parameters} & \\
    T_0\;[BJD_{TDB}] & 2458354.215006_{-0.000162}^{+0.00016} \\
    P \;[days] & 0.792064182\pm0.000000116 \\
    b & 0.9132 _{-0.0205}^{+0.0215}\\
    R_\star/a & 0.2545_{-0.0184}^{+0.0259} \\
    R_p/R_\star & 0.0435_{-0.0015}^{+0.00127}\\
    \text{Limb-darkening coefficients} & \\
    C_1 & 0.261_{-0.172}^{+0.202}\\
    C_2 & 0.293_{-0.183}^{+0.239}\\
    \text{Derived parameters} & \\
    T_{14}\;[hr] & 0.8022_{-0.0186}^{+0.0203} \\
    a/R_\star & 3.929_{-0.363}^{+0.306} \\
    i \;[deg] & 76.55_{-1.75}^{+1.27} \\
    T_{eq}\;[K] & 1941.7_{-71.7}^{+96.1} \\
    a\;[AU] & 0.01734_{-0.0016}^{+0.00136} \\
    M_p\;[M_J] & 0.09328_{-0.00242}^{+0.00245} \\
    M_p\;[M_\oplus] & 29.646_{-0.768}^{+0.777} \\
    R_p\;[R_J] & 0.4016_{-0.014}^{+0.0121} \\
    R_p\;[R_\oplus] & 4.501_{-0.157}^{+0.135}\\
    \hline
    \end{array}$
\end{table}

\subsection{CHEOPS secondary eclipse observations}

We have observed 20 secondary eclipse events LTT9779b from CHEOPS (PI: James Jenkins), the first 10 of which were obtained during September-October 2020 \citep[also see][]{2023A&A...675A..81H}, followed by 10 further observations during August-October 2022. The data corresponding to these observations are now publicly available at the CHEOPS archive\footnote{https://cheops-archive.astro.unige.ch/archive$\_$browser/}. Table \ref{tab:tab1} lists the details of these observations. The epochs are with respect to the first transit of LTT9779b detected by TESS. The table also provides the percentage of the whole secondary eclipse events covered by the CHEOPS observations, calculated using the orbital ephemeris and transit duration estimated from the transit modeling of the TESS lightcurves in this work (details in Section \ref{sec:sec31}). The interruptions in CHEOPS observations are caused by high levels of stray light (SL) and South Atlantic Anomaly (SAA) crossings \citep{2021ExA....51..109B}. The detail of the analyses and modeling of the CHEOPS data is described in Section \ref{sec:sec32}.

\begin{figure}
	\centering
	\includegraphics[width=\linewidth]{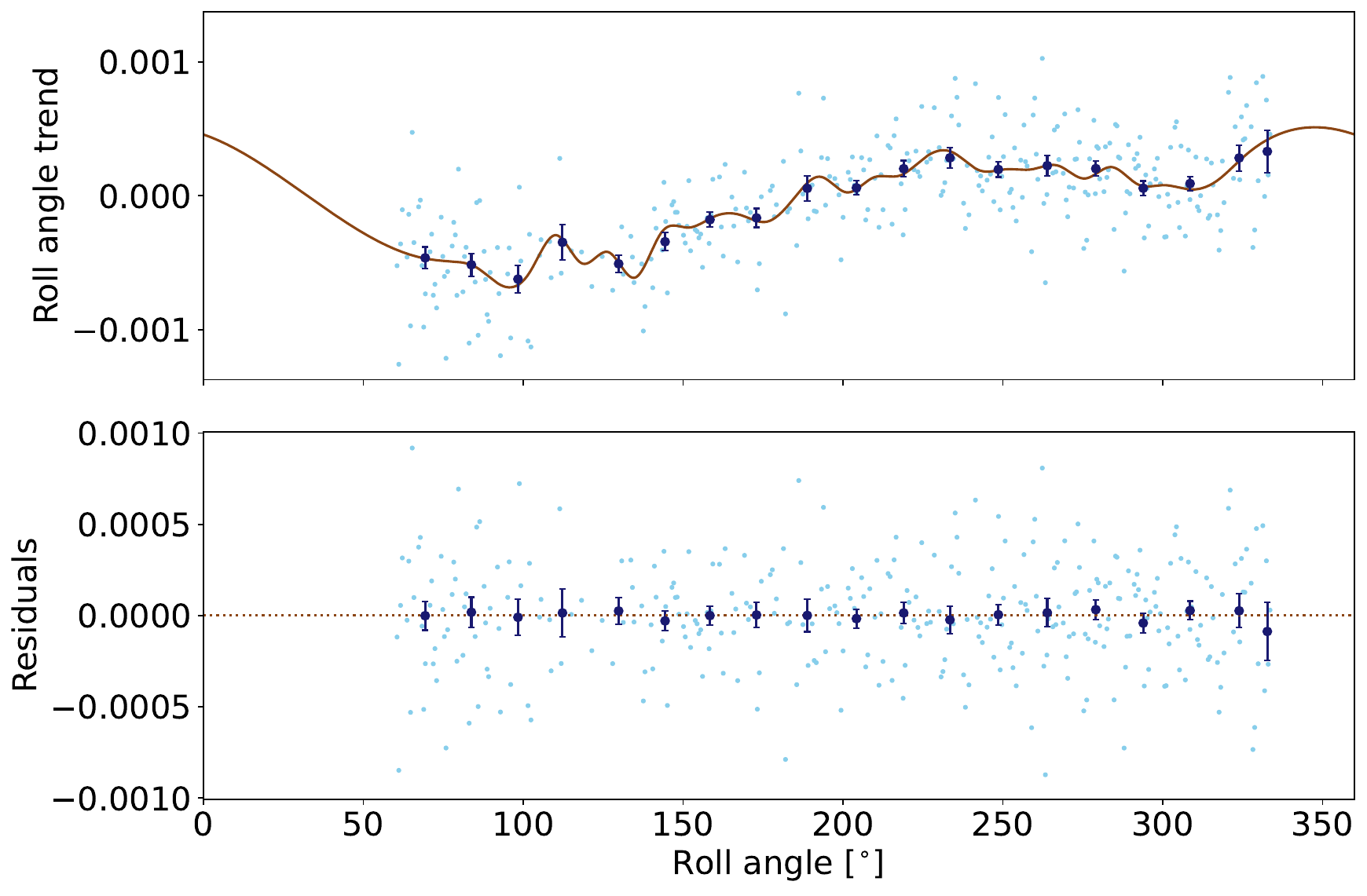}\vspace{0.3cm}
    \includegraphics[width=\linewidth]{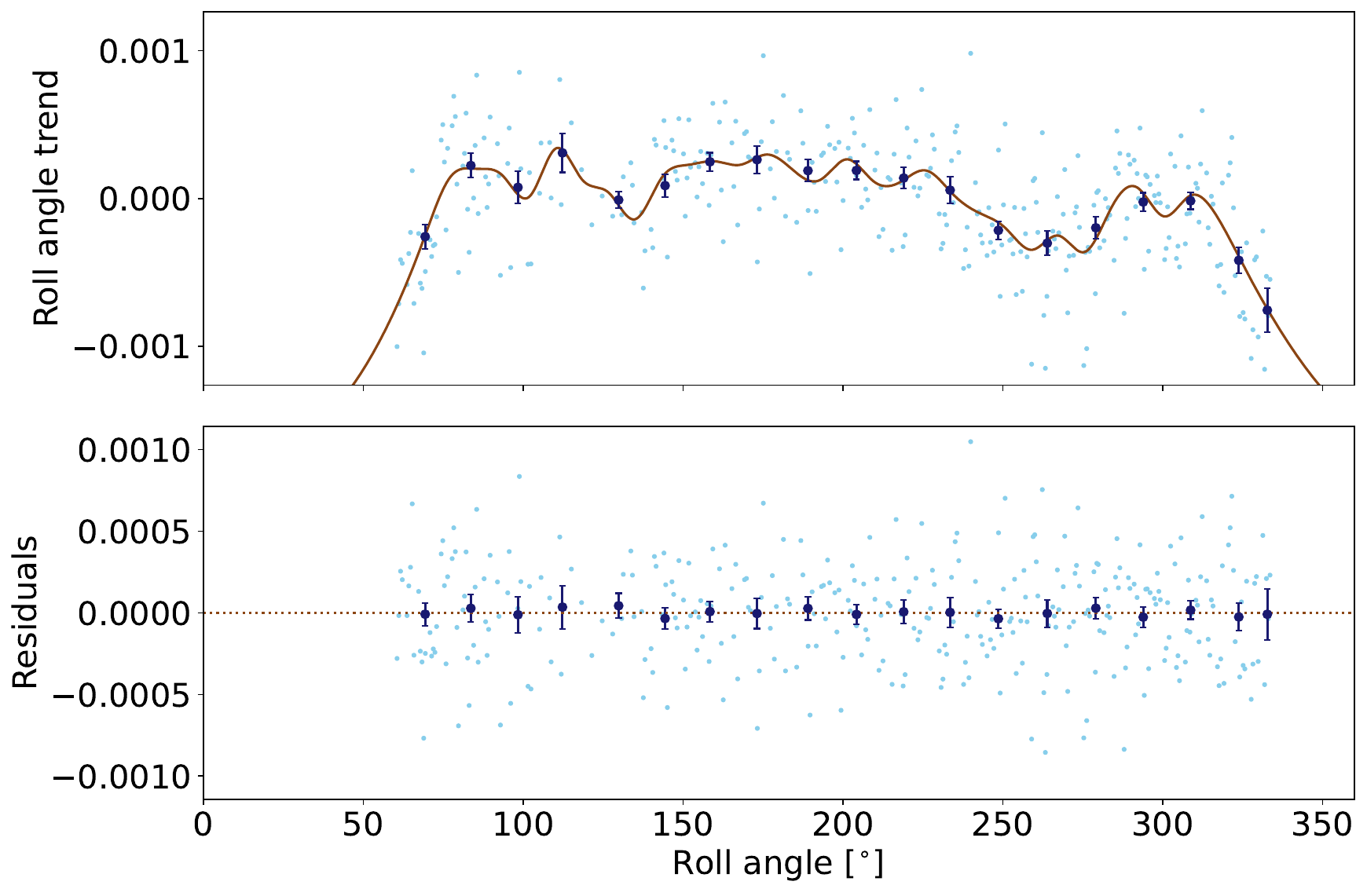}\vspace{0.2cm}
	\caption{The trends in the background flux variations with respect to the roll-angle for the CHEOPS observation corresponding to epoch = 1889, from the lightcurves reduced by the DRP (top) and PIPE (bottom). The light blue points show individual observations (without uncertainties), and the blue points show data binned over 15\textdegree\ intervals. It is worth noting that the trend of variation is different between the lightcurves reduced using the DRP and PIPE, since PIPE has a different sensitivity to background stars that contaminate the photometric apertures in the DRP outputs.}
	\label{fig:fig20}
\end{figure}

\section{Analyses and Modeling}\label{sec:sec3}

\subsection{TESS lightcurves}\label{sec:sec31}

The long TESS lightcurves were sliced, each part containing the full-transit observations, along with significant baselines (at least 3 times the transit duration) on either side of the transits, to obtain 87 transit lightcurves. The large-scale variations in these transit lightcurves, originating from the long-term variation in the stellar flux and instrumental factors, were reduced using a baseline correction method \citep[e.g.][]{saha2022precise, 2023ApJS..268....2S}, where the out-of-transit sections of the lightcurves were modeled using a first-order and a second-order polynomial, and the best-fit models with the least Bayesian Information Criterion \cite[BIC, e.g.][]{neath2012bayesian} were subtracted from the entire lightcurves.

To reduce the time-uncorrelated fluctuations in the lightcurves, the wavelet denoising \citep{Donoho1994IdealDI, 806084, WaveletDenoise2012, 2021AJ....162...18S, 2021AJ....162..221S} technique was used. This is particularly useful for TESS data, where signals are more susceptible to photometric contamination from background sources. Wavelet denoising does not affect the high-frequency terms (such as ingress or egress of a transit) in the signals, unlike other smoothing techniques such as binning or Gaussian moving averages. Instead, it selectively targets lower-amplitude noise components in the lightcurves, which are distinct from the large transit signatures, thereby providing the flexibility to enhance photometric precision without distorting them \citep[e.g.][]{2021AJ....162..221S}. Our implementation of wavelet denoising is similar to \cite{2024arXiv240720846S, 2025MNRAS.tmp..524S}, where the Symlet family of wavelets \citep{daubechies1988orthonormal}, which are the least asymmetric modified versions of the Daubechies wavelets \citep{daubechies1992ten, 1995ComPh...9..635R}, were used. For the Discrete Wavelet Transforms (DWTs), the PyWavelets \citep{Lee2019} package was used. We have used a level-1 wavelet denoising in order to avoid oversmoothing, along with the implementation of the Universal Thresholding Law \citep{Donoho1994IdealDI}.

In order to reduce the effect of the time-correlated noise components in the transit lightcurves, which originate from the small-scale variability and pulsations of the host star, as well as instrumental effects, we have modeled the time-correlated noise using the Gaussian process (GP) regression \citep{2006gpml.book.....R, 2015ApJ...810L..23J, 2019MNRAS.489.5764P, 2021AJ....162...18S, 2021AJ....162..221S} method, simultaneously while modeling the transit signatures in the lightcurves. We have used a similar procedure as in \cite{2024arXiv240720846S, 2025MNRAS.tmp..524S} for the GP regression, which uses the Matern class covariance function with $\nu$ = $3/2$, with two free parameters, i.e., the signal standard deviation and characteristic time scale. The analytical transit formalism by \citet{2002ApJ...580L.171M} was used for the transit modeling. The simultaneous transit modeling with GP regression was performed by the Markov-chain Monte Carlo (MCMC) sampling technique, with the implementation of the Hastings-Metropolis algorithm \citep{1970Bimka..57...97H}. Figure \ref{fig:fig1} shows a representative subset of the modeled transit lightcurves, with the rest shown in Figure \ref{fig:fig51} in the appendix. The estimated mid-transit times from the transit modeling were used to estimate the linear orbital ephemeris. Combined with the stellar and radial velocity parameters from \cite{2020NatAs...4.1148J}, other derivable parameters for the system have been estimated. All the estimated and derived parameters are listed in Table \ref{tab:tab2}.

\begin{figure*}
	\centering
    \begin{tabular}{cc}
    \includegraphics[width=0.5\linewidth]{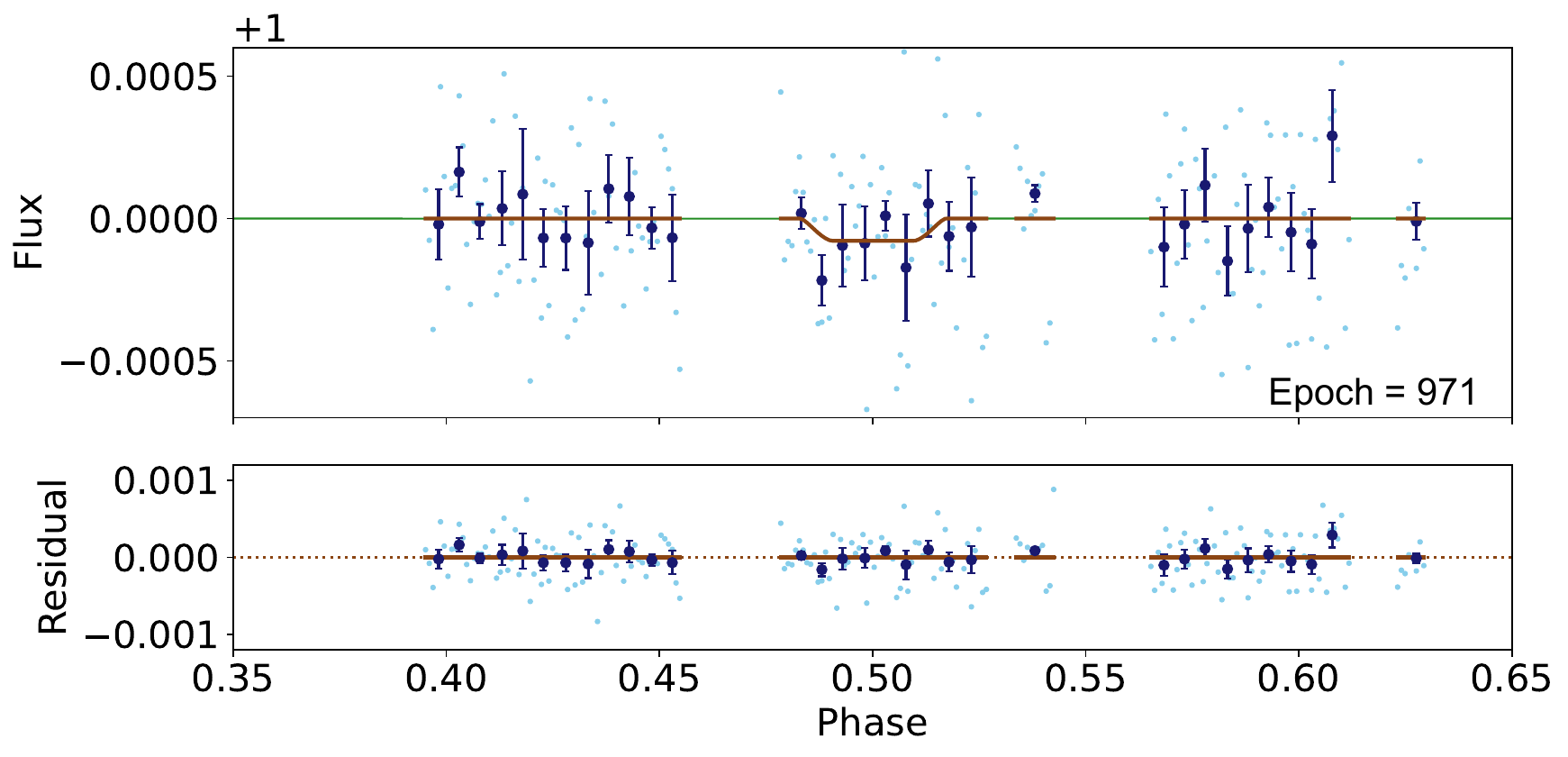} & \includegraphics[width=0.5\linewidth]{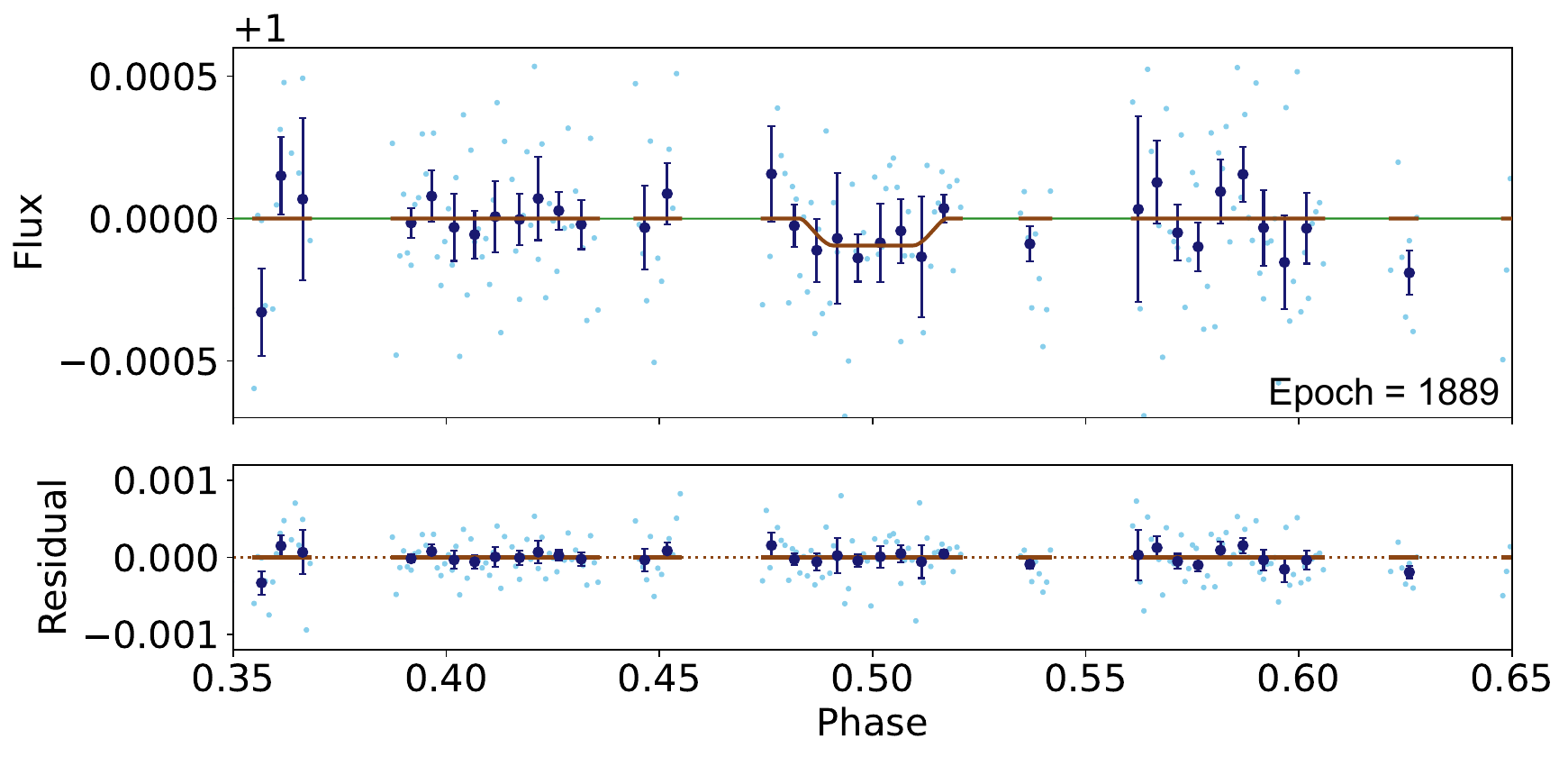}\\
	\includegraphics[width=0.5\linewidth]{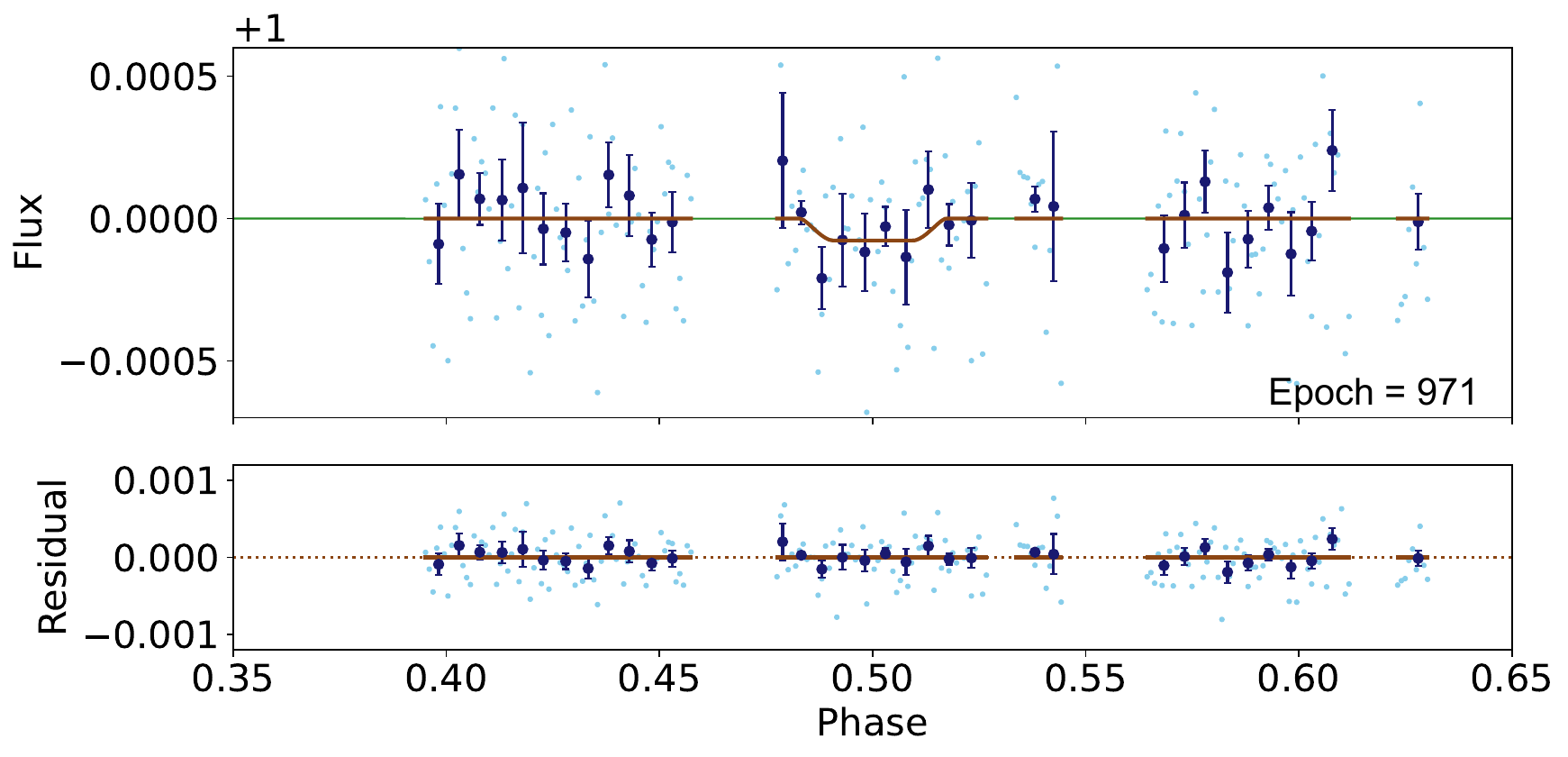} & \includegraphics[width=0.5\linewidth]{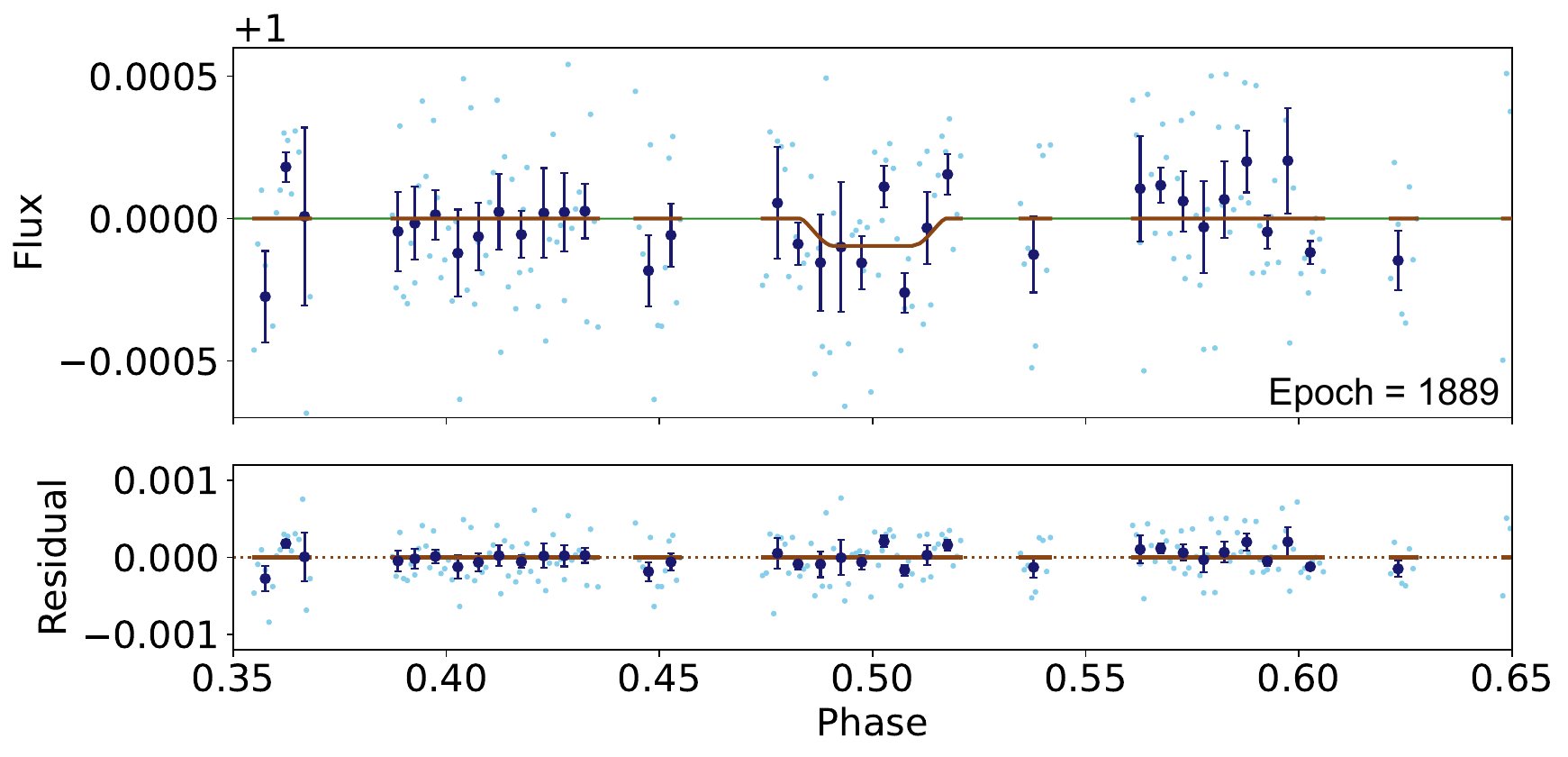}
    \end{tabular}
	\caption{The detrended individual (light blue, without uncertainties), and binned (blue, binned over 7.2 min intervals) secondary eclipse lightcurves corresponding to epochs = 971 and 1889, reduced by using the DRP (top) and PIPE (bottom). The best-fit transit models are shown as the brown/green lines, where the green lines show the disruptions in the lightcurves, and the residuals from the best-fit models in the bottom panels. The lightcurves reduced by the DRP and PIPE are in statistical agreement with each other.}
	\label{fig:fig2}
\end{figure*}

\subsection{CHEOPS secondary eclipse observations}\label{sec:sec32}

The CHEOPS data products downloadable from the official CHEOPS archive consist of raw data, as well as time-series data reduced using the official Data Reduction Pipeline \citep[DRP,][]{2020A&A...635A..24H}. The time-series data are available for aperture photometry performed using different aperture sizes, ranging from 15 to 40 pixels. We have examined the lightcurves generated from these different aperture sizes and found the data corresponding to the aperture size of 24 pixels to be the best in terms of lightcurve dispersion for the majority of the cases. Since the secondary eclipse depths are expected to be small when compared with various sources of contamination possible in the photometric data, it is prudent to use a second independent data reduction, so that the analyses and results can be cross-verified. Thus, we have also performed an independent reduction of the raw CHEOPS data using the PSF Imagette Photometric Extraction \cite[PIPE,][]{2024ascl.soft04002B} package. As the name suggests, PIPE uses the PSF (point spread function) photometry method and has been well adopted by several previous studies \citep[e.g.,][]{2022A&A...667A...1S, 2024A&A...683A...1S}.

For the analyses and modeling of the secondary eclipse lightcurves, we have used the PYCHEOPS \citep{2022MNRAS.514...77M} package, which has been exclusively developed for the analyses of CHEOPS data. First, we have removed the large outliers in the lightcurves resulting from cosmic ray hits. The inhomogeneous and asymmetric shape of the CHEOPS point spread function \citep{2021ExA....51..109B} adds a roll-angle dependent variation in the background flux to the lightcurves. The presence of two close-by stars with G = 19 and G = 15 at distances of 19" and 21" to the source, respectively, \citep[see][]{2023A&A...675A..81H} contributed primarily to this variation for our secondary eclipse lightcurves. We have used a glint function to simultaneously model this background variation, correlated with the roll-angle, whilst modeling the secondary eclipses. We have used the glint function with a fixed number of splines (N$\mathrm{_s}$ = 32) and an independent glint scale parameter (see Figure \ref{fig:fig20}).

\begin{table}
    \centering
    \caption{Estimated secondary eclipse depths from the lightcurves reduced by DRP and PIPE}
    \label{tab:tab3}
    \begin{tabular}{lcc}
    \hline
    \hline
    \multirow{2}{*}{Epoch} & \multicolumn{2}{c}{Eclipse Depth [ppm]} \\
    \cline{2-3}
    & DRP & PIPE \\
    \hline
    964 & $107.7 \pm 62.6$ & $113.4 \pm 62.6$ \\
    965 & $89.3 \pm 63.9$ & $52.5 \pm 45.7$ \\
    966 & $133.6 \pm 74.2$ & $125.4 \pm 67.9$ \\
    967 & $121.7 \pm 69.7$ & $118.5 \pm 63.5$ \\
    968 & $50.9 \pm 43.8$ & $65.5 \pm 50.3$ \\
    971 & $78.6 \pm 56.8$ & $77.3 \pm 53.9$ \\
    986 & $38 \pm 36.2$ & $43.4 \pm 38.7$ \\
    991 & $177.7 \pm 98.5$ & $114.6 \pm 80.4$ \\
    993 & $83.8 \pm 66$ & $87.9 \pm 65.9$ \\
    1003 & $127.3 \pm 70.3$ & $66.7 \pm 51.8$ \\
    1841 & $57.7 \pm 52.2$ & $56.8 \pm 50$ \\
    1878 & $92.9 \pm 61.8$ & $61.1 \pm 47$ \\
    1881 & $123.1 \pm 74.8$ & $120.1 \pm 70.6$ \\
    1882 & $46.3 \pm 42.1$ & $33.7 \pm 32.9$ \\
    1883 & $197.2 \pm 73.7$ & $171.2 \pm 68.2$ \\
    1889 & $95.2 \pm 57.9$ & $96.3 \pm 57.4$ \\
    1892 & $86.7 \pm 74.5$ & $136 \pm 89.9$ \\
    1895 & $21_{-21}^{+23.2}$ & $57.4 \pm 45.5$ \\
    1900 & $40.1 \pm 37.4$ & $51.2 \pm 41.8$ \\
    1906 & $28.3_{-28.3}^{+28.7}$ & $55.4 \pm 48.7$ \\
    \hline
    Mean & $89.85 \pm 13.68$ & $85.2 \pm 13.05$ \\
    \hline
    \end{tabular}
\end{table}

\begin{figure}
	\centering
	\includegraphics[width=\linewidth]{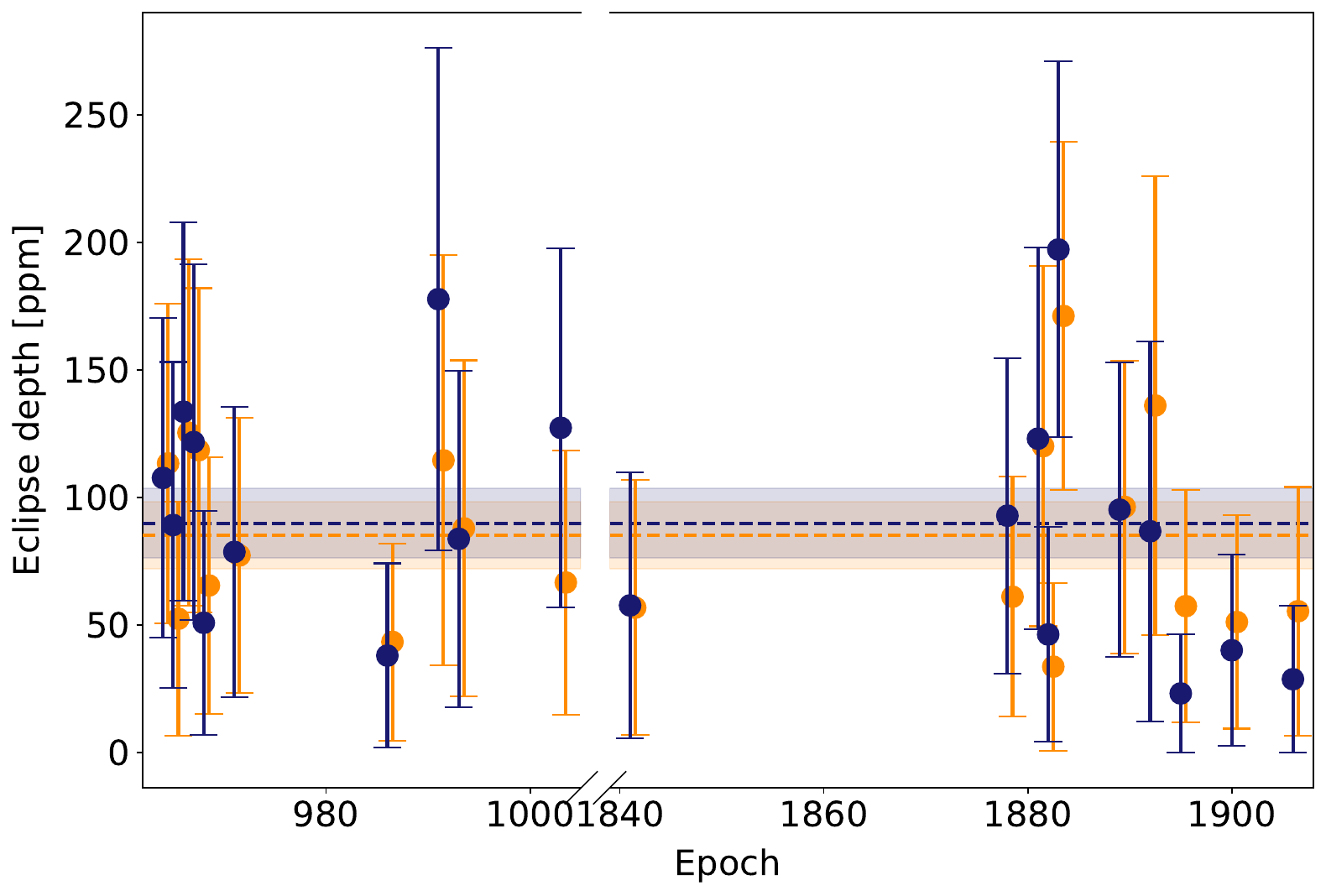}
	\caption{The secondary eclipse depths estimated for each individual epoch from the lightcurves reduced by the DRP (blue) and PIPE (orange, with a slight offset along the x-axis for visibility), along with the calculated mean secondary eclipse depths (blue and orange dotted lines for the lightcurves reduced by DRP and PIPE respectively, with the shaded regions showing the uncertainties). It is interesting to note that the eclipse depths estimated from the two sets of lightcurve data reduced by independent pipelines (i.e., DRP and PIPE) are in close statistical agreement with each other.}
	\label{fig:fig3}
\end{figure}

The modeling of the secondary eclipses was performed using the MCMC sampling technique, where we used the updated transit parameters estimated from our TESS transit lightcurve analyses, including the more precise ephemeris, as fixed parameters, keeping the eclipse depths as the independent variables. The depths of the secondary eclipses from each epoch were estimated independently for both sets of lightcurves reduced using the DRP and PIPE. We have found that the detrended lightcurves from both DRP and PIPE outputs are in statistical agreement with each other (see Figure \ref{fig:fig2}, and Figure \ref{fig:fig41} in the appendix). This helps cross-verify our data reduction and pre-processing steps. The estimated depths of all the secondary eclipses from the lightcurves reduced by both the DRP and PIPE, along with the calculated mean depths, are given in Table \ref{tab:tab3}. Figure \ref{fig:fig3} is a graphical representation of these tables, where it can be seen that the estimated depths from the lightcurves reduced by the DRP and PIPE are in statistical agreement with each other. We also modeled all the lightcurves simultaneously with a single eclipse depth, which resulted in larger residual trends in the fitted lightcurves. This suggests a dynamically varying nature for these eclipses, with some eclipses dropping below our detection threshold and others well above. However, we were not able to confirm the atmosphere's variable nature statistically because of limited SNRs and large systematics in our observed data.

\begin{figure}
	\centering
	\includegraphics[width=\linewidth]{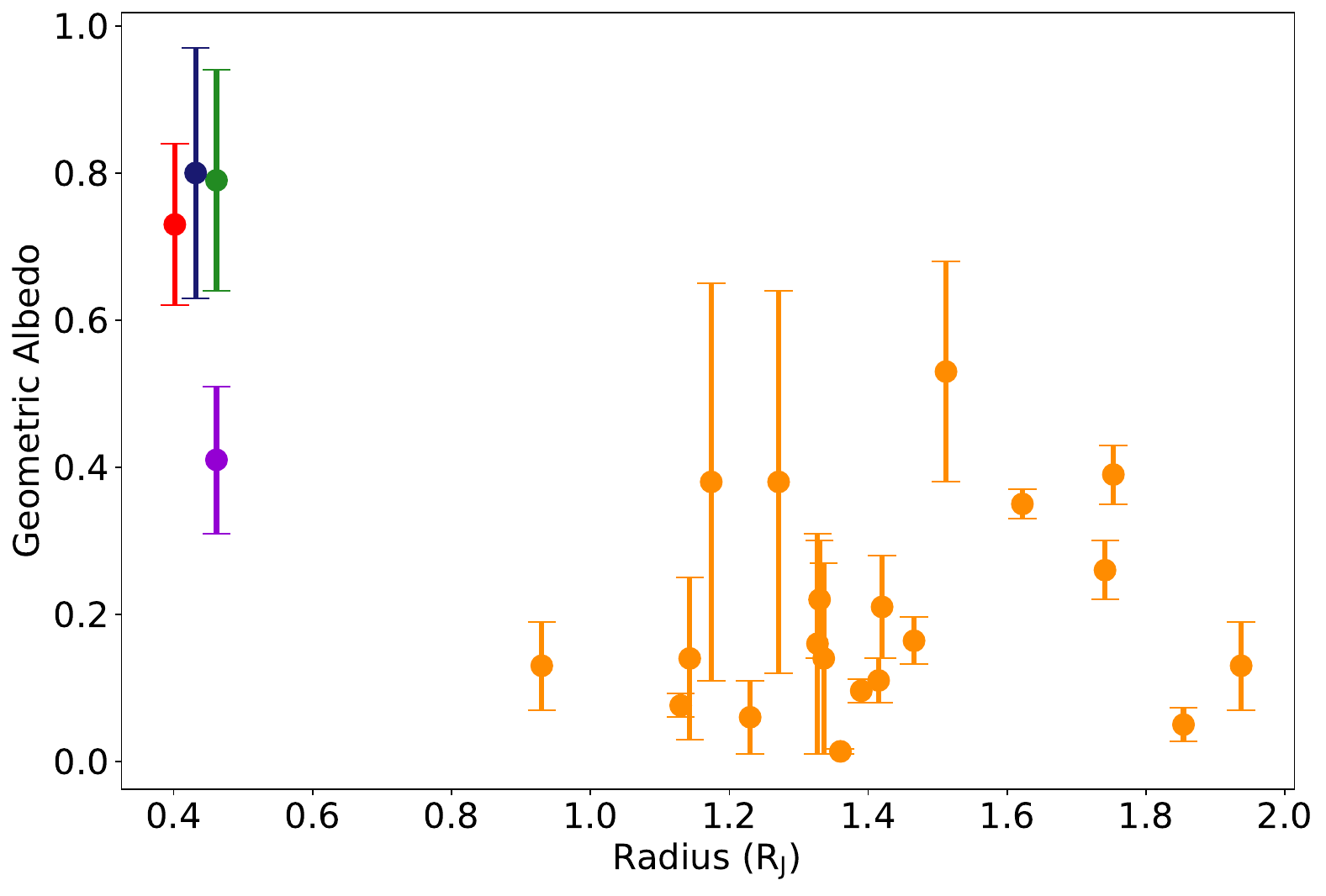}
	\caption{The geometric albedo vs radius plot for LTT9779b estimated from this work (DRP, red), from \citealp{2023A&A...675A..81H} (blue), and from \citealp{2025arXiv250114016C} (western day-side in green and eastern day-side in violet) also showing the known well-constrained geometric albedos of other hot- and ultra-hot Jupiters from the literature (orange).}
	\label{fig:fig:ag}
\end{figure}

We estimated mean eclipse depths from the DRP and PIPE lightcurves to be 89.85$\pm$13.68 ppm and 85.2$\pm$13.05 ppm, respectively. Based on these values, we calculated the corresponding geometric albedos of LTT9779b to be $0.73\pm0.11$ and $0.70\pm0.11$, from the DRP and PIPE reductions, respectively (also see Figure \ref{fig:fig:ag}). We note that we report arithmetic means here, since if we perform a weighted mean, we arrive at a lower eclipse depth, e.g., 58.1$\pm$10.6 ppm from DRP, which still remains in statistical agreement with the non-weighted value. However, by weighting, we are heavily biasing towards the `non-detections' (those with depths below around 55 ppm), as these appear to have more precise measurements due to the fact that the uncertainty spread can not be negative, (essentially they all have $\sim$100\% uncertainties), setting our detection limit. Therefore, we choose to conservatively take the arithmetic mean, which is in excellent statistical agreement with H23, but with substantially improved statistical precision. The revised eclipse depth is also consistent with \citet{2025MNRAS.538.1853R}, who noted that the H23 value was somewhat larger than predicted by their models. In addition, a very recent study using JWST/NIRISS phase curve observations \citep{2025arXiv250114016C} suggests a varying albedo of 0.41$\pm$10 to 0.79$\pm$0.15 from the eastern to the western dayside of LTT9779b. While the upper limits from these results are in good statistical agreement with our estimated values, they also point towards a dynamically changing atmosphere of this planet, as we had suggested in H23 and have alluded to here.

\section{Interpretation}\label{sec:sec4}
\subsection{Model setup}
In order to interpret our measured eclipse depth, we use the radiative/convective/chemical equilibrium code scCHIMERA~\citep{Line2021,Wiser2024}. The code solves for the radiative/convective state of the atmosphere while assuming that chemical equilibrium is maintained at all times. The chemical equilibrium is calculated with the  CEA Gibbs free energy minimization code. The code takes into account both titanium-based clouds and silicate clouds ($\mathrm{MgSiO_3}$). However, we find that only the silicate clouds have a large enough opacity to affect the albedo of the planet. When a given species condenses out of the atmosphere, we assume that it is not locally available to react chemically with the atmosphere. We further assume that all layers above the condensation level have an abundance of the condensed species that are set by the partial pressure of the species at the cloud formation level (vertical rainout approximation). 

When chemical equilibrium predicts cloud condensation, we solve the vertical structure of the cloud using the formalism from~\citep{Ackerman2001}. This formalism assumes a balance between the settling rate of the particles (and thus their size) and the vertical mixing.

Metallicity is changed in the atmosphere by first multiplying all elemental abundances of species other than H and He by the metallicity factor, and then normalizing all elements so that the sum of their volume mixing ratio equals 1. This approach ensures that the ratio of all elements remains solar, but that all are enriched compared to hydrogen. 

We have already shown in \citet{2023A&A...675A..81H} that to obtain a high albedo through cloud formation for LTT9779b, a high metallicity was required, but also the exact metallicity was degenerate with the heat transport efficiency. We therefore updated the scCHIMERA models to better model the reflected light of LTT9779b. When solving for the equilibrium thermal structure of the planet's dayside, we adjust the incoming irradiation to take into account the fact that some energy has been transported towards the dayside. Modeling-wise, this is accomplished by modulating the received stellar flux by a redistribution factor  $f=(T_{\rm day}/T_{\rm eq})^4$. However, we realized that the reflected light component should not be scaled the same way. Indeed, while some of the absorbed stellar energy is transported towards the nightside by the atmospheric circulation, none of the reflected light photons have the same fate. As a consequence, the reflected light component needs to be solved for a no-redistribution scenario (so $f_{\rm ref}=2.66$), even when the thermal component is calculated with a different value of the heat redistribution parameter. This adjustment has the consequence of naturally increasing the reflected-to-thermal component compared to the results in \citeauthor{2023A&A...675A..81H}. We further tracked an incorrect $\pi$ factor that was multiplying into the fluxes in \citeauthor{2023A&A...675A..81H} and corrected for it here.

We further look at a special case where TiO/VO abundances are not set by local equilibrium considerations. TiO/VO are important species because they can strongly change the energy balance of the atmospheres of irradiated planets, leading to the presence of thermal inversions and thus emission features in their spectra~\citep{Fortney2008}. Indeed, it has been shown from hot Jupiter literature that some depletion mechanism is probably reducing the concentration of TiO and VO compared to their equilibrium values ~\citep{Line2016a, Parmentier2016, Mansfield2021, Roth2024}, with one possibility being their cold-trapping into the cooler nightside~\citep{Parmentier2013}. Thus, we consider both models at chemical equilibrium and models where we remove altogether TiO and VO from the mix. 

\subsection{Model results}

\begin{figure}
    \centering
    \includegraphics[width=\linewidth]{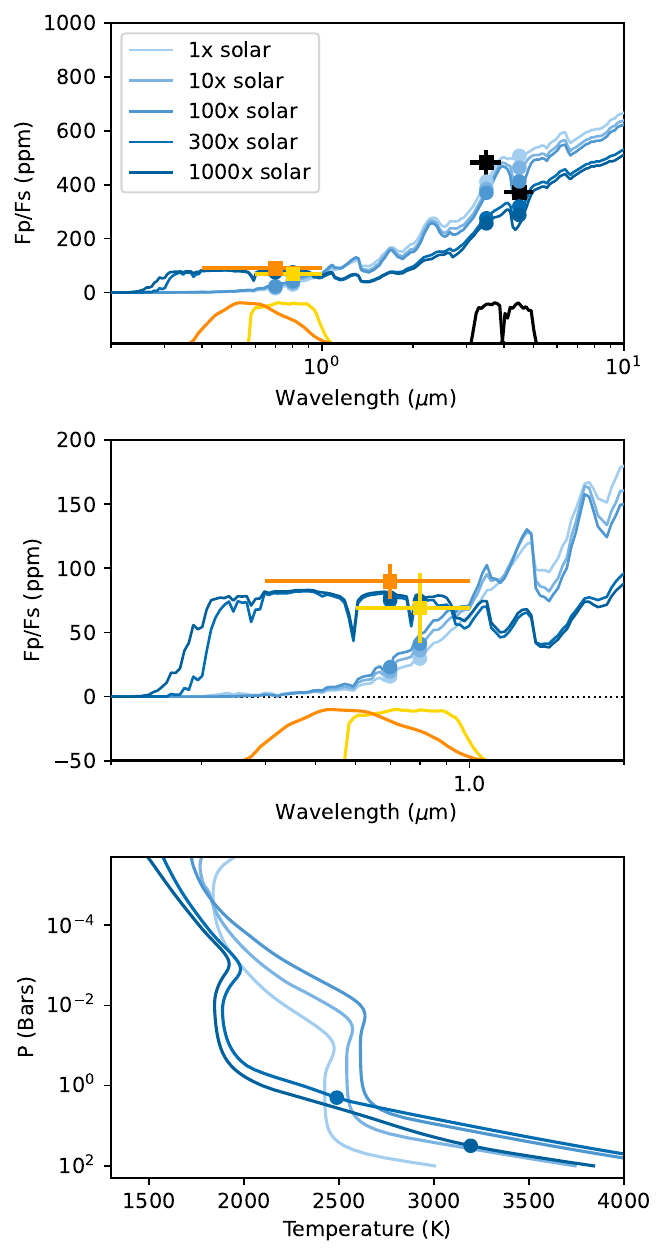}
    \caption{scCHIMERA radiative/convective/chemical equilibrium models where TiO/VO are assumed to have rained out of the atmosphere, along with the planet-to-star flux ratios from CHEOPS (orange), TESS (yellow), and Spitzer (black). The models are all for dayside-only redistribution (f=2) and varying metallicity. High albedo is obtained only at high metallicity through the condensation of $\mathrm{MgSiO_3}$. The base of the silicate clouds is shown as a dot in the PT profile. When no dots are shown, the silicate does not form. Kzz was set to $\mathrm{10^9 cm^2/s}$ and $f_{\rm sed}$ was set to 0.1, favoring small, more reflective particles.}
    \label{fig:spectra_NoTiO}
\end{figure}

\begin{figure}
    \centering
    \includegraphics[width=\linewidth]{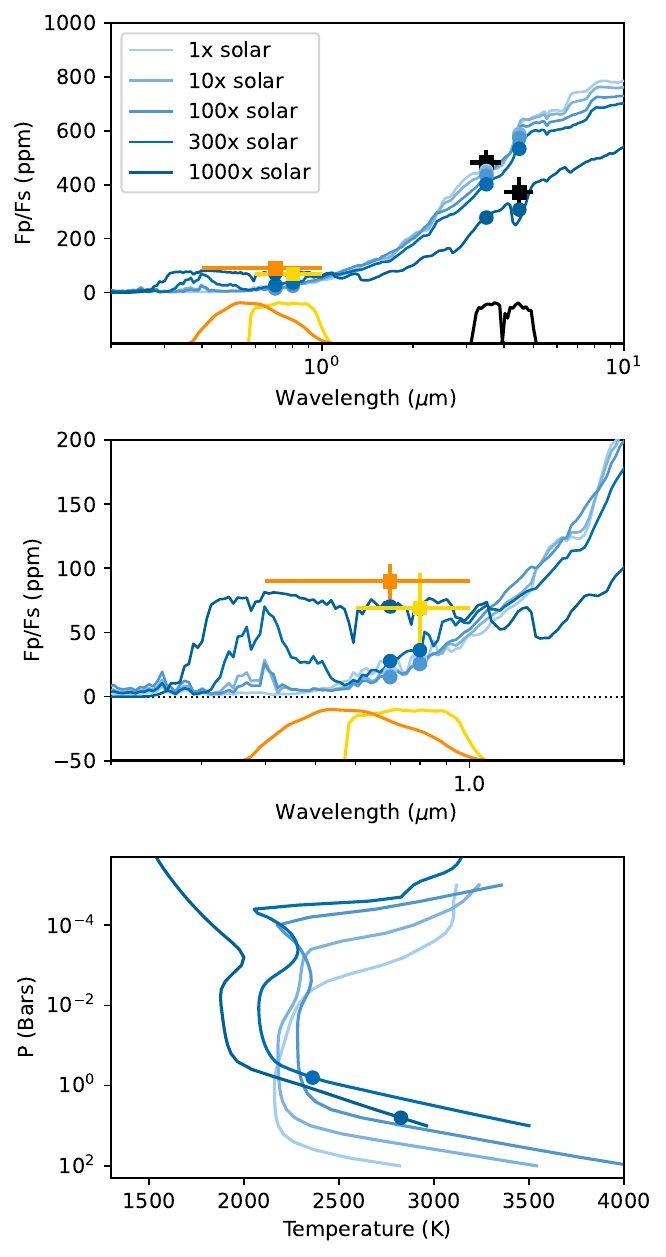}
    \caption{scCHIMERA radiative/convective/chemical equilibrium models. These models do include TiO and VO when local chemical equilibrium predicts their presence. The other model parameters and the the planet-to-star flux ratios shown are the same as in Figure~\ref{fig:spectra_NoTiO}}
    \label{fig:spectra_TiO}
\end{figure}

The models are shown in Figures~\ref{fig:spectra_NoTiO} and \ref{fig:spectra_TiO}, and they confirm that a strong reflected light component can only be achieved at high metallicity. We choose favorable parameters to form reflecting clouds by setting the diffusion parameter, $K_{zz}$, to a relatively high value of $\mathrm{10^9cm^2/s}$, whereas the sedimentation efficiency is set to a low value of $f_{\rm sed}=0.1$, favoring smaller and more reflective particles. 

In Figure~\ref{fig:spectra_TiO}, we show the thermal profiles and spectra for the chemical equilibrium case. We see that a thermal inversion is predicted in most cases. Even if some condensation happens in the atmosphere, a strong thermal inversion can be maintained due to $\mathrm{H_{2}O}$ thermal dissociation at low pressures. Only for the very high metallicities does the cloud condense in sufficient amount to cool sufficiently the atmosphere and remove all signs of thermal inversion. 

\begin{figure}
    \centering
    \includegraphics[width=\linewidth]{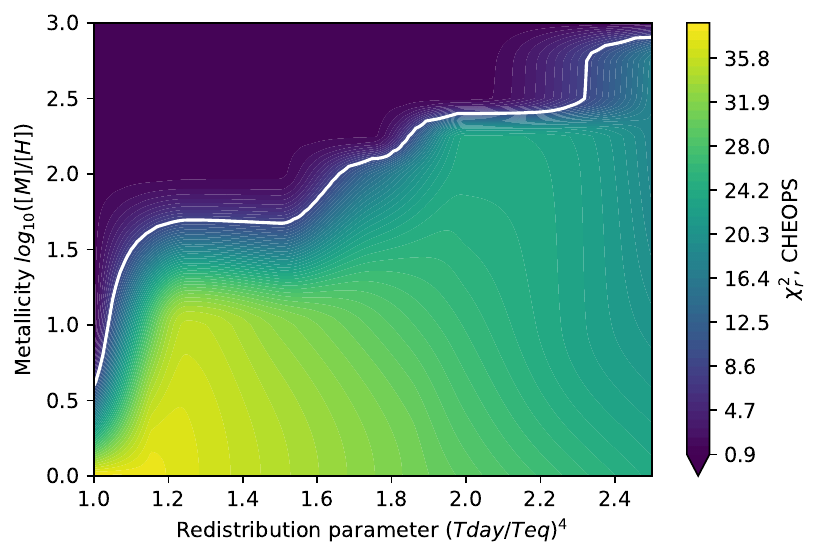}    \includegraphics[width=\linewidth]{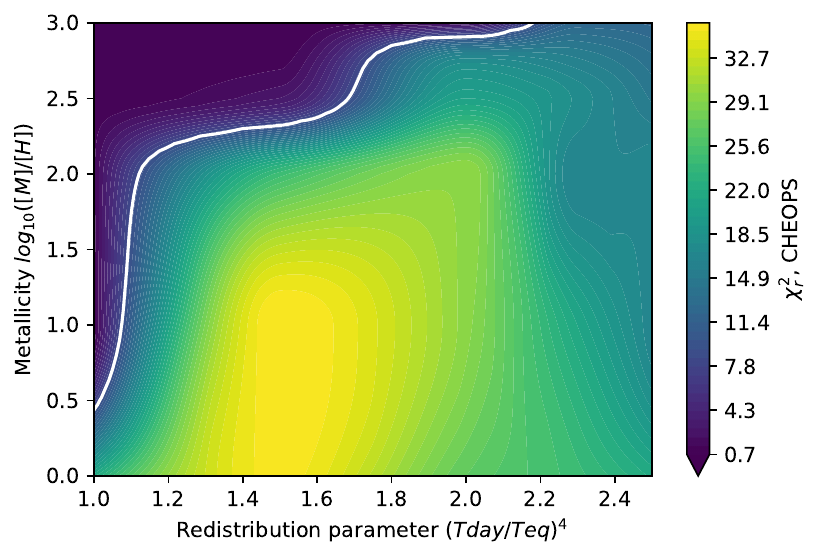}
    \caption{$\chi^2$ maps between our models and the CHEOPS datapoint for chemical equilibrium models (bottom) and models where TiO and VO have rained out of the atmosphere (top). The contour shows the region that agrees with the data within $3-\sigma$. The rainout of TiO/VO allows for an easier condensation of the silicate clouds and thus a high albedo at lower metallicities than for the case in chemical equilibrium.}
    \label{fig:ig::Xi2maps}
\end{figure}

As shown in Figure~\ref{fig:spectra_NoTiO}, if some external mechanisms, such as day/night cold trap (\citealp{Parmentier2013}), remove TiO and VO from the atmosphere, then no thermal inversion is predicted by the models. 

In both cases, our models show that silicate clouds can form between 1 and 10~bars (circles in the pressure temperature profiles of the bottom panel in Figure~\ref{fig:spectra_TiO}). The chosen value of $f_{\rm sed}$ leads to an efficient transport of clouds upwards, up to the photospheric pressures. However, the models have a hard time fitting both the high value of the CHEOPS data and the Spitzer points. In our models, when clouds form, their Bond albedo is too large, and the thermal profile becomes too cold to fit the Spitzer data. 

In order to investigate more thoroughly the possible range of metallicities that can allow for cloud formation, we extended our models to a grid of models with the same parameters, varying only the heat redistribution parameter and the metallicity of the planet. The reduced $\chi^2$ maps of the grid are shown in Figure~\ref{fig:ig::Xi2maps}. For the models in chemical equilibrium, we see that the model favors either a poor heat redistribution (e.g., f>2) and very high metallicities (>500x solar) or a very efficient heat transport (f<2) and a lower metallicity (<300x solar). However, both solutions are problematic. Heat transport in hot planets has been observed to be inefficient~\citep{2017PASP..129a4001S} due to their small radiative timescale (<10h at 0.1bar for LTT-9779b based on eq. 10 of ~\citep{2002A&A...385..166S}). Second, metallicity values larger than 500x solar correspond to a few percent or less of $\mathrm{H_2}$ in the atmosphere, which is complex to reconcile with the core-accretion formation paradigm.

For the case without TiO/VO, more reasonable values of metallicities are possible. In particular, silicate clouds can form for metallicities larger than 30$\times$ solar for efficient heat transport (f=1.2) and larger than 200x solar for dayside only heat redistribution (f=2).

The fact that our models can not reproduce both the optical and thermal measurements could be solved by exploring a wider range of scattering phase functions for the particles. Indeed, a phase function with strong backscattering could, in principle, allow for a high geometric albedo without too strong a Bond albedo. In other words, it would allow an increase in the optical flux without decreasing the temperature of the atmosphere. However, regardless of the scattering phase function that could be tuned to match the optical data, clouds do have to form on the planet's dayside, and thus our conclusion of high-metallicity does not depend on the assumption about scattering

Our models consider the averaged state of the atmosphere. Recent JWST phase curve observations by~\citet{2025arXiv250114016C} show that the cloud cover was inhomogeneous over the planet's dayside, with a western limb having a somewhat larger albedo than the eastern limb. However, as shown in their Fig. 3, the data are compatible with clouds all across the dayside, with variation in their albedo. In such a case, our approach, considering the averaged dayside thermal profile, remains valid. We additionally note that clouds in such hot planets are very unlikely to remain far from their equilibrium because their evaporation timescales are extremely small~\citep[see Fig. 10 of][]{2003A&A...399..297W}. As a consequence, the conditions to form on LTT-9779b have to remain on the dayside, even if the clouds first form on the cooler nightside. Finally, the retrieved thermal profiles from~\citet{2025arXiv250114016C} are barely cold enough on the dayside to condense out silicate clouds when solar metallicity is assumed. Our conclusion that the metallicity is large, based on our radiative/convective framework rather than a retrieval, appears to be robust when confronted with the current JWST results of LTT-977b.

\section{Discussion and Conclusions}\label{sec:sec5}

This work is aimed at constraining the physical and chemical properties of the extreme Neptunian desert planet LTT9779b, using secondary eclipse observations taken using CHEOPS. First, we used the TESS observations from three sectors, i.e., sectors 1, 29, and 69, to reanalyze the transit properties and update all physical parameters, including the ephemeris. Since these parameters are directly used for modeling the secondary eclipse observations, as well as emission and transmission spectroscopic modeling of the atmospheric properties, they are essential updates. Given that LTT9779b is one of the most unique and interesting targets for a plethora of future follow-up studies, these updated parameters will be beneficial not only for the present work but also for several future studies. Our analysis of the transit lightcurves included a robust algorithm focused on reducing the impact of various noise components in the parameter estimations. This included techniques such as wavelet denoising and GP regression, which can effectively reduce the time-uncorrelated and correlated noise components in the lightcurves.

In the next step, our ten new secondary eclipse observations from CHEOPS were analyzed, in combination with ten already published observations from CHEOPS, which were previously reported by our team \citet{2023A&A...675A..81H}. Apart from using the reduced time-series data provided by the official CHEOPS data reduction pipeline (DRP), we have also reduced the raw data using a completely independent pipeline, PIPE, which used the PSF photometry method. The roll-angle-dependent background contaminations in the CHEOPS lightcurves were modeled using an N-order glint function implemented within PYCHEOPS, and subtracted from the lightcurves simultaneously with the secondary eclipse modeling. The two sets of detrended lightcurves and the estimated secondary eclipse depths, obtained through the two independent data reduction procedures, were found to be in excellent statistical agreement. Thus, it helped us to cross-verify the robustness of our analyses, which is essential for this type of study involving the detection of small and noise-sensitive signatures from observed data. Our analysis with the extended dataset helped us to constrain the secondary eclipse depth of LTT9779b with significantly higher precision than was reported previously in \citet{2023A&A...675A..81H}, which also resulted in a more robust and precise estimation of its geometric albedo.

Armed with this new and precise eclipse measurement, we were able to re-run our atmospheric models with more realistic physical processes included, all of which confirmed the presence of a high-metallicity atmosphere for the planet.  Furthermore, we found difficulties when trying to model the highly reflective optical output and the strong thermal NIR output simultaneously, and we suggest possible ways forward to circumvent these issues.  However, regardless of whether more detailed models are applied, the atmosphere has to be significantly enriched in metals when compared to the Sun, and likely the planet hosts a high-altitude silicate cloud layer. These new constraints help further ellucidate the formation and evolution history of this planet, since hosting a metallic atmosphere with silicate clouds that reflect away 70\% of the infalling radiation likely helps to maintain the planet cooler than would be expected if it was as opaque as the vast majority of hot Jupiters for example (\citealp{2023A&A...675A..81H,radica24}).  Furthermore, as discussed in \citet{2024ApJ...962L..19V}, a highly metallic atmosphere argues for a smaller atmospheric scale height, decreasing the evaporation rate whilst increasing the cooling rate of any atmospheric outflow.  Such cooling processes could help explain why the planet has not lost all of its atmosphere at this time, becoming another bare hot rock super-Earth. As the first planet of its type to be studied in such detail, LTT9779b continues to be the benchmark example with which new physical processes are being teased out from the detailed follow-up observations that can be performed to study its atmosphere.

\begin{acknowledgements}
We thank the anonymous reviewer for their valuable suggestions, which helped improve the manuscript. SS acknowledges Fondo Comité Mixto-ESO Chile ORP 025/2022 to support this research. The computations presented in this work were performed using the Geryon-3 supercomputing cluster, which was assembled and is maintained using funds provided by the ANID-BASAL Center FB210003, Center for Astrophysics and  Associated Technologies, CATA. This paper includes data collected by the TESS mission, which are publicly available from the Mikulski Archive for Space Telescopes (MAST). I acknowledge the use of public TOI Release data from pipelines at the TESS Science Office and at the TESS Science Processing Operations Center. Funding for the TESS mission is provided by NASA’s Science Mission directorate. Support for MAST is provided by the NASA Office of Space Science via grant NNX13AC07G and by other grants and contracts. This paper includes data collected by the CHEOPS mission. CHEOPS is an ESA mission in partnership with Switzerland with important contributions to the payload and the ground segment from Austria, Belgium, France, Germany, Hungary, Italy, Portugal, Spain, Sweden, and the United Kingdom. The CHEOPS Consortium would like to gratefully acknowledge the support received by all the agencies, offices, universities, and industries involved. Their flexibility and willingness to explore new approaches were essential to the success of this mission. CHEOPS data analyzed in this article will be made available in the CHEOPS mission archive (https://cheops.unige.ch/archive$\_$browser/).
\end{acknowledgements}

\bibliography{ms}{}

\onecolumn
\begin{appendix}

\section{Figures showing all modeled lightcurves}

\begin{figure*}[!h]
	\centering
    \begin{tabular}{cc}
	\includegraphics[width=0.5\linewidth]{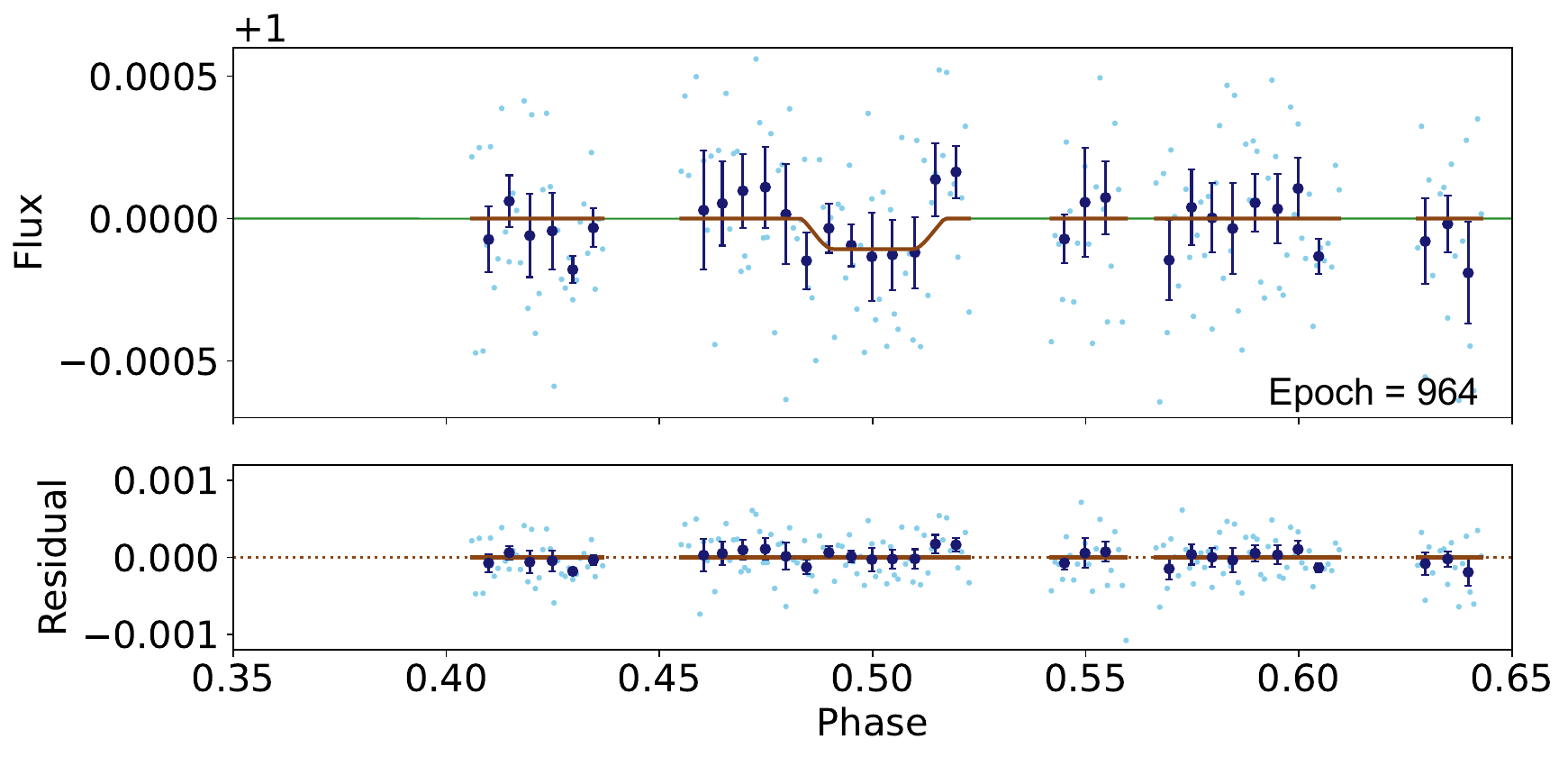} & \includegraphics[width=0.5\linewidth]{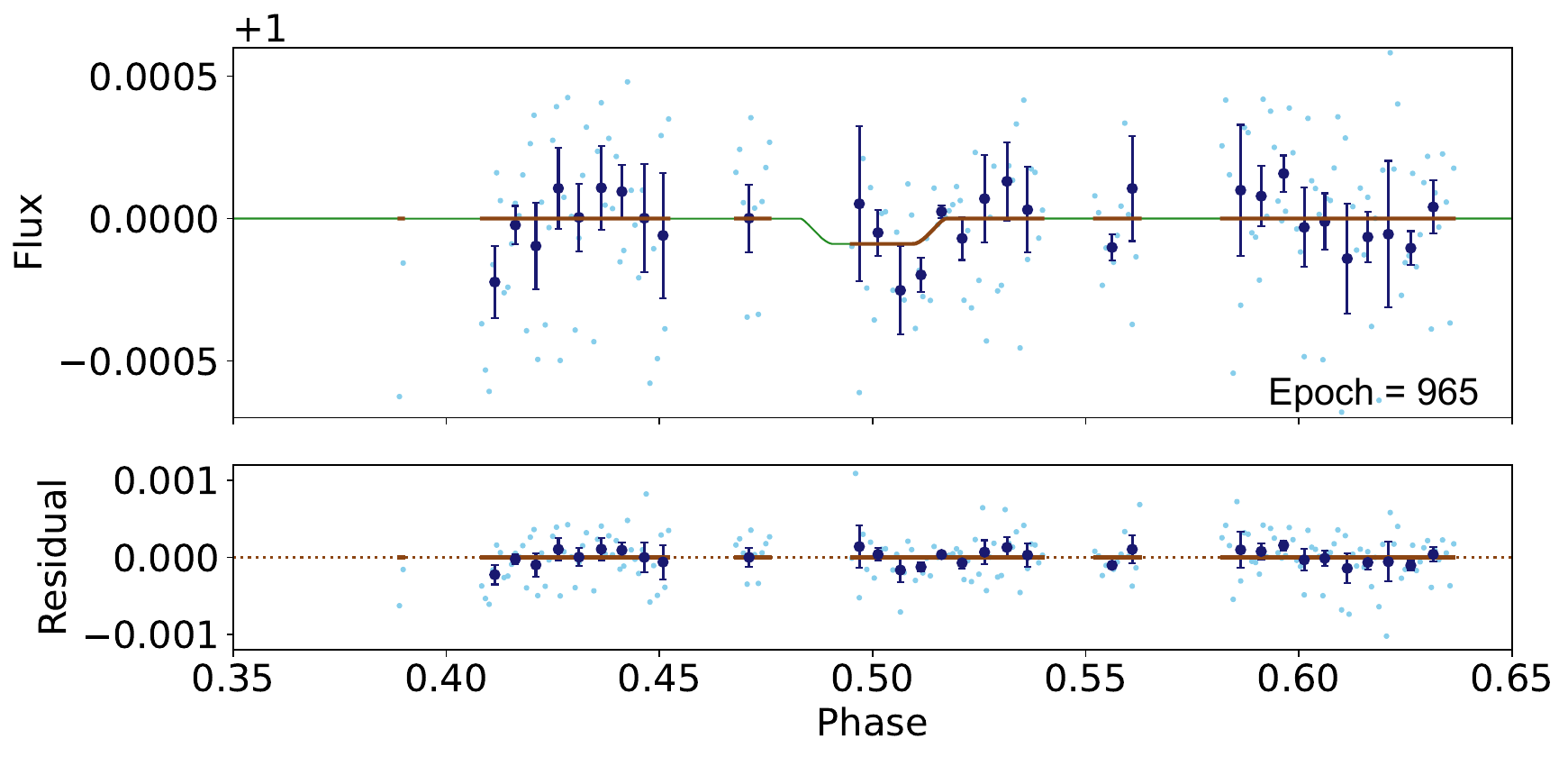} \\
    \includegraphics[width=0.5\linewidth]{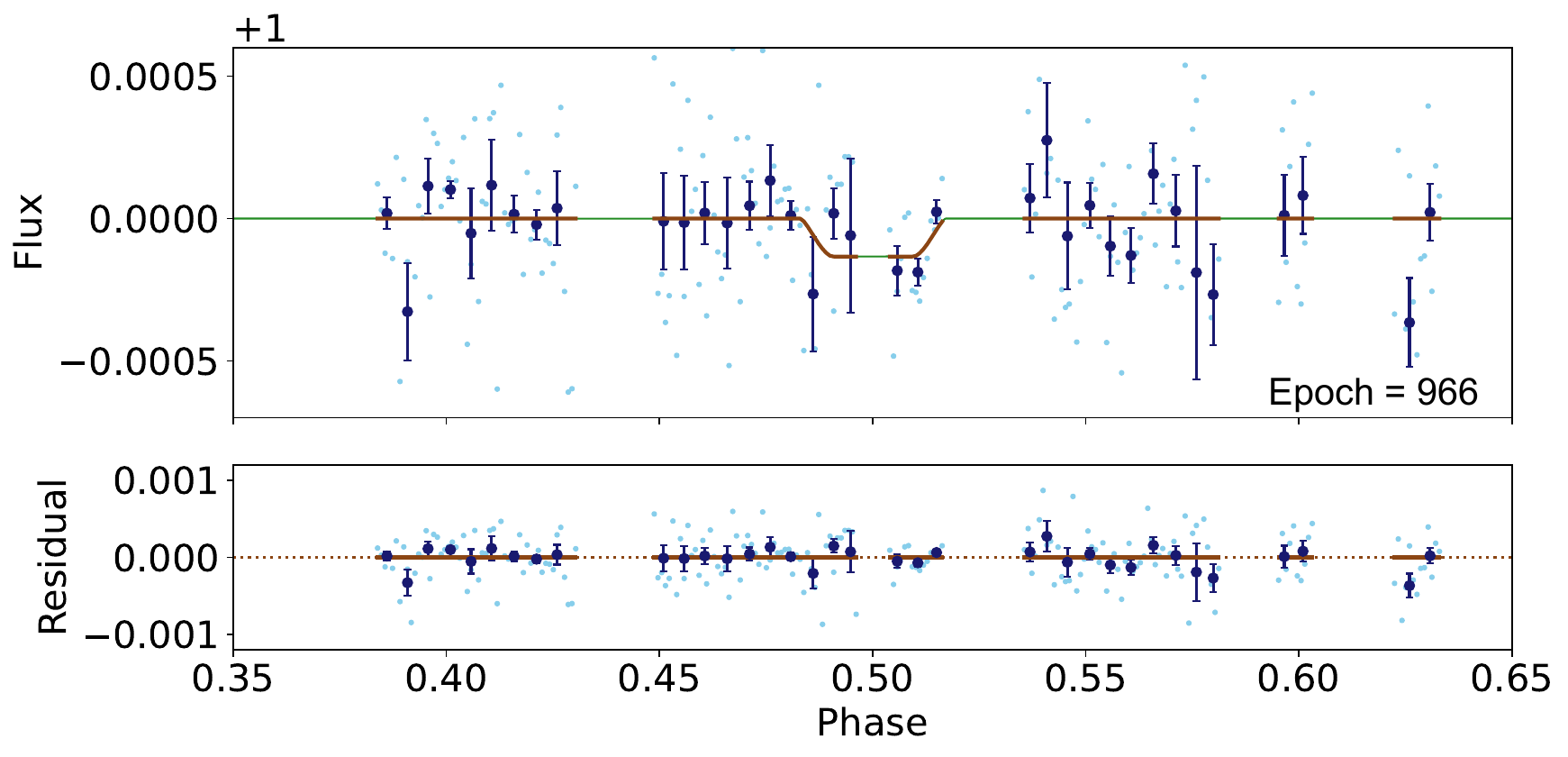} & \includegraphics[width=0.5\linewidth]{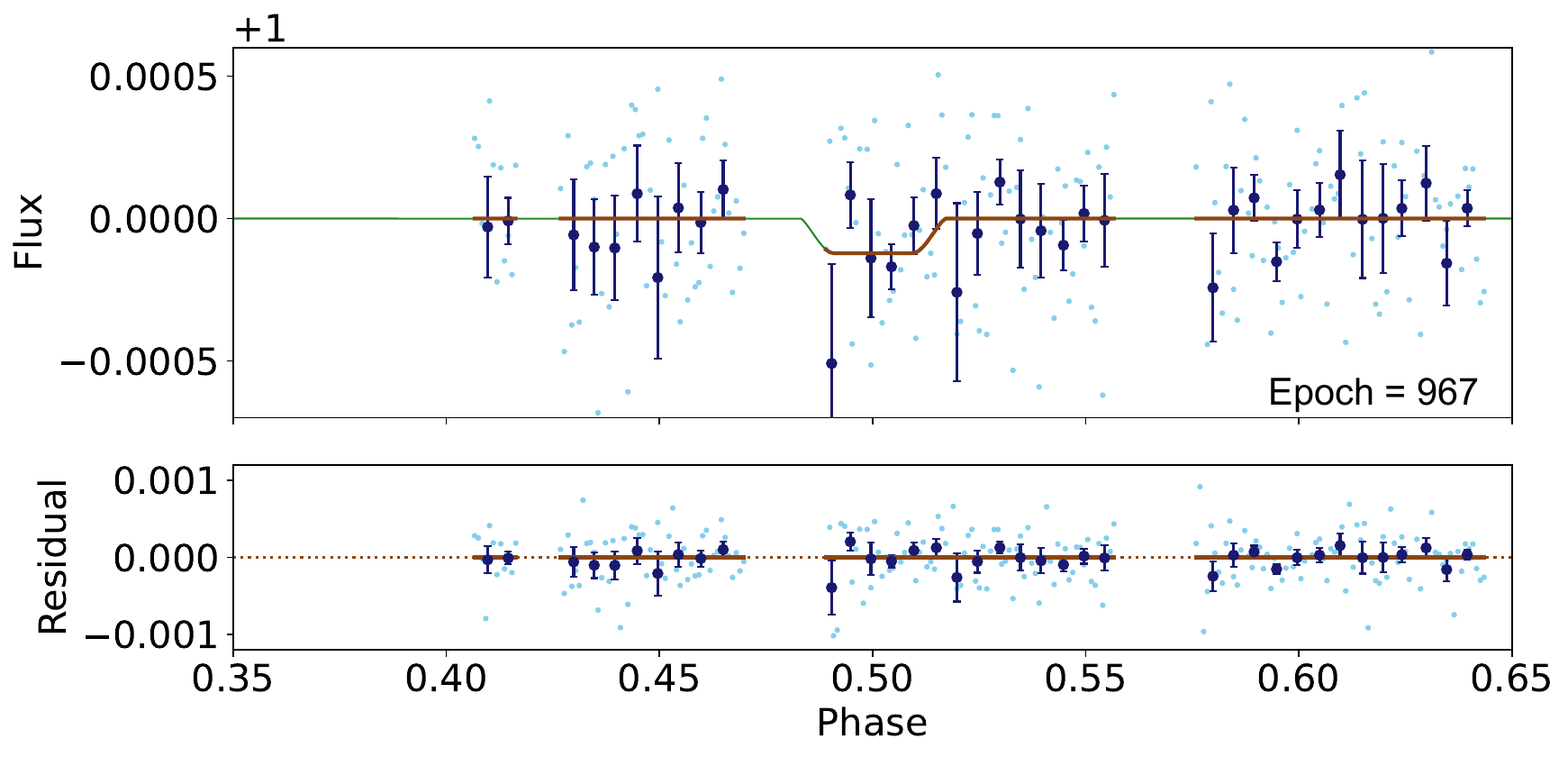} \\
    \includegraphics[width=0.5\linewidth]{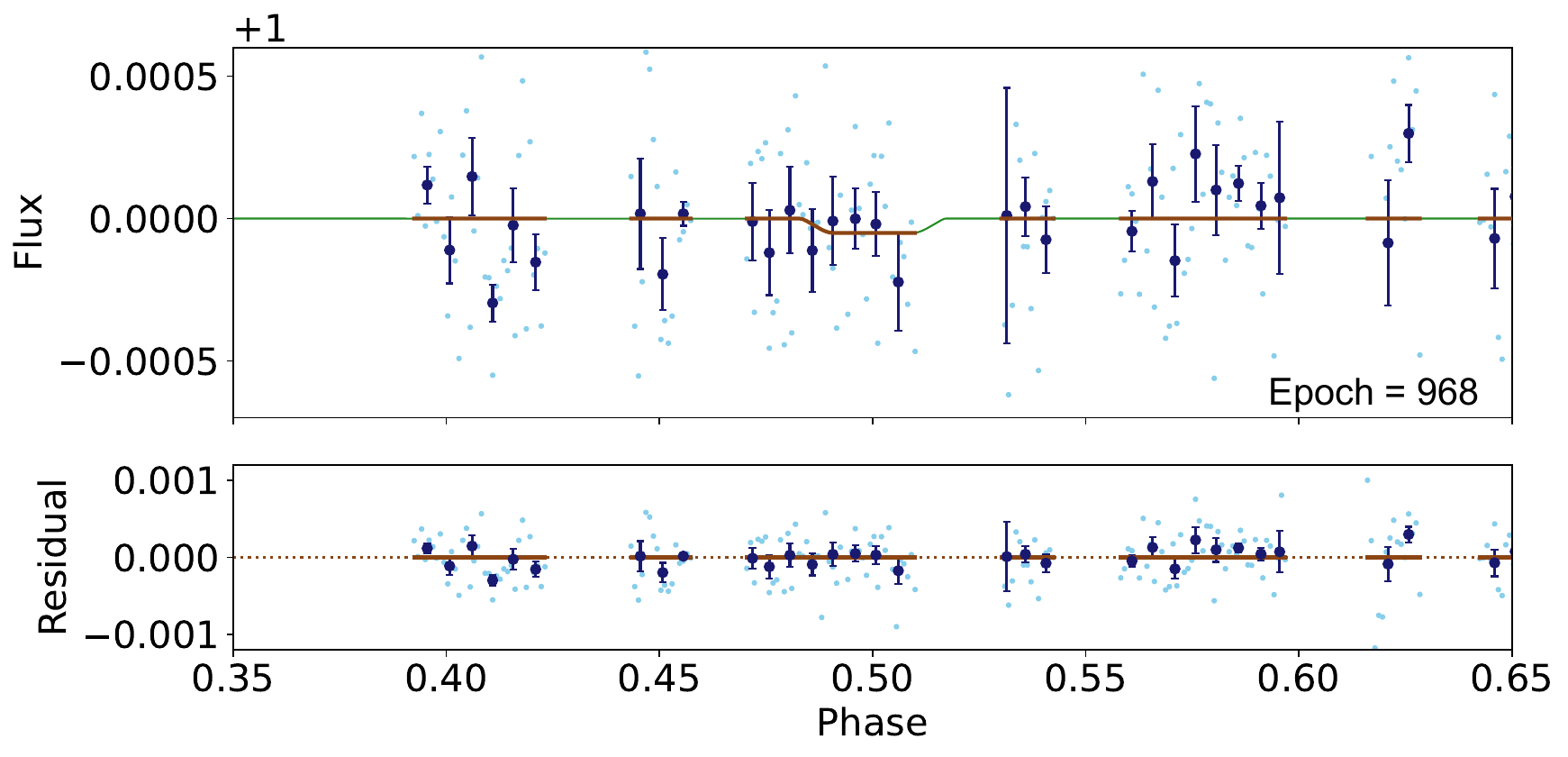} & \includegraphics[width=0.5\linewidth]{F306.pdf} \\
    \includegraphics[width=0.5\linewidth]{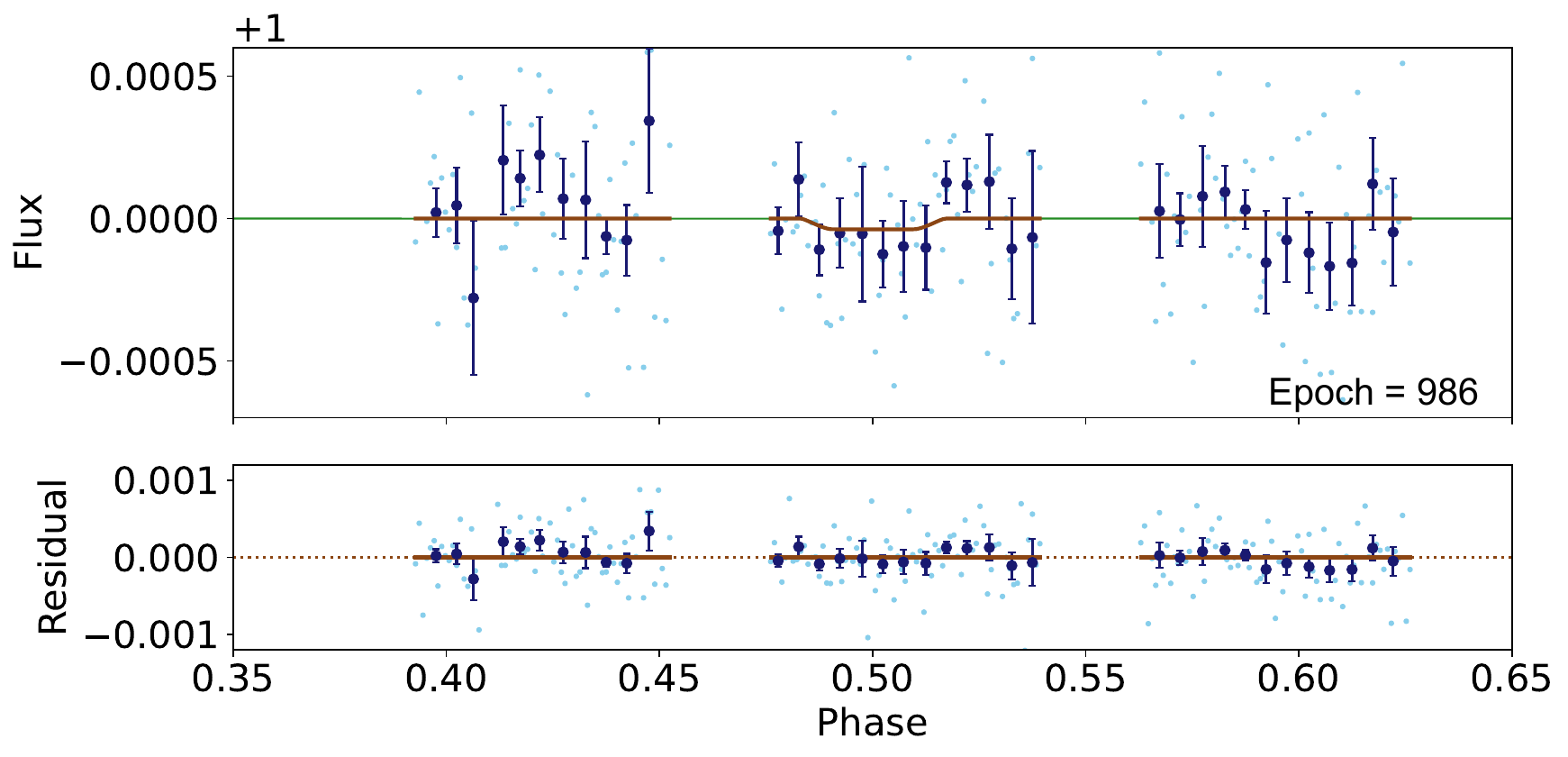} & \includegraphics[width=0.5\linewidth]{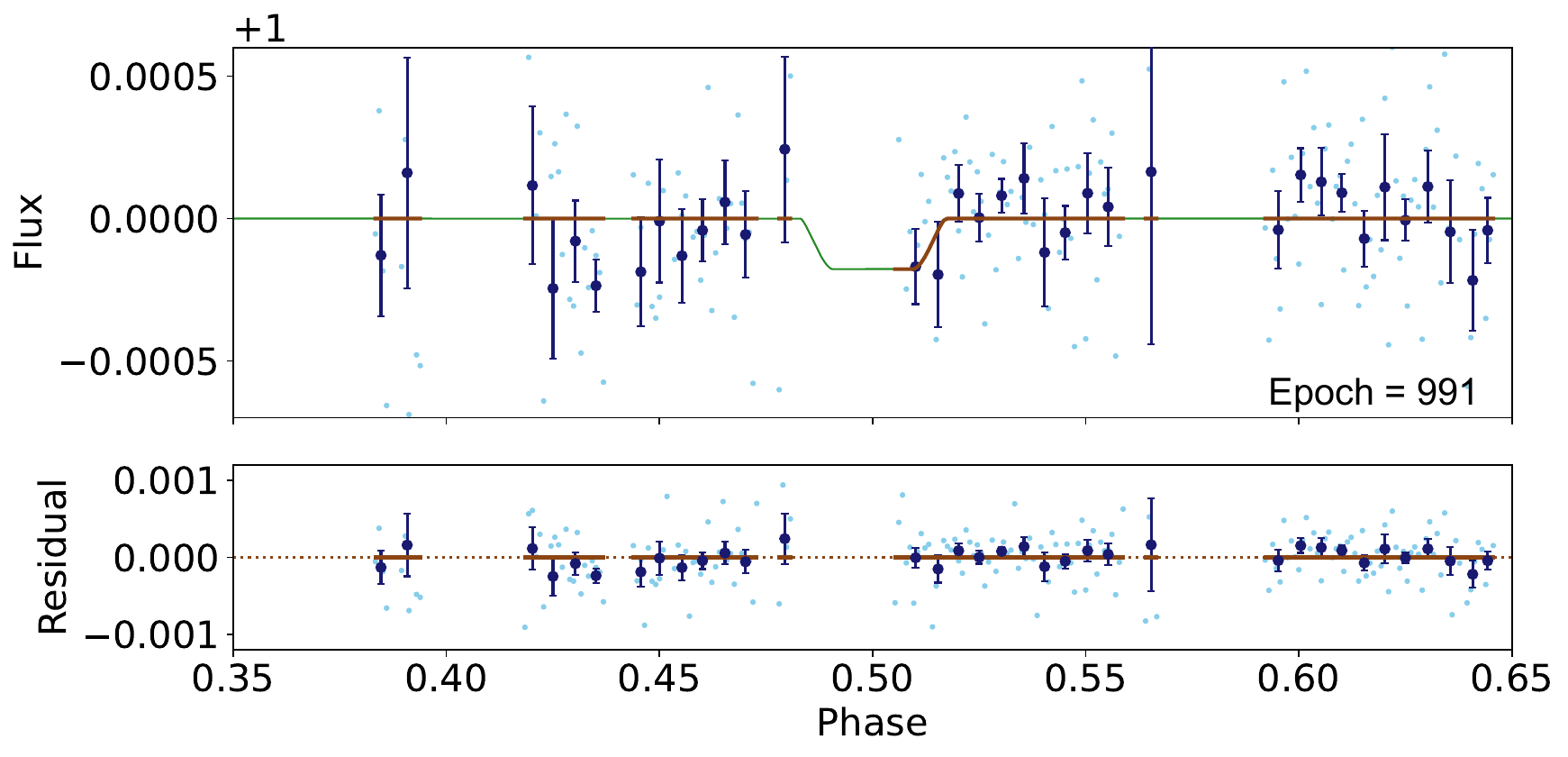} \\
    \includegraphics[width=0.5\linewidth]{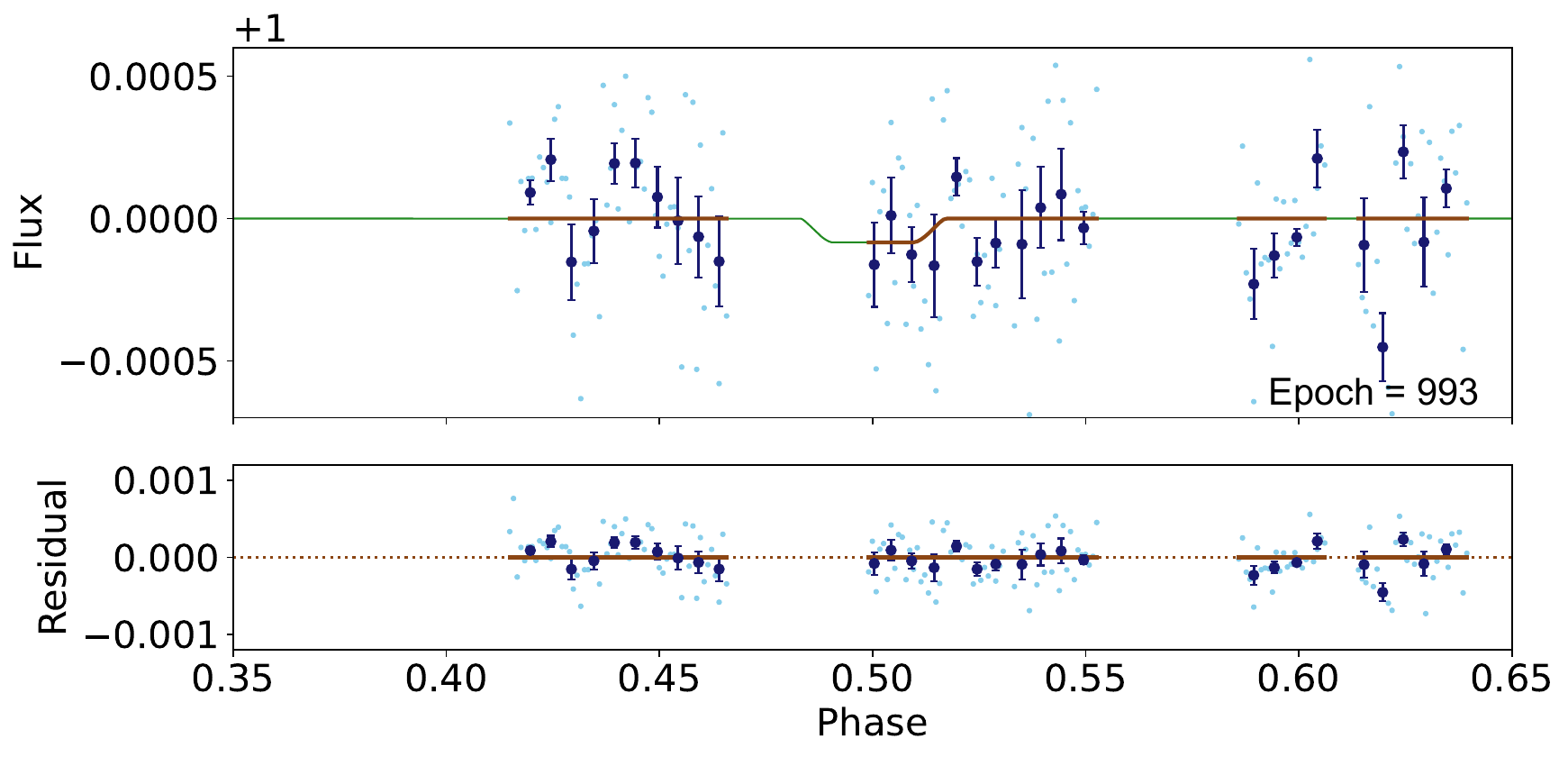} & \includegraphics[width=0.5\linewidth]{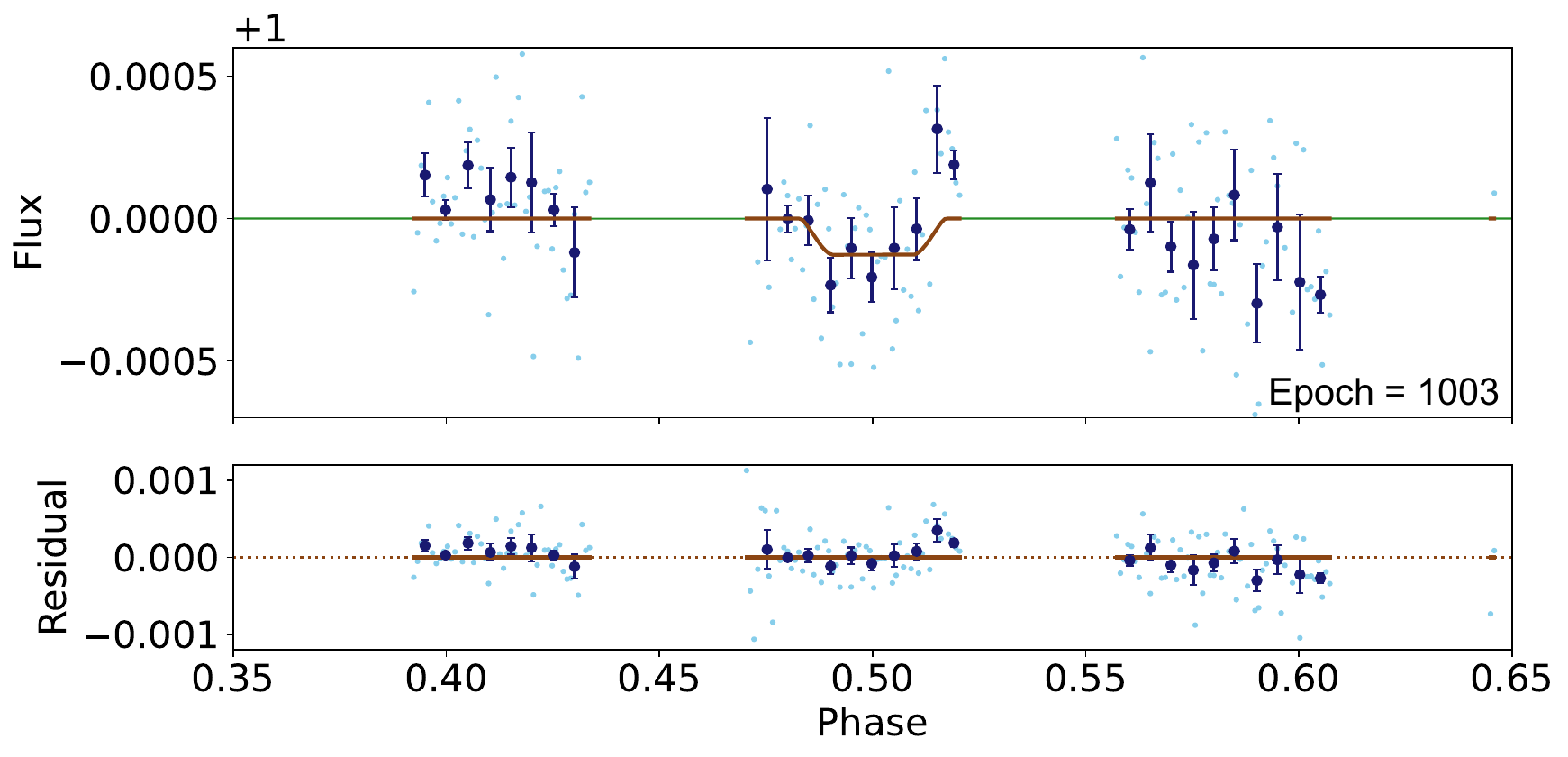} \\
    \end{tabular}
	\caption{Same as Figure \ref{fig:fig2}, but for all the lightcurves reduced by DRP (part 1 of 2).}
	\label{fig:fig41}
\end{figure*}

\addtocounter{figure}{-1}

\begin{figure*}
	\centering
    \begin{tabular}{cc}
    \includegraphics[width=0.5\linewidth]{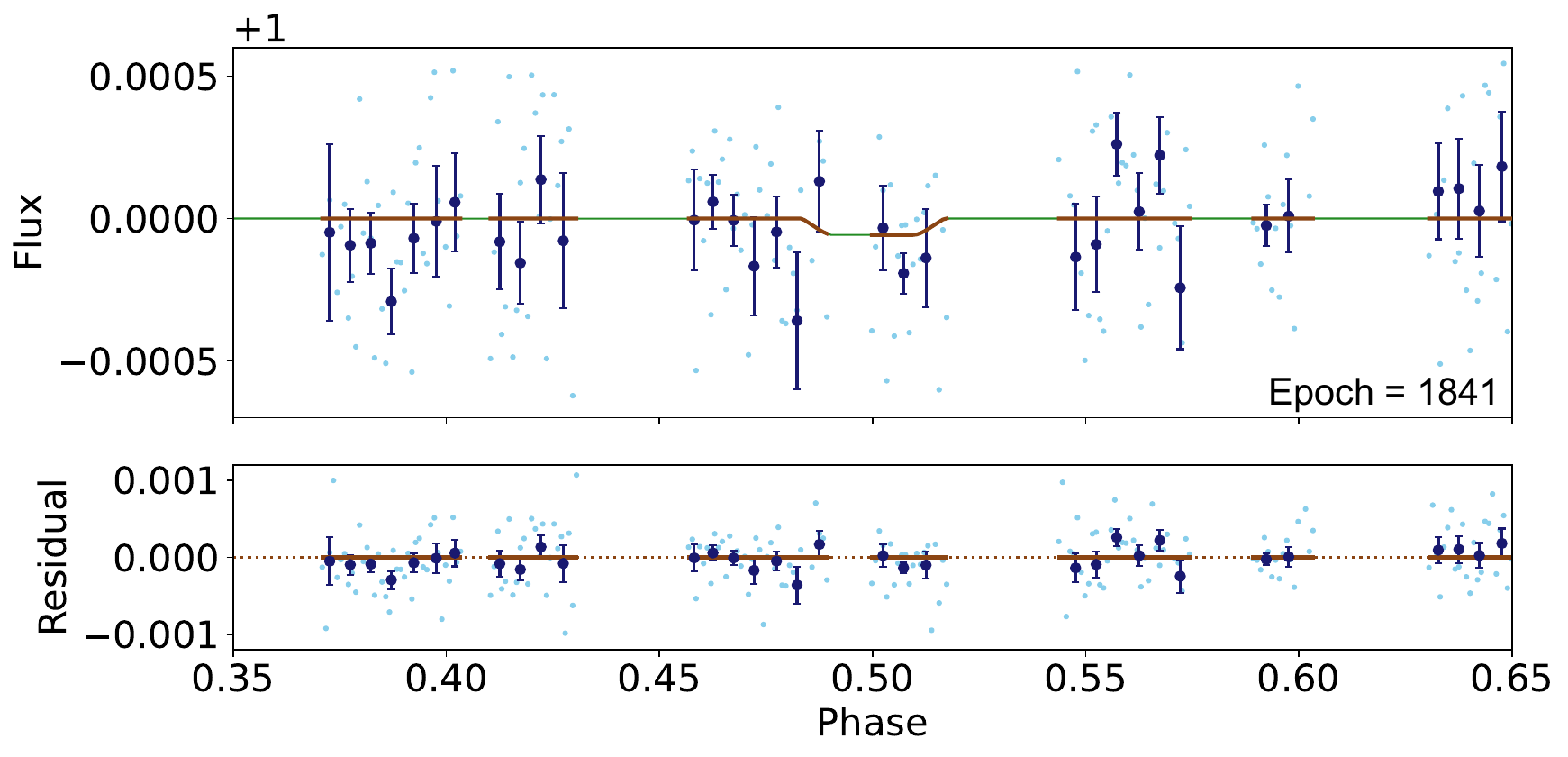} & \includegraphics[width=0.5\linewidth]{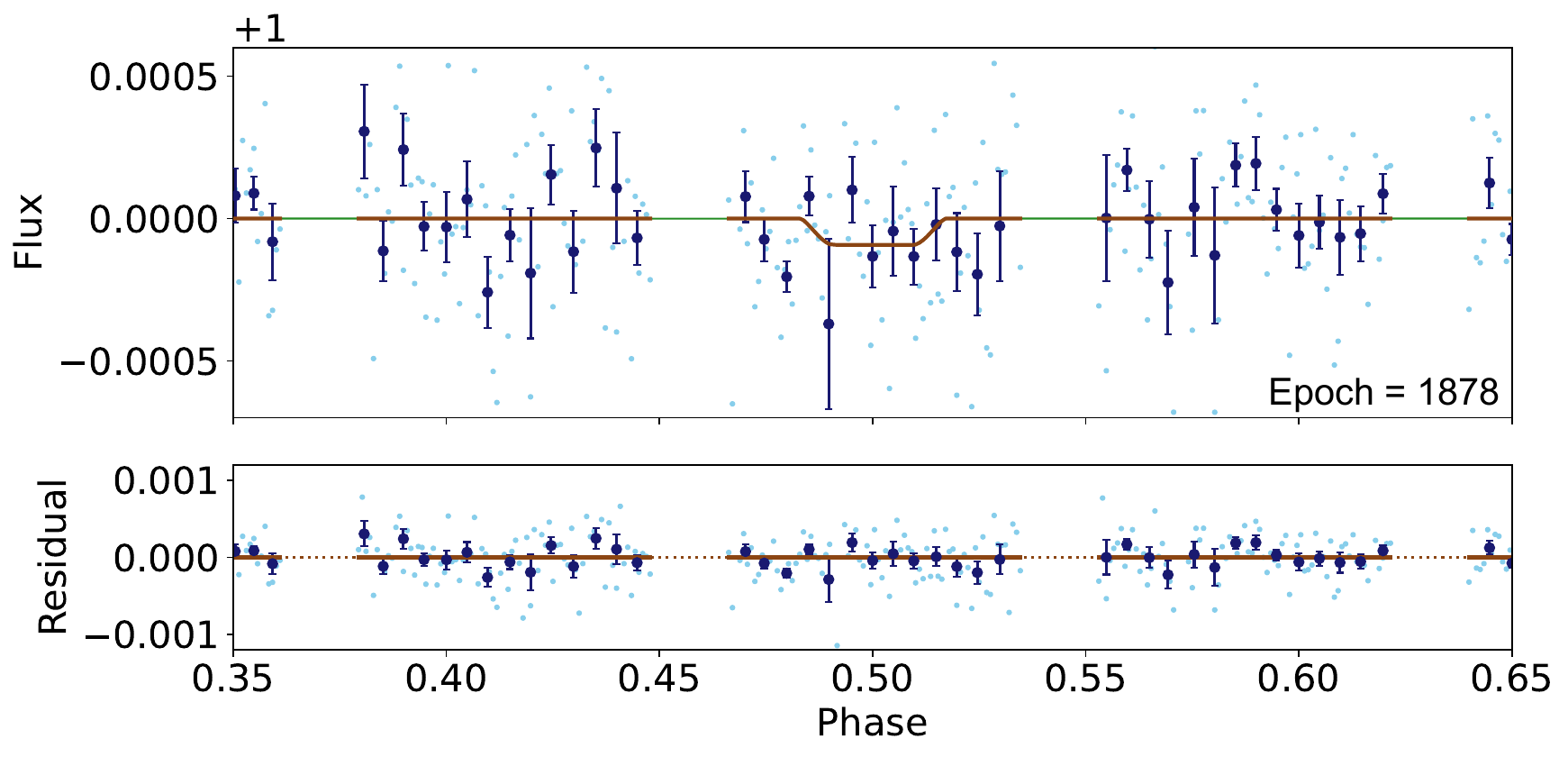} \\
    \includegraphics[width=0.5\linewidth]{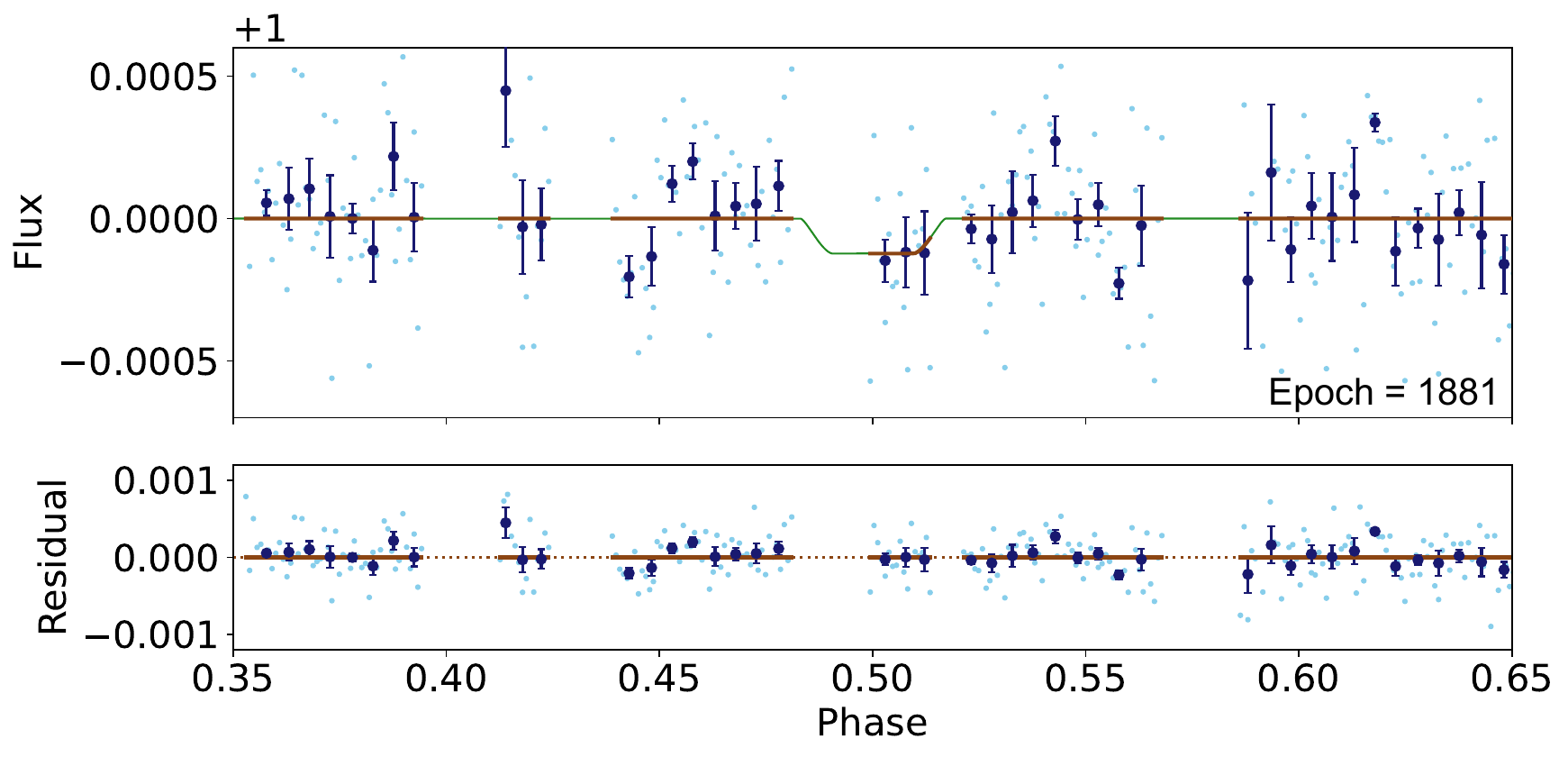} & \includegraphics[width=0.5\linewidth]{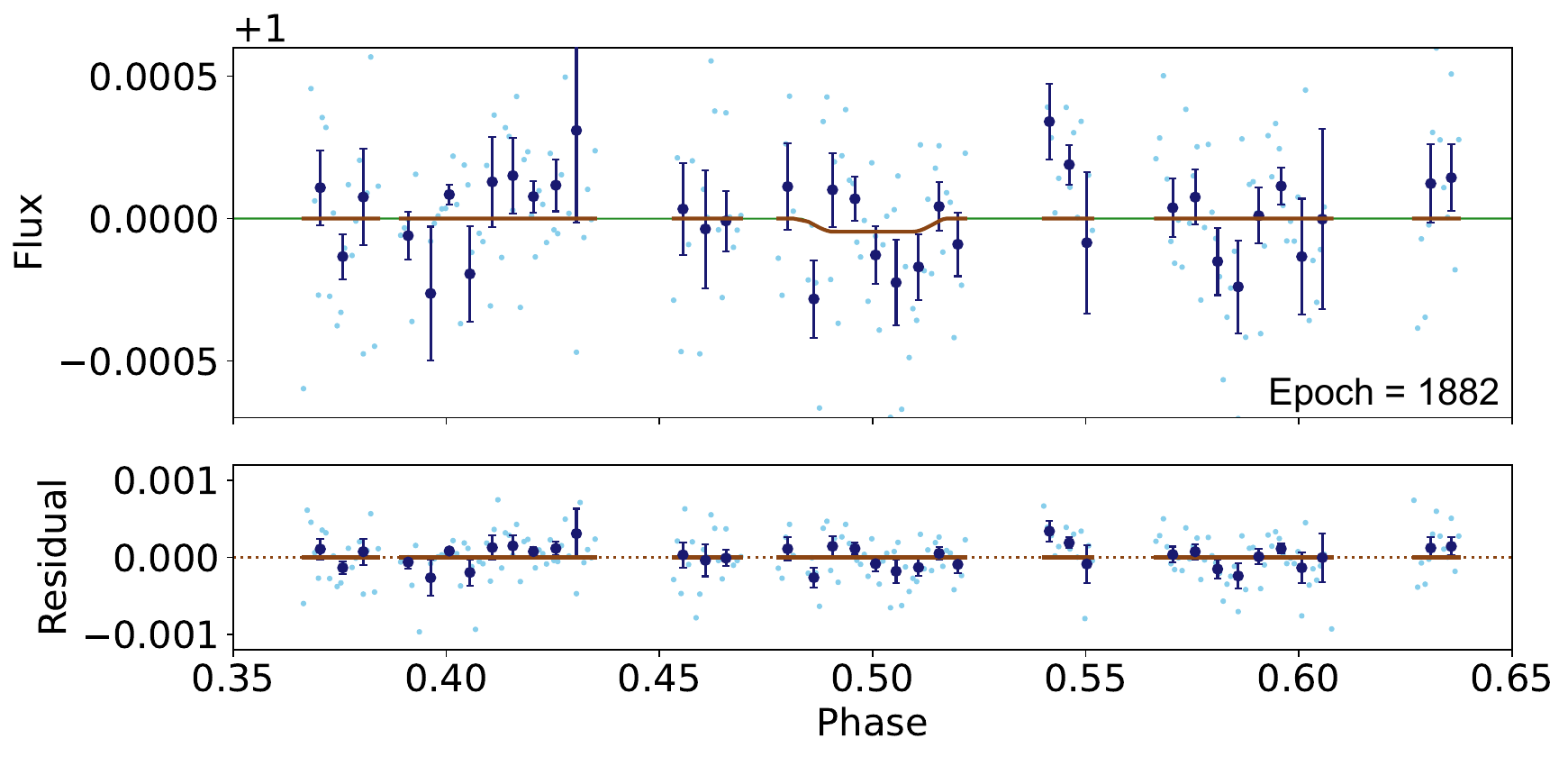} \\
    \includegraphics[width=0.5\linewidth]{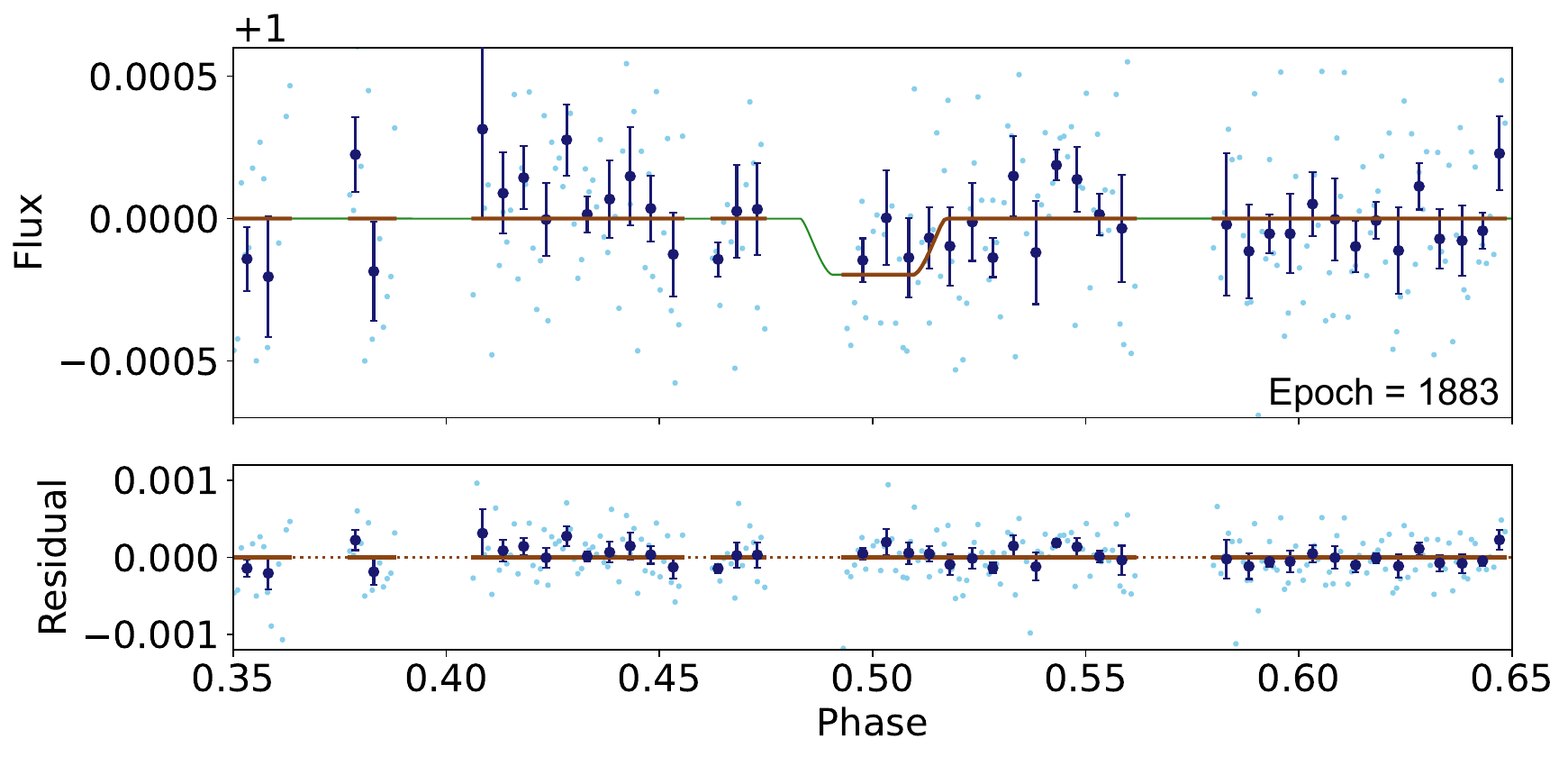} & \includegraphics[width=0.5\linewidth]{F316.pdf} \\
    \includegraphics[width=0.5\linewidth]{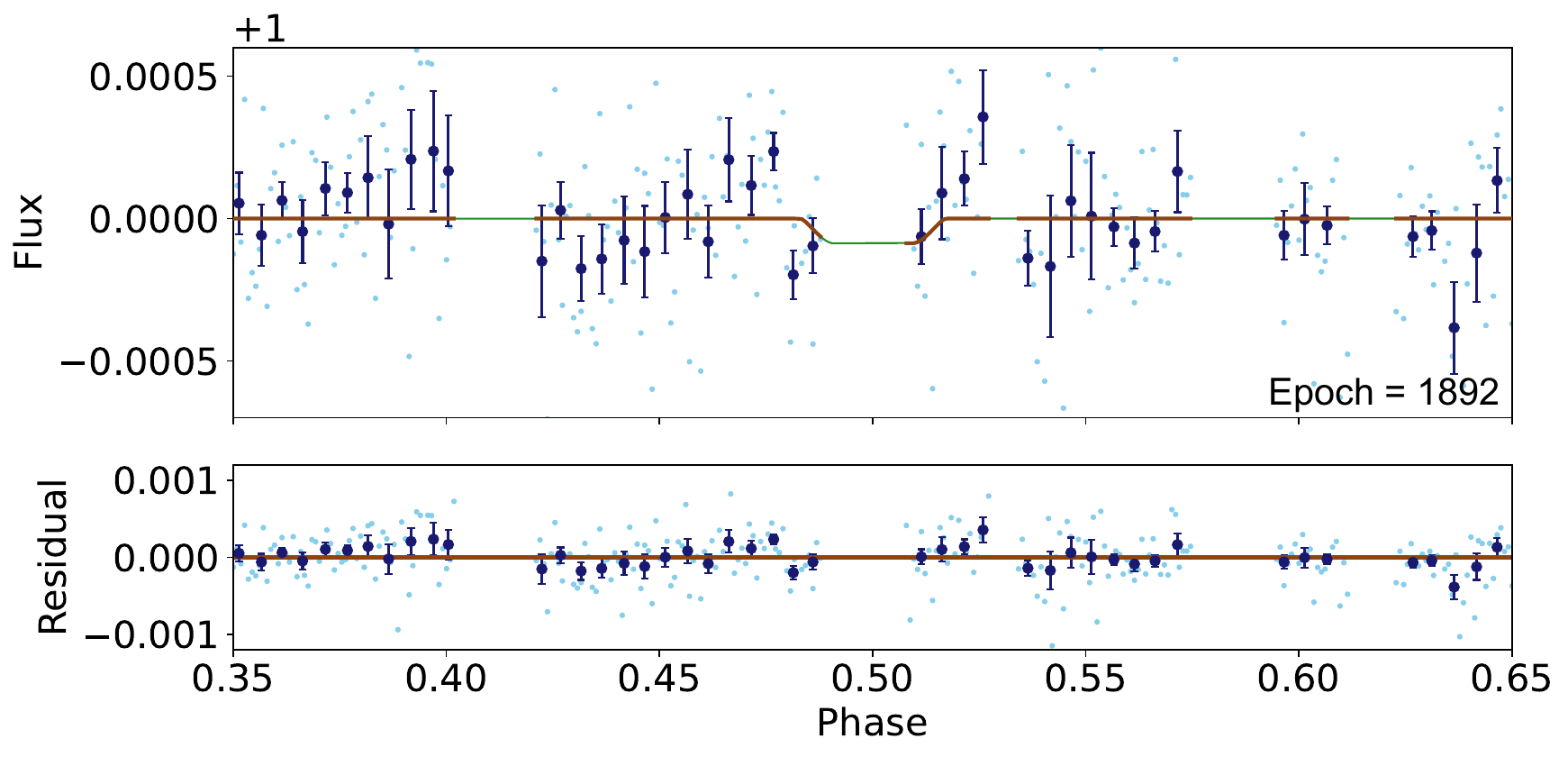} & \includegraphics[width=0.5\linewidth]{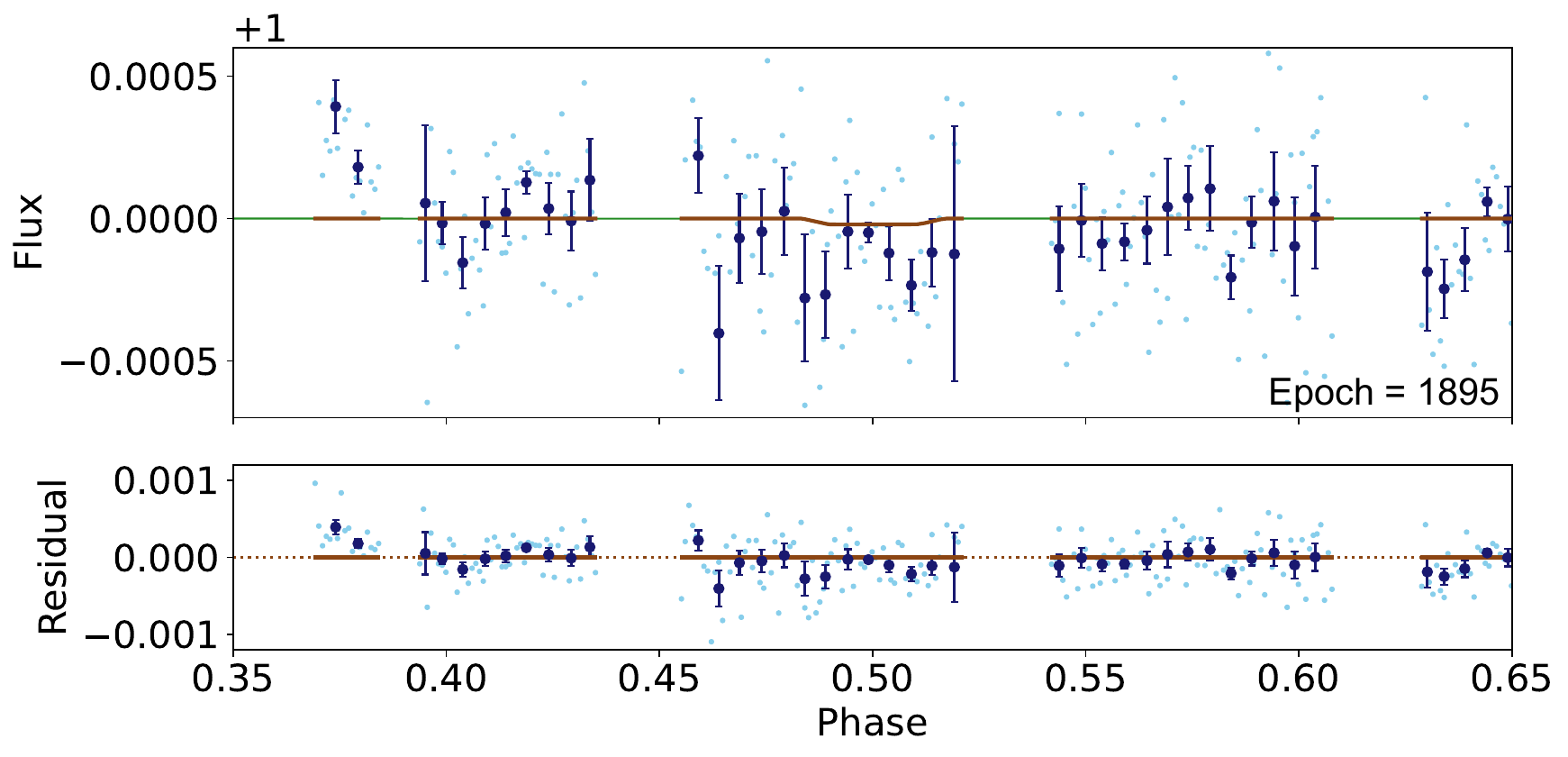} \\
    \includegraphics[width=0.5\linewidth]{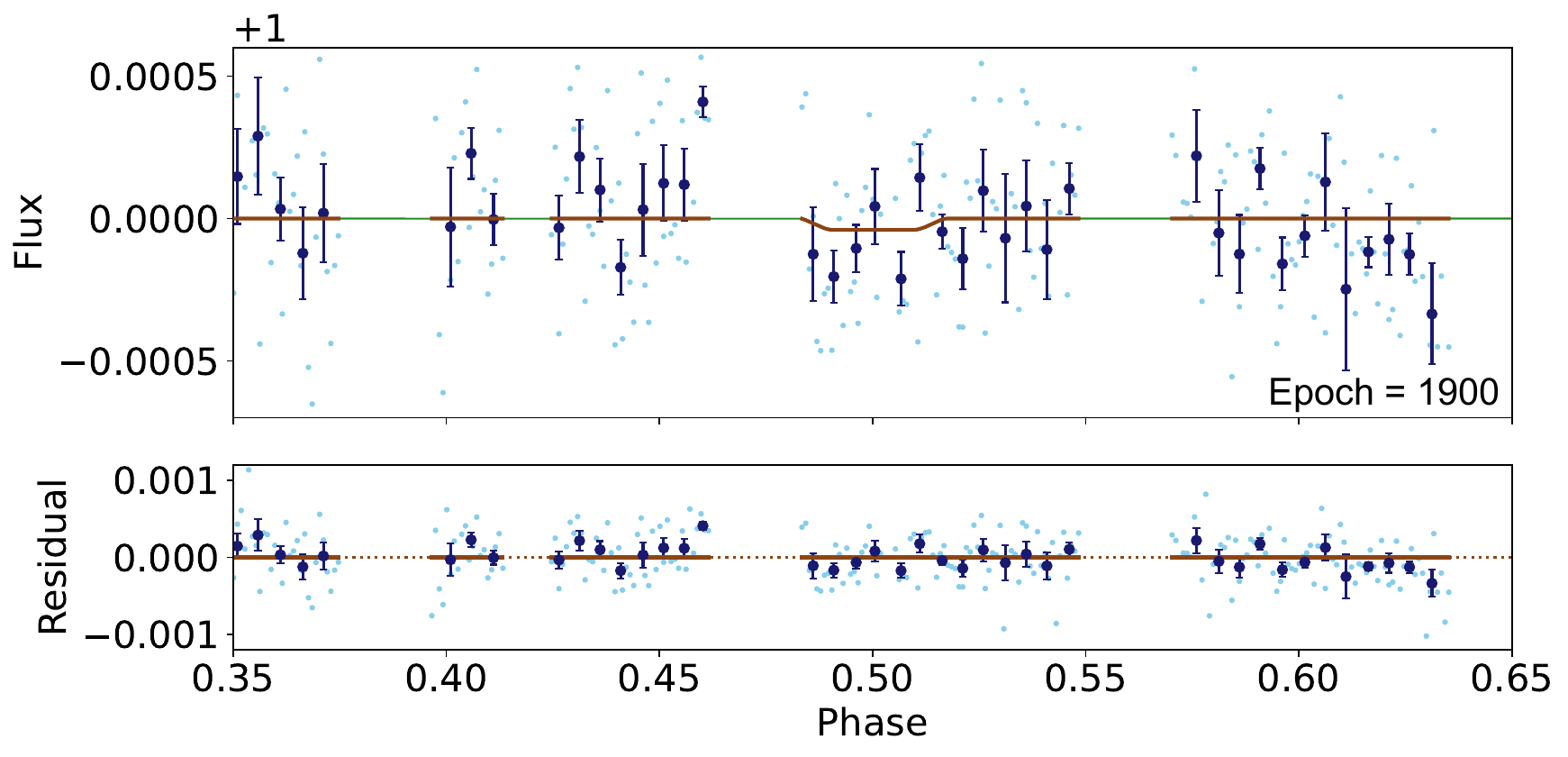} & \includegraphics[width=0.5\linewidth]{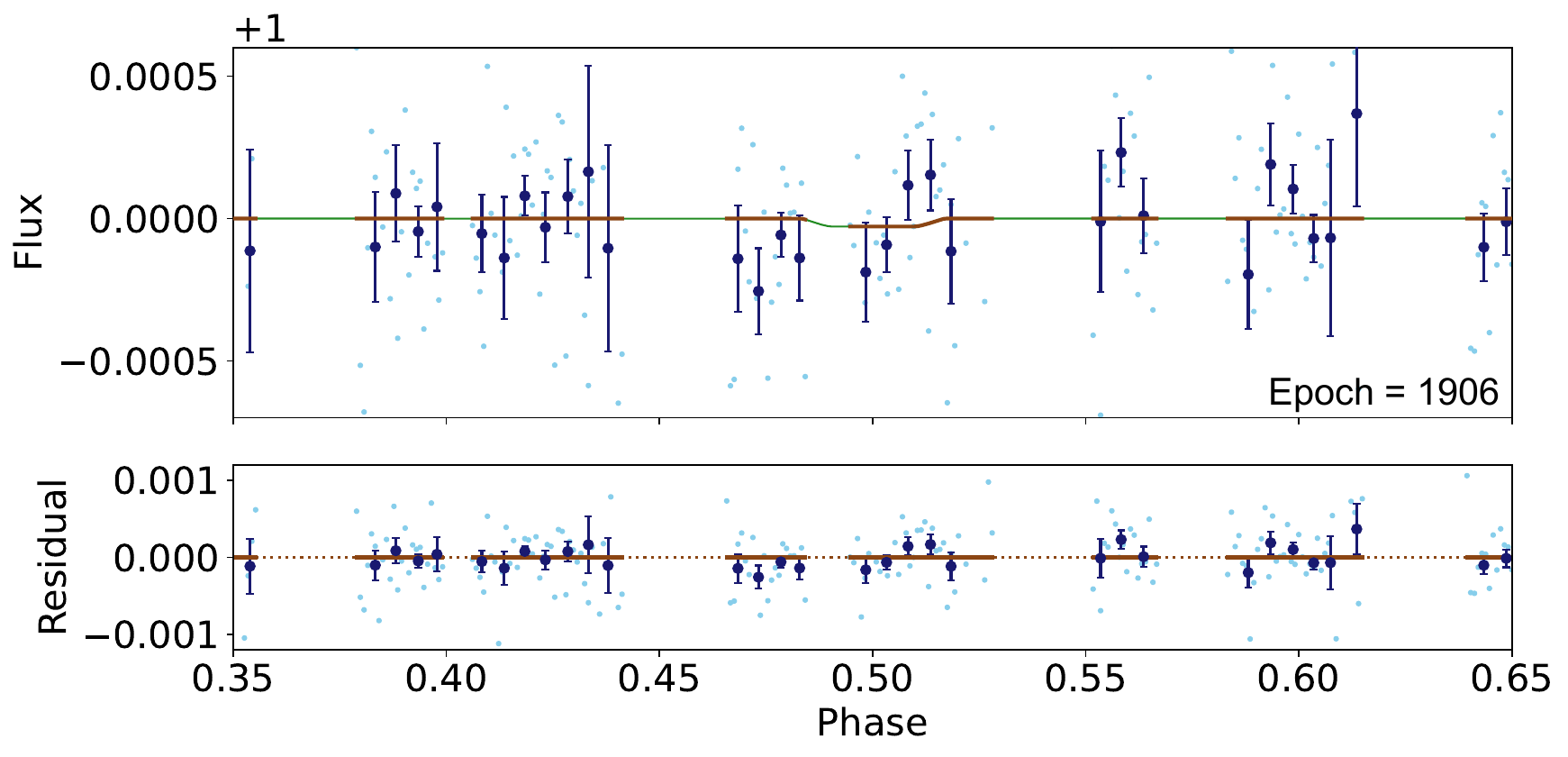} \\
    \end{tabular}
	\caption{(continued, part 2 of 2).}
	\label{fig:fig42}
\end{figure*}

\begin{figure*}
	\centering
    \begin{tabular}{cc}
	\includegraphics[width=0.5\linewidth]{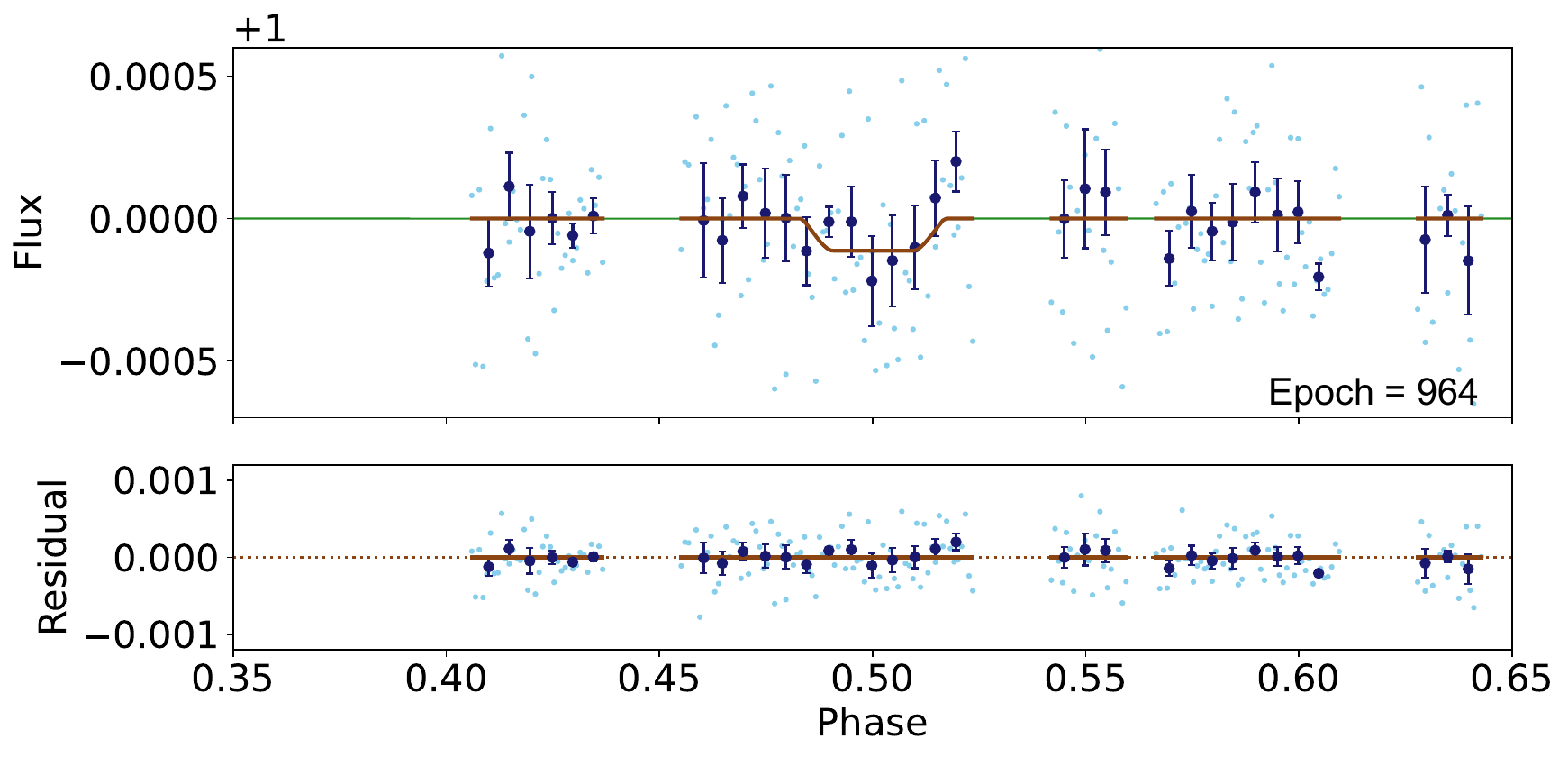} & \includegraphics[width=0.5\linewidth]{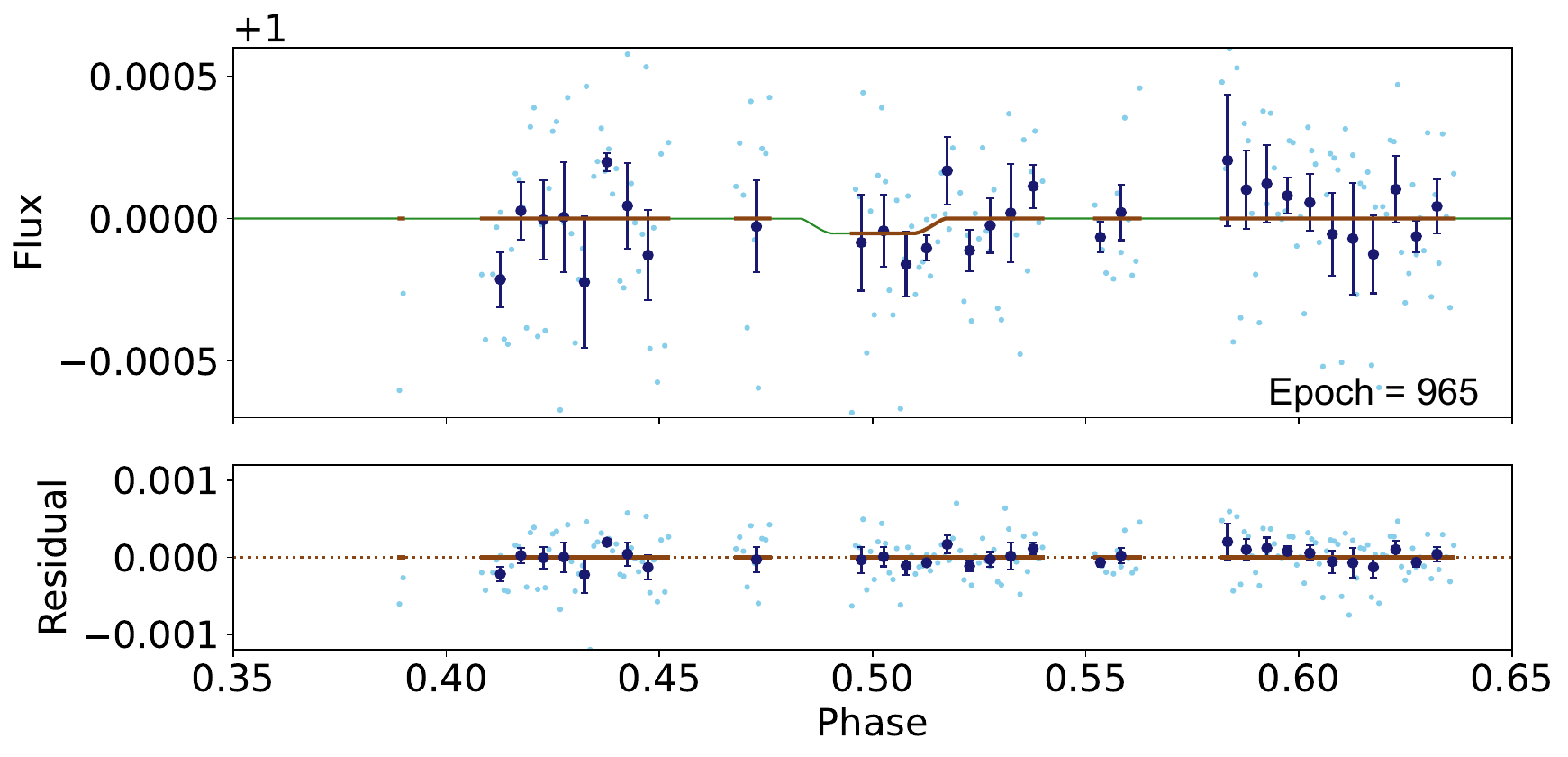} \\
    \includegraphics[width=0.5\linewidth]{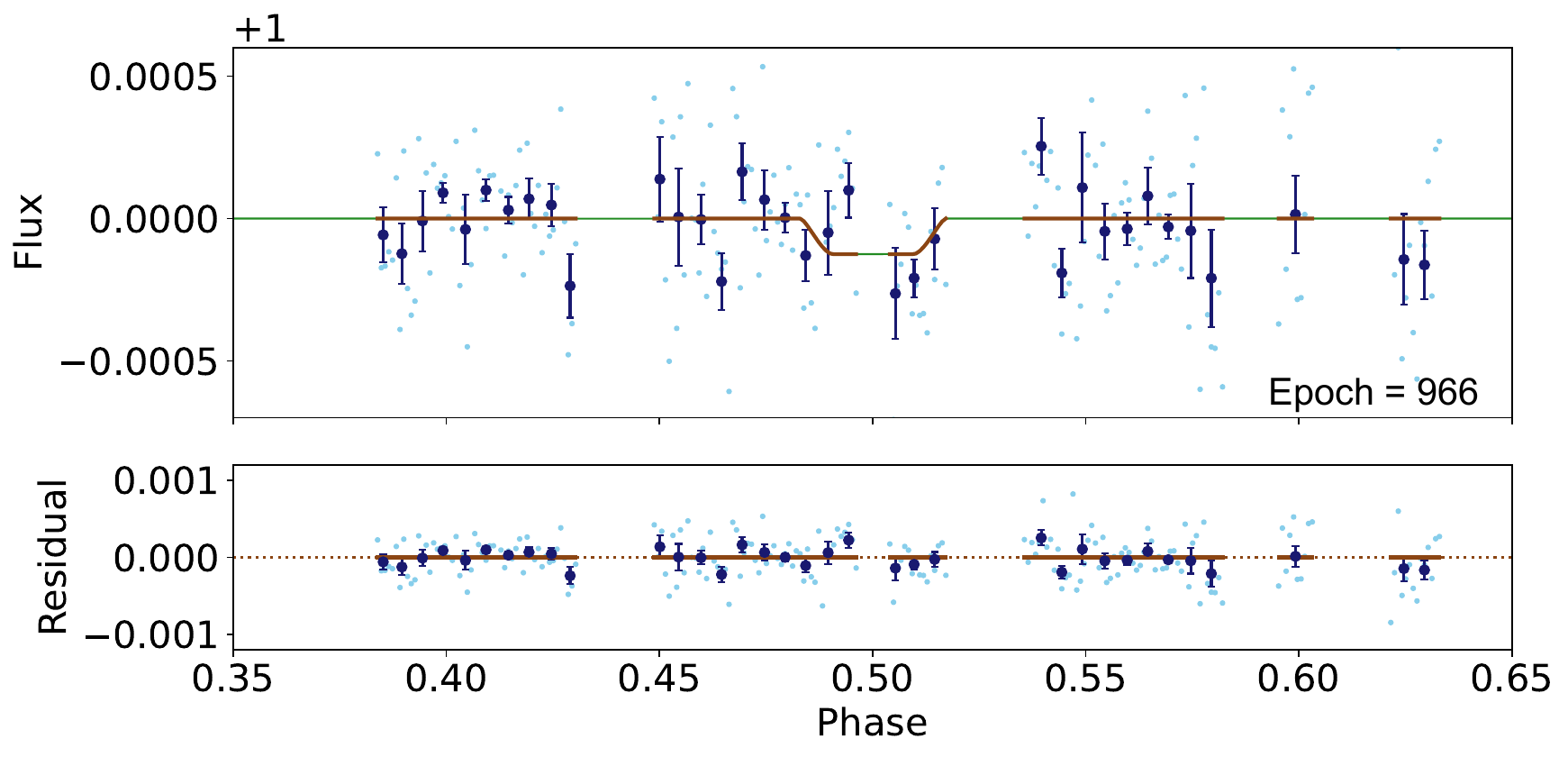} & \includegraphics[width=0.5\linewidth]{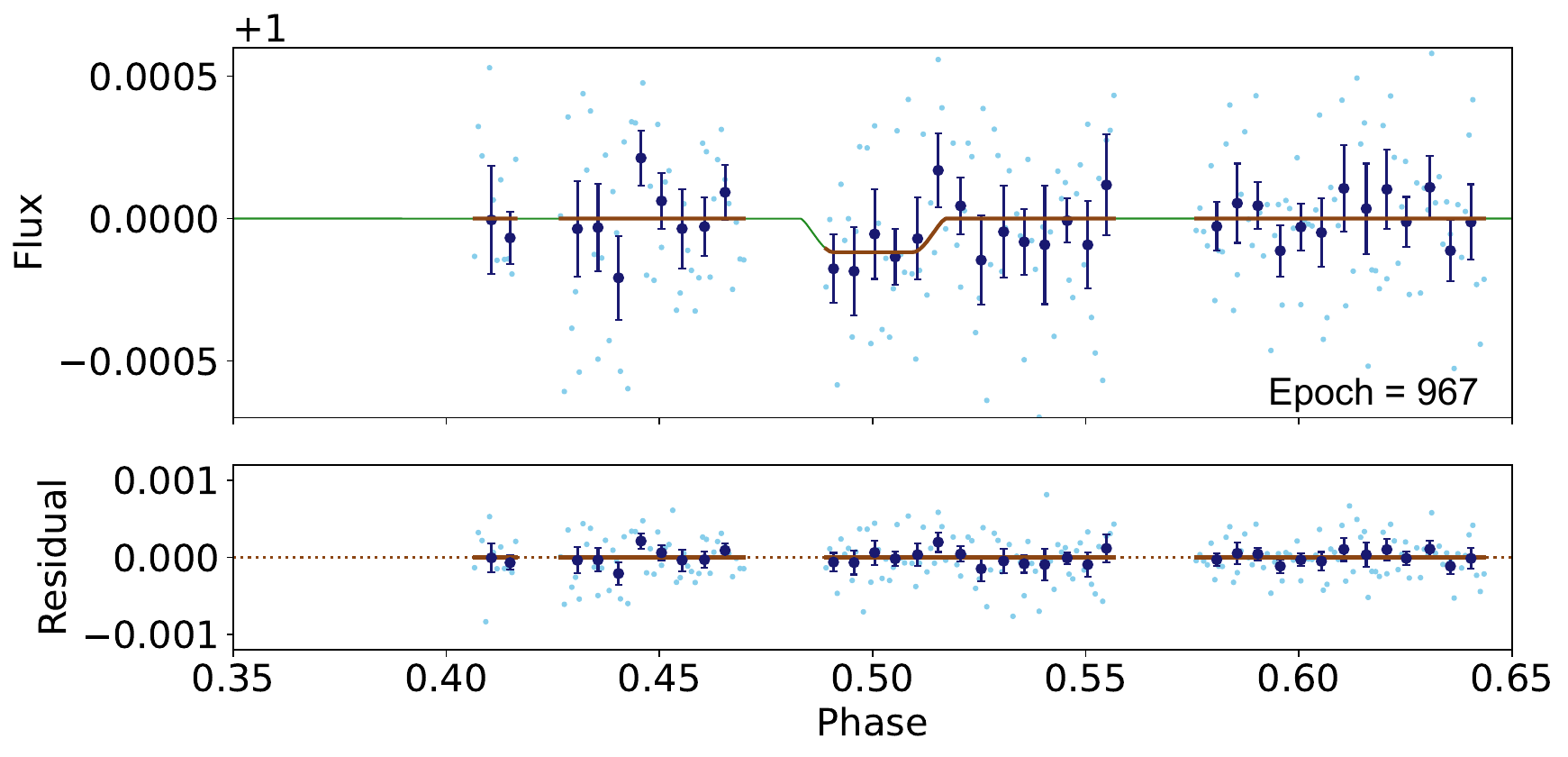} \\
    \includegraphics[width=0.5\linewidth]{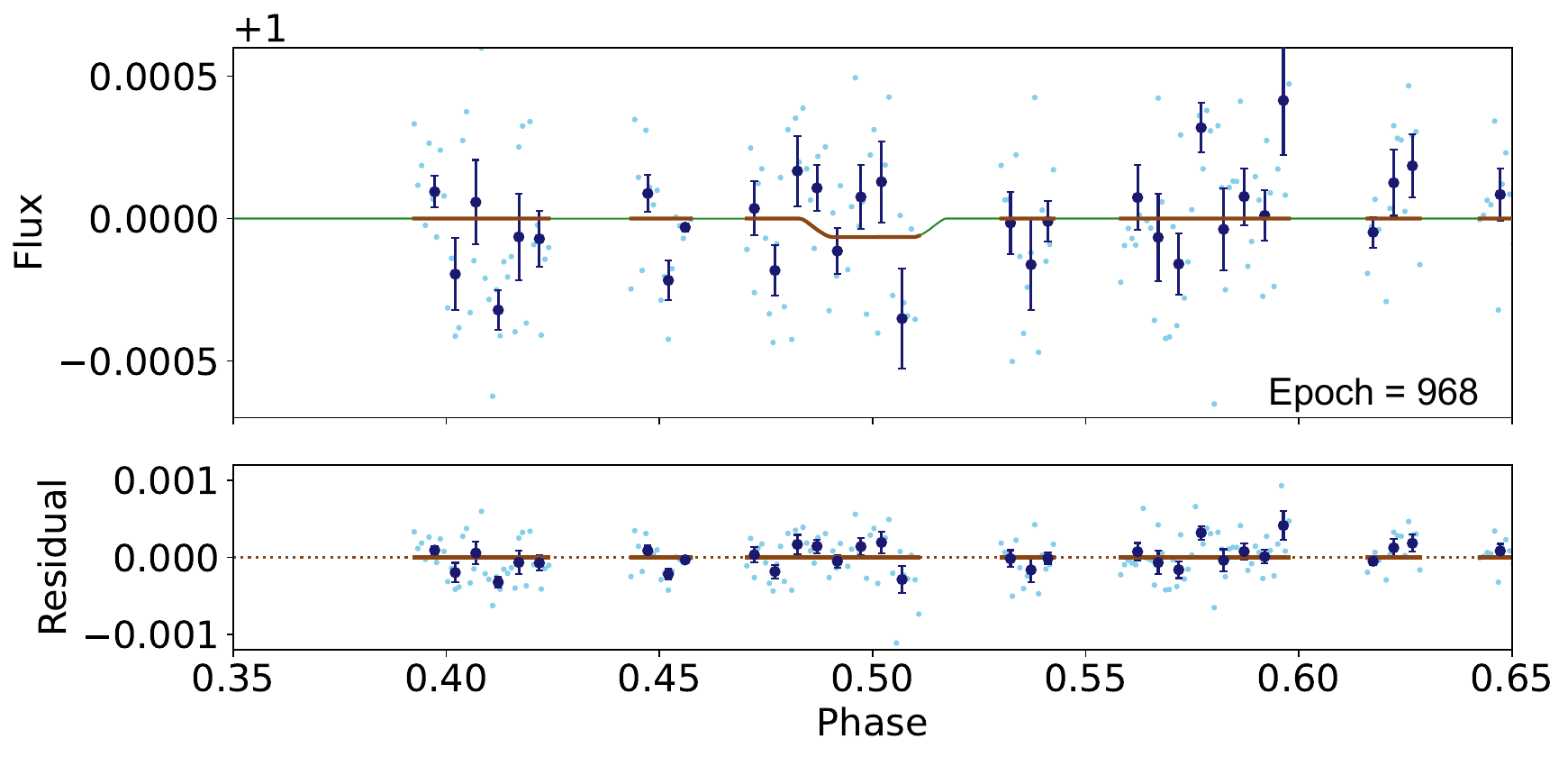} & \includegraphics[width=0.5\linewidth]{F406.pdf} \\
    \includegraphics[width=0.5\linewidth]{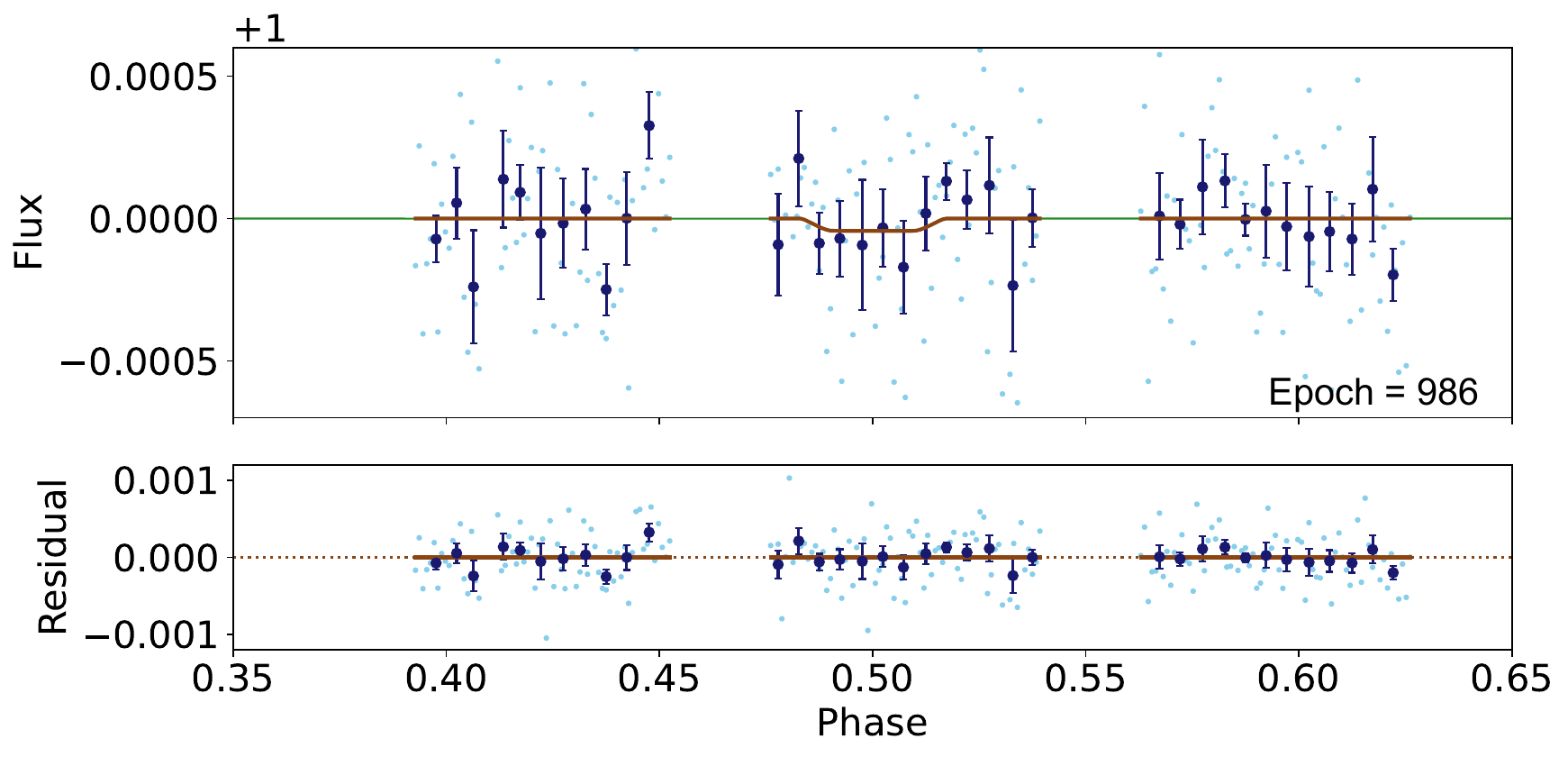} & \includegraphics[width=0.5\linewidth]{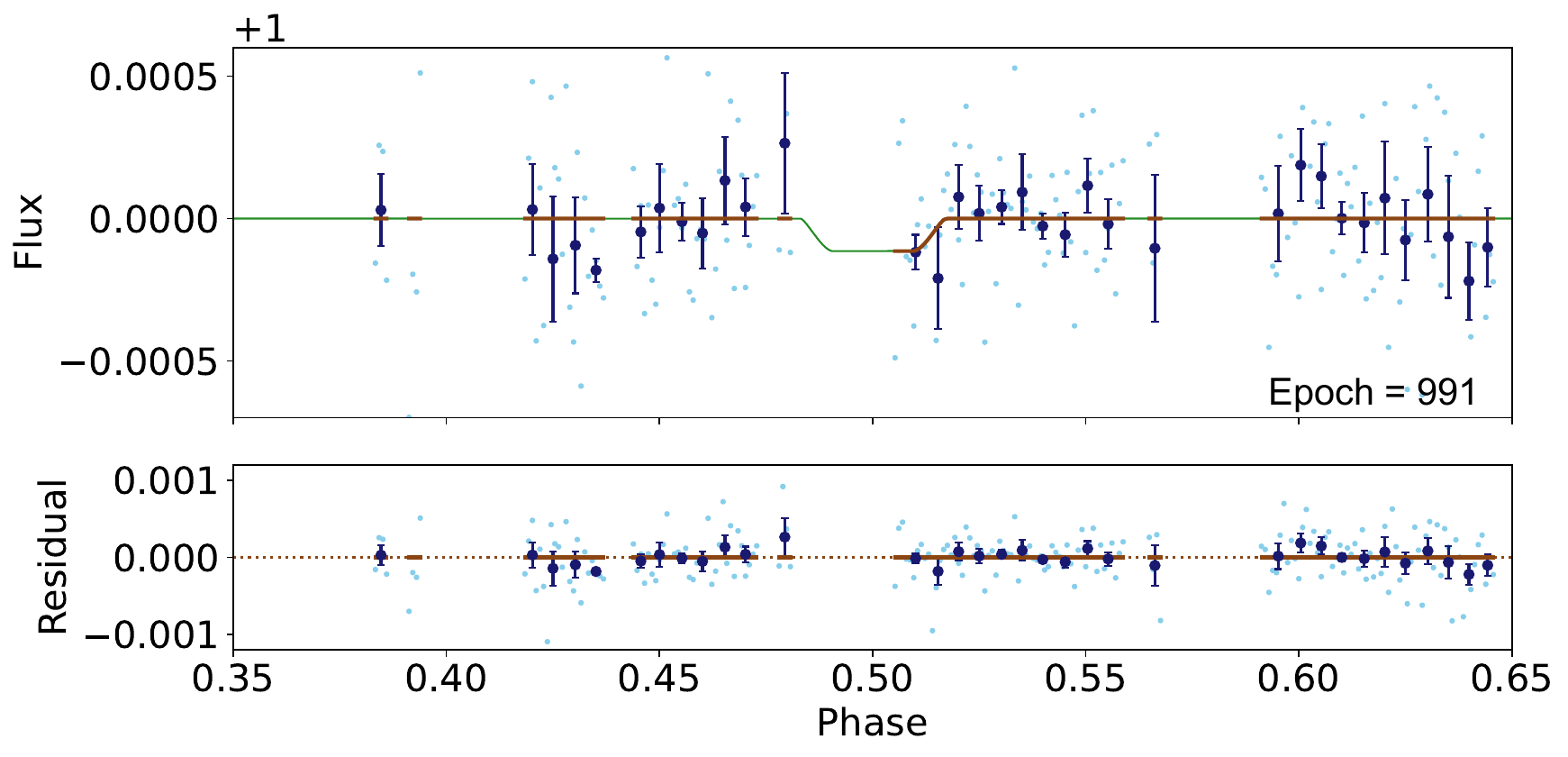} \\
    \includegraphics[width=0.5\linewidth]{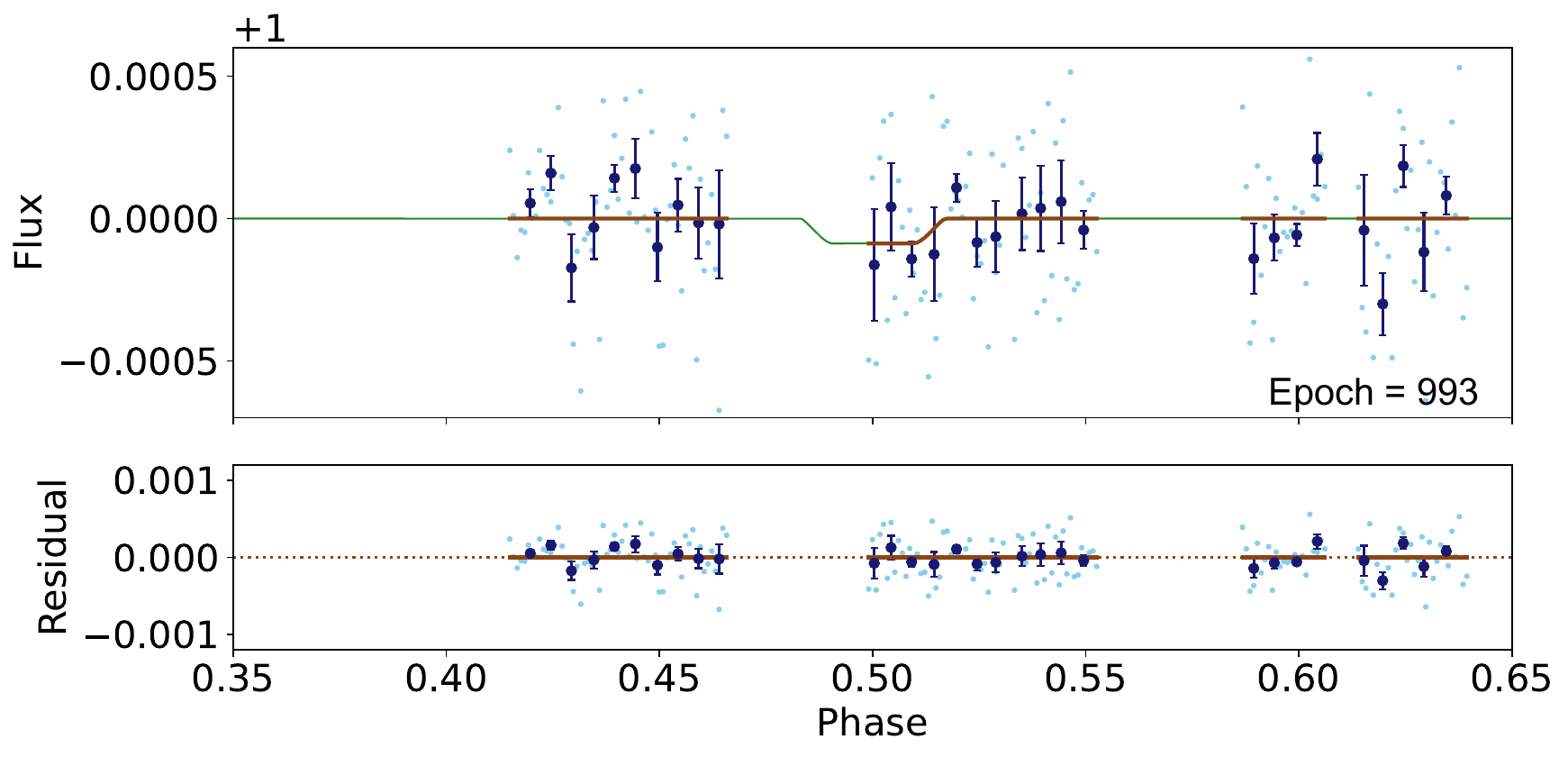} & \includegraphics[width=0.5\linewidth]{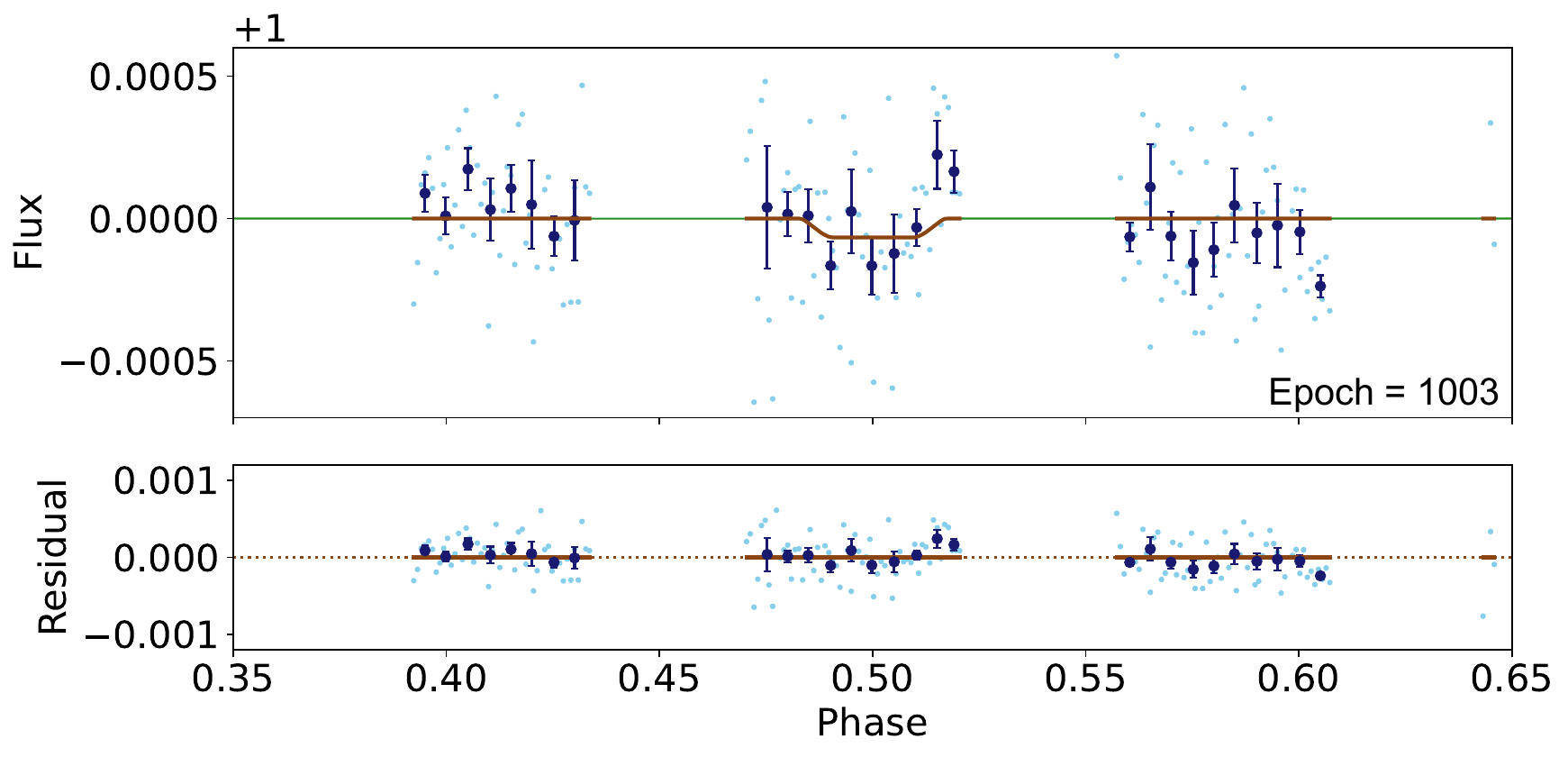} \\
    \end{tabular}
	\caption{Same as Figure \ref{fig:fig2}, but for all the lightcurves reduced by PIPE (part 1 of 2).}
	\label{fig:fig43}
\end{figure*}

\addtocounter{figure}{-1}

\begin{figure*}
	\centering
    \begin{tabular}{cc}
    \includegraphics[width=0.5\linewidth]{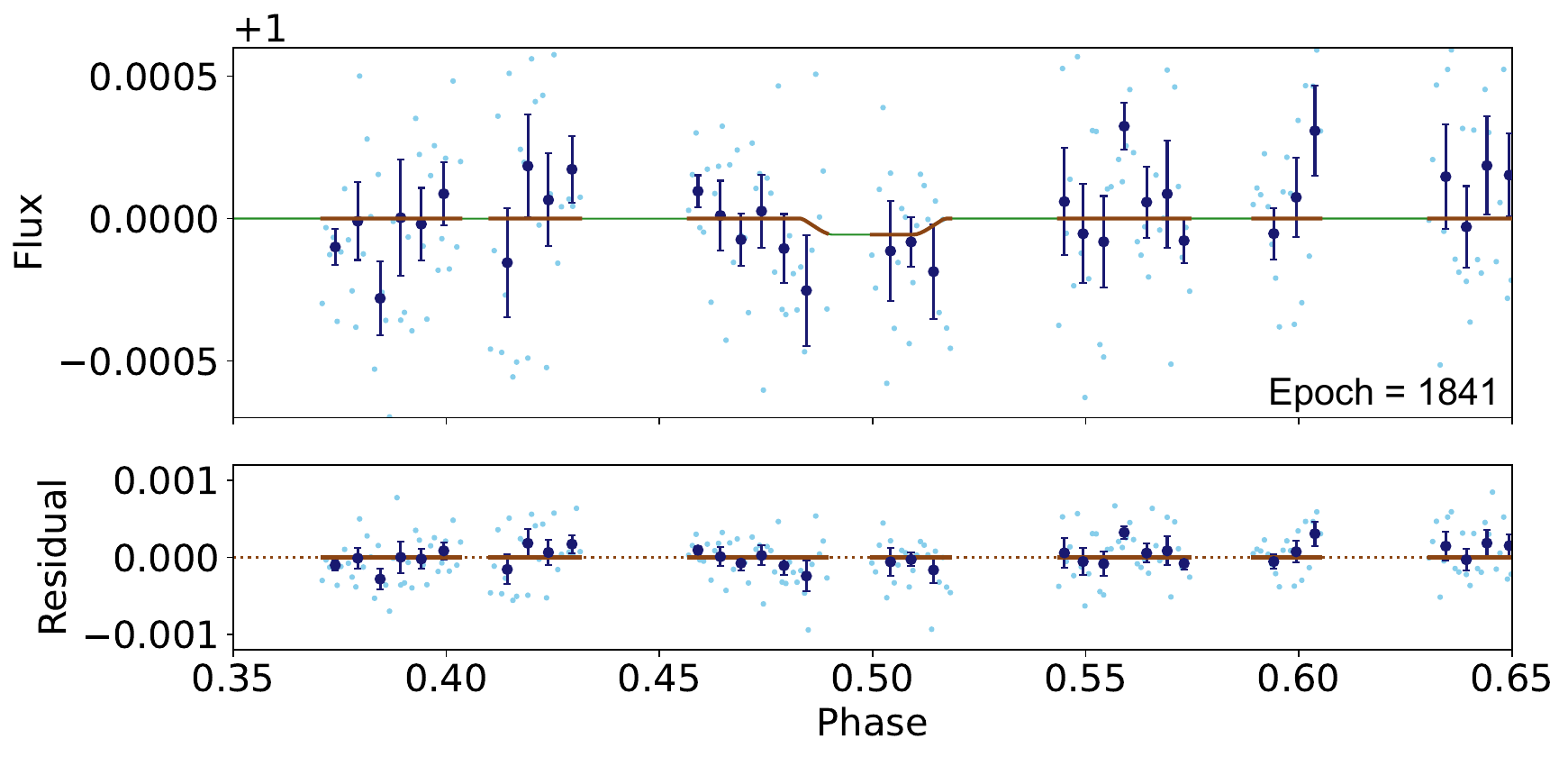} & \includegraphics[width=0.5\linewidth]{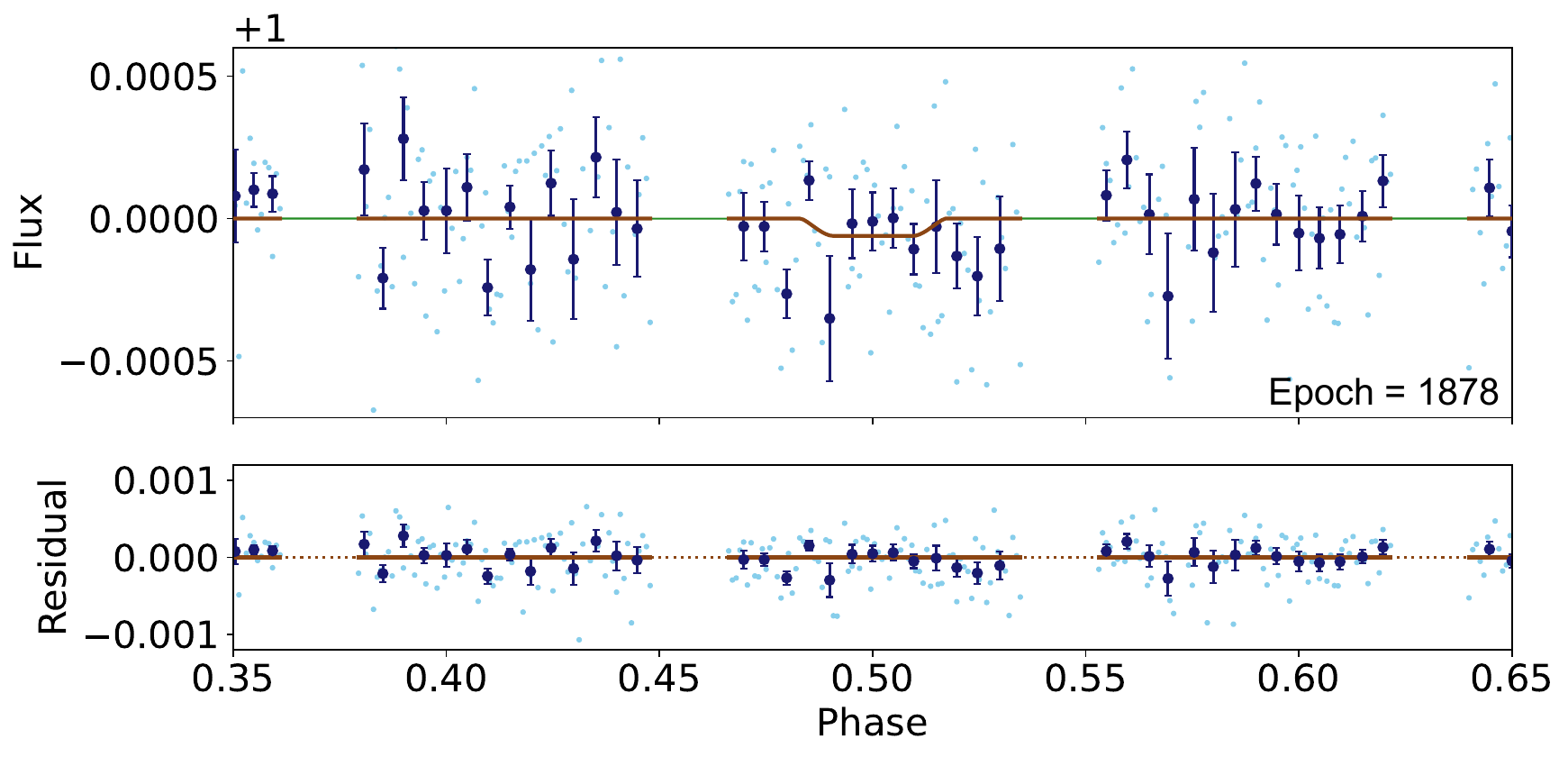} \\
    \includegraphics[width=0.5\linewidth]{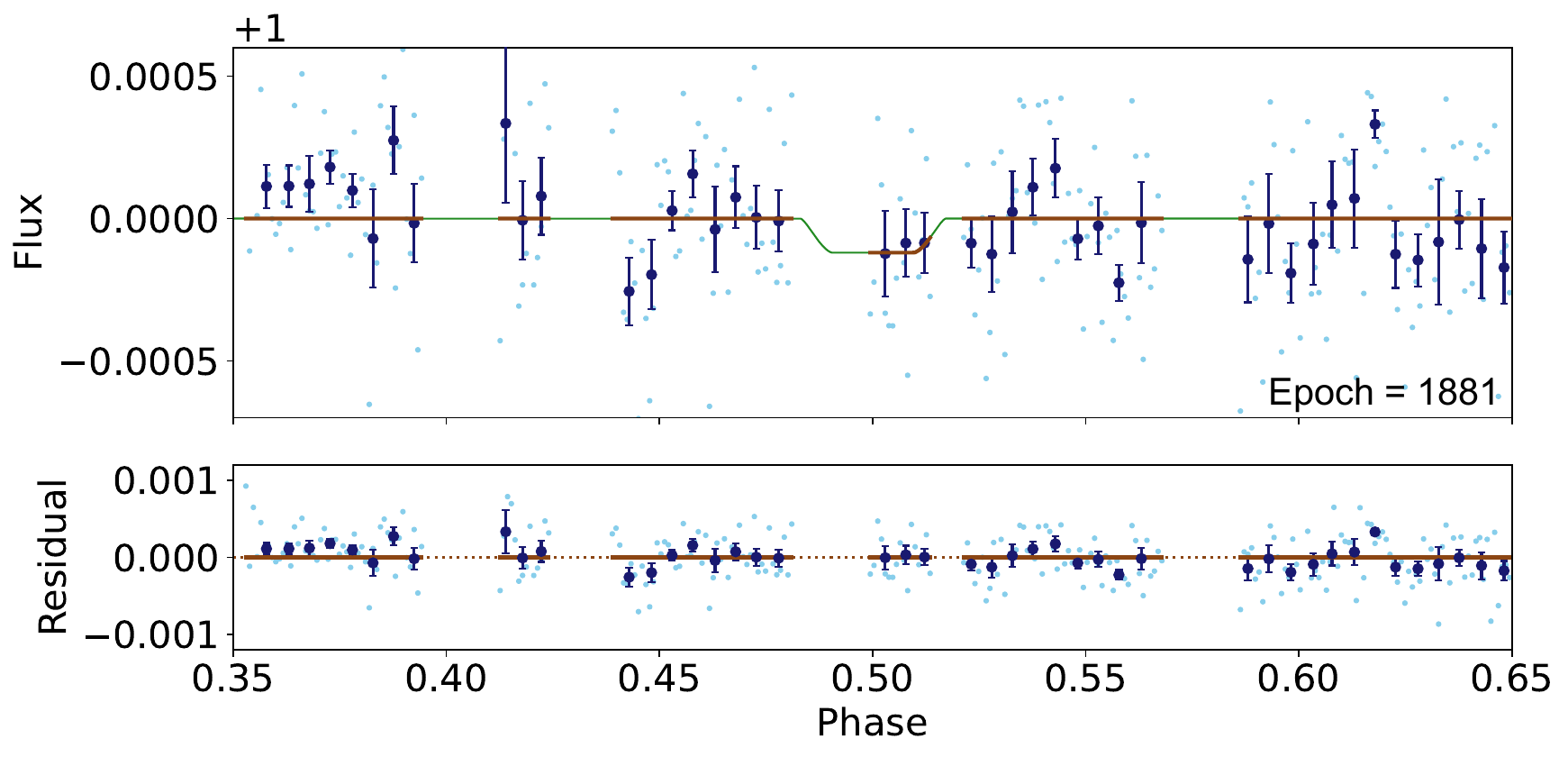} & \includegraphics[width=0.5\linewidth]{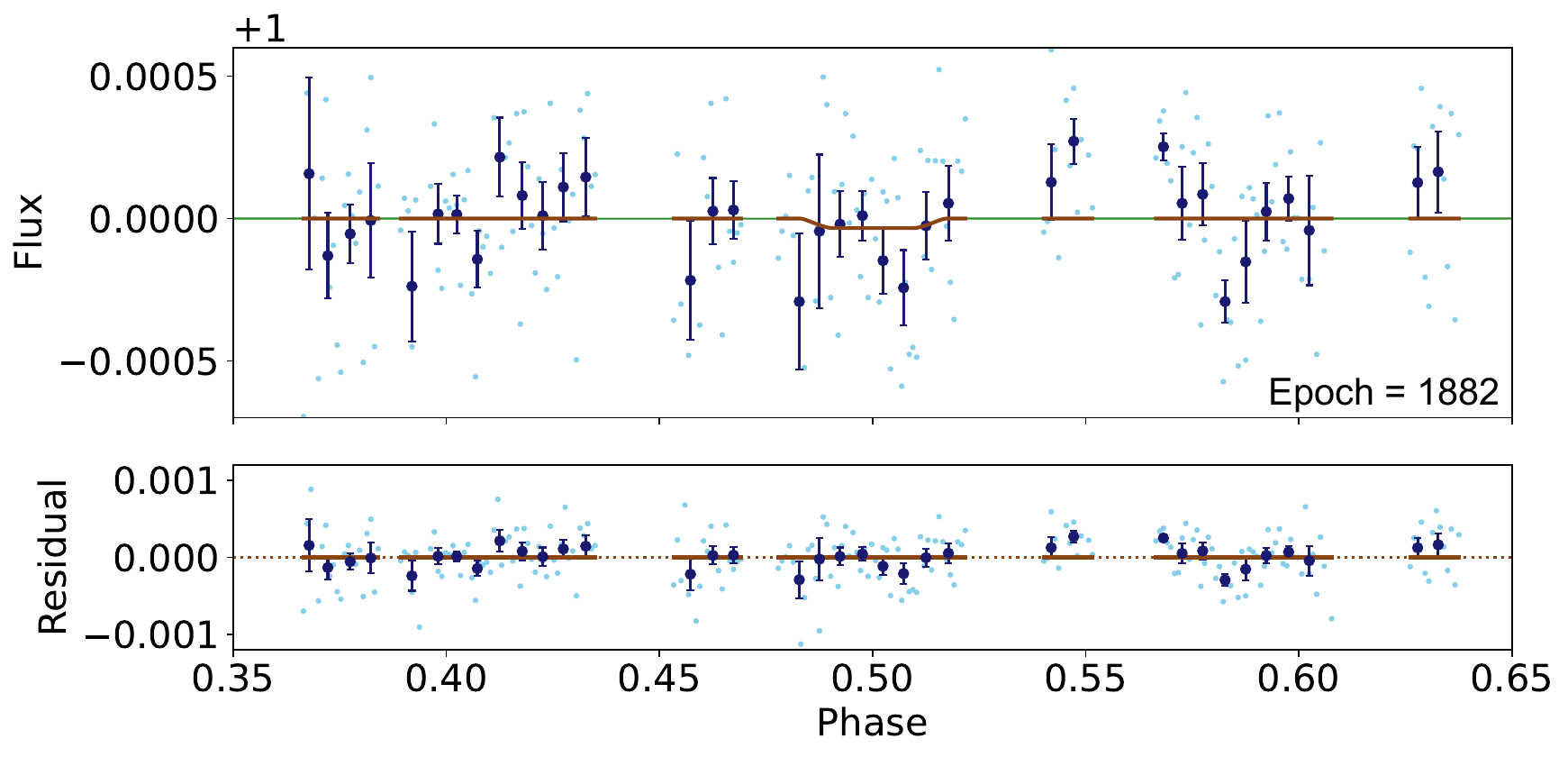} \\
    \includegraphics[width=0.5\linewidth]{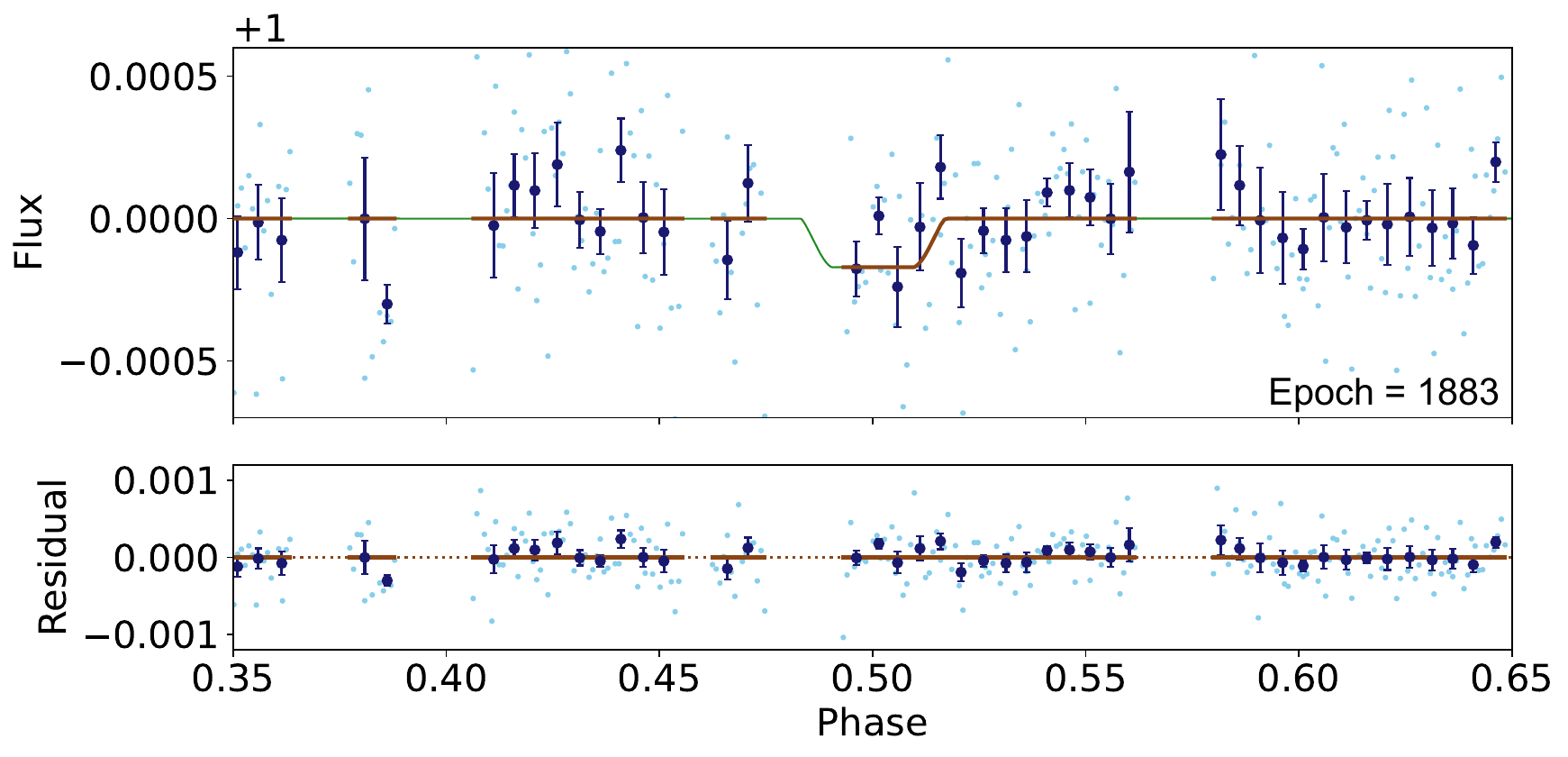} & \includegraphics[width=0.5\linewidth]{F416.pdf} \\
    \includegraphics[width=0.5\linewidth]{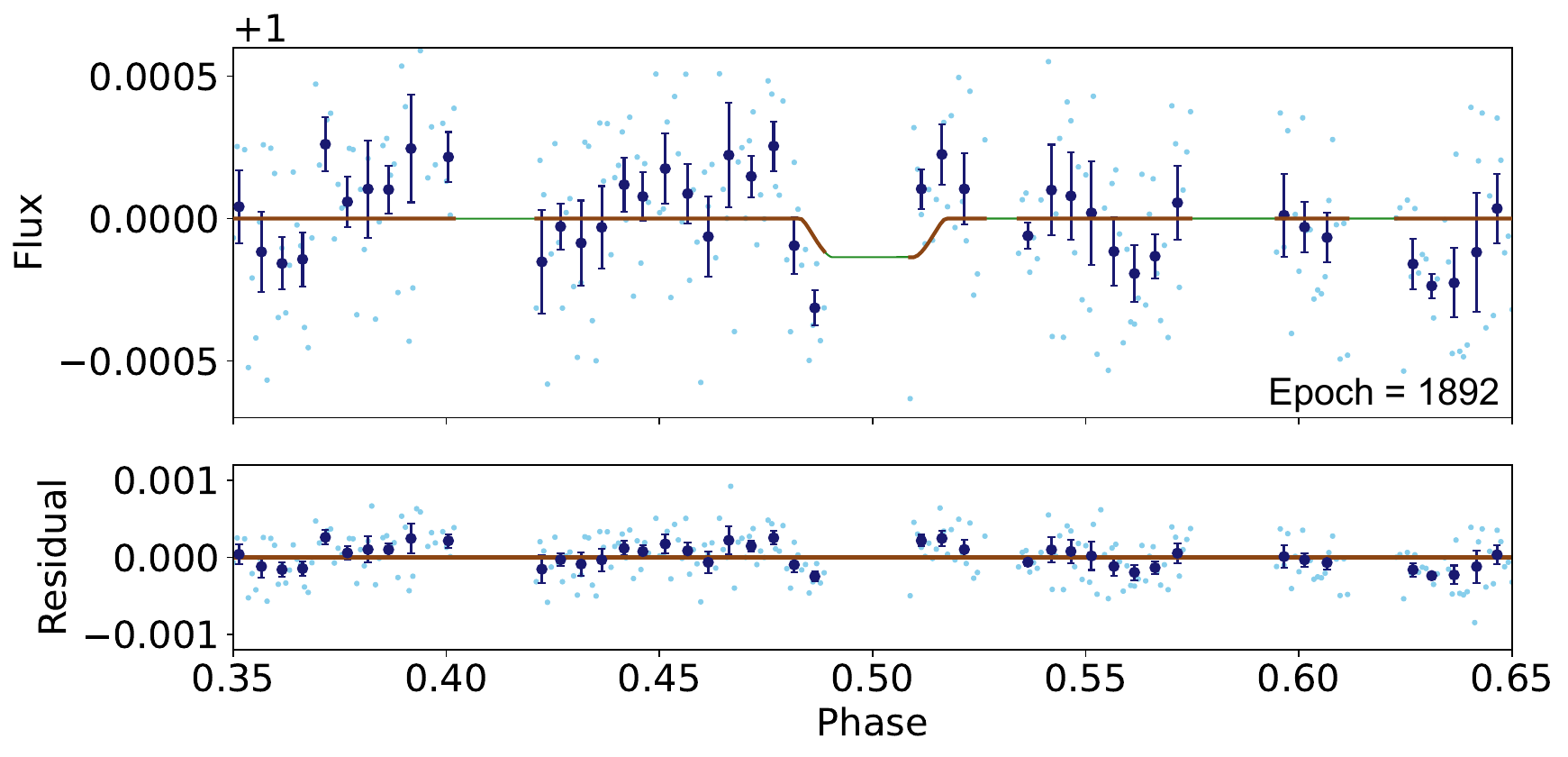} & \includegraphics[width=0.5\linewidth]{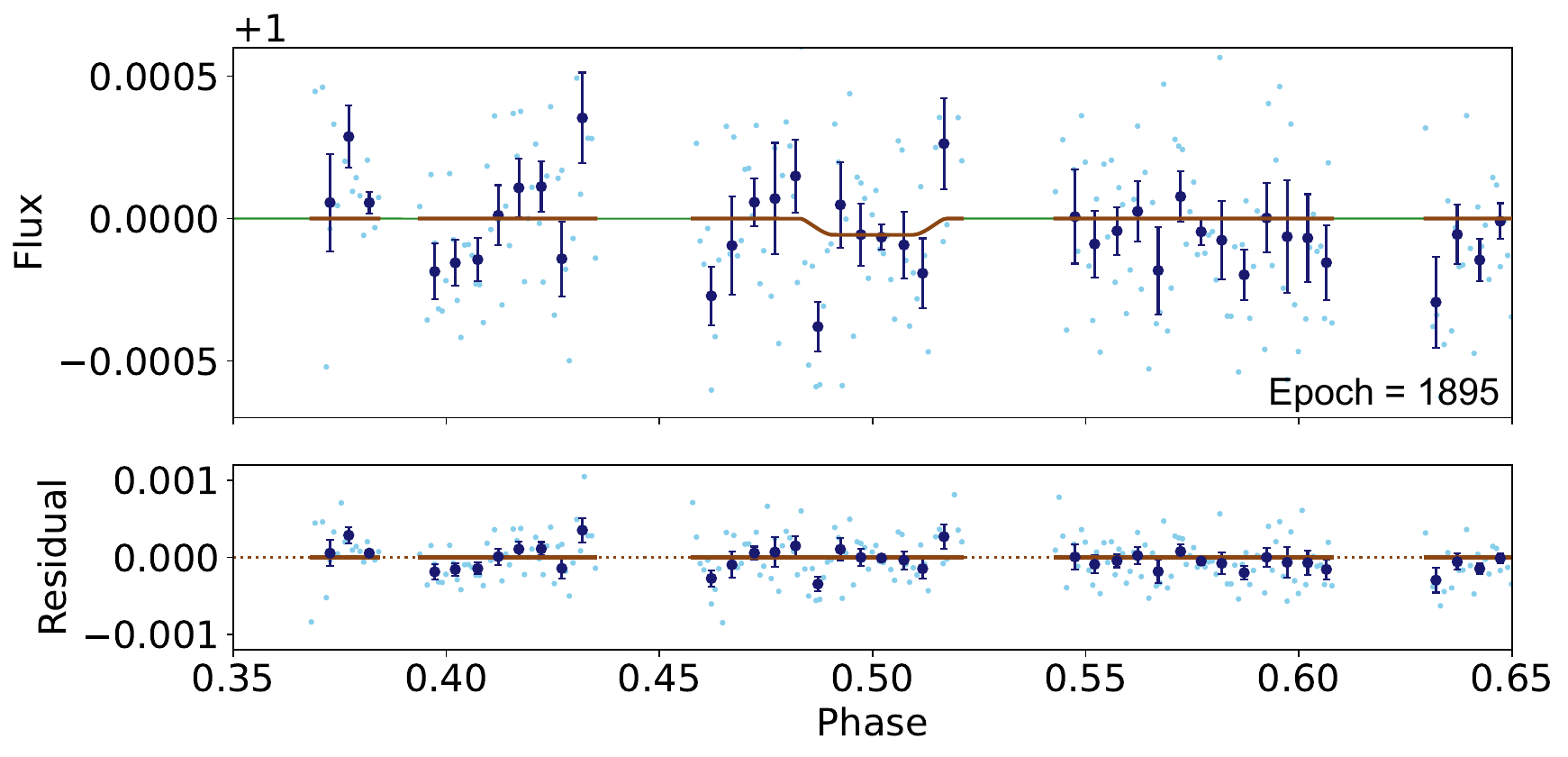} \\
    \includegraphics[width=0.5\linewidth]{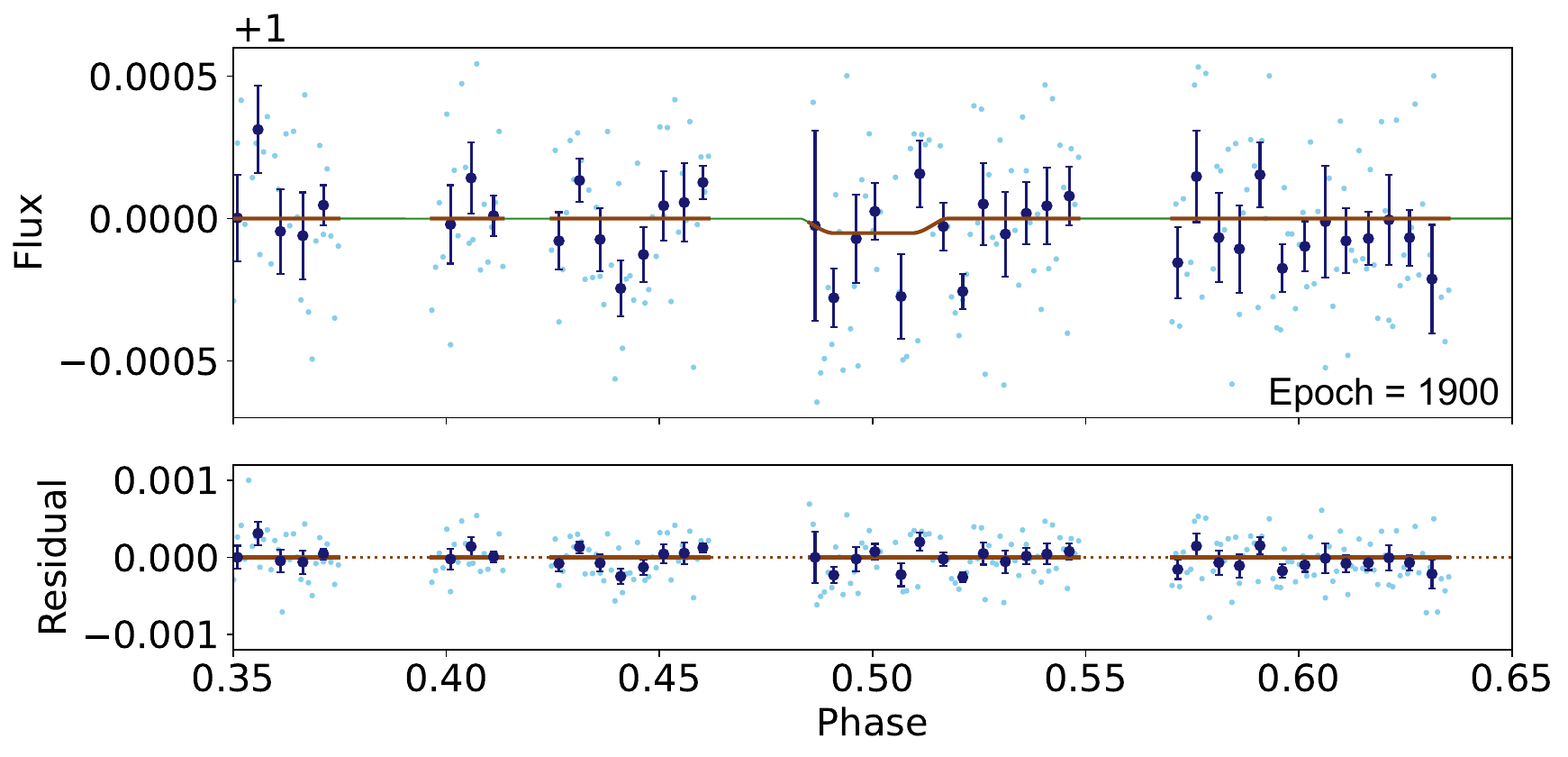} & \includegraphics[width=0.5\linewidth]{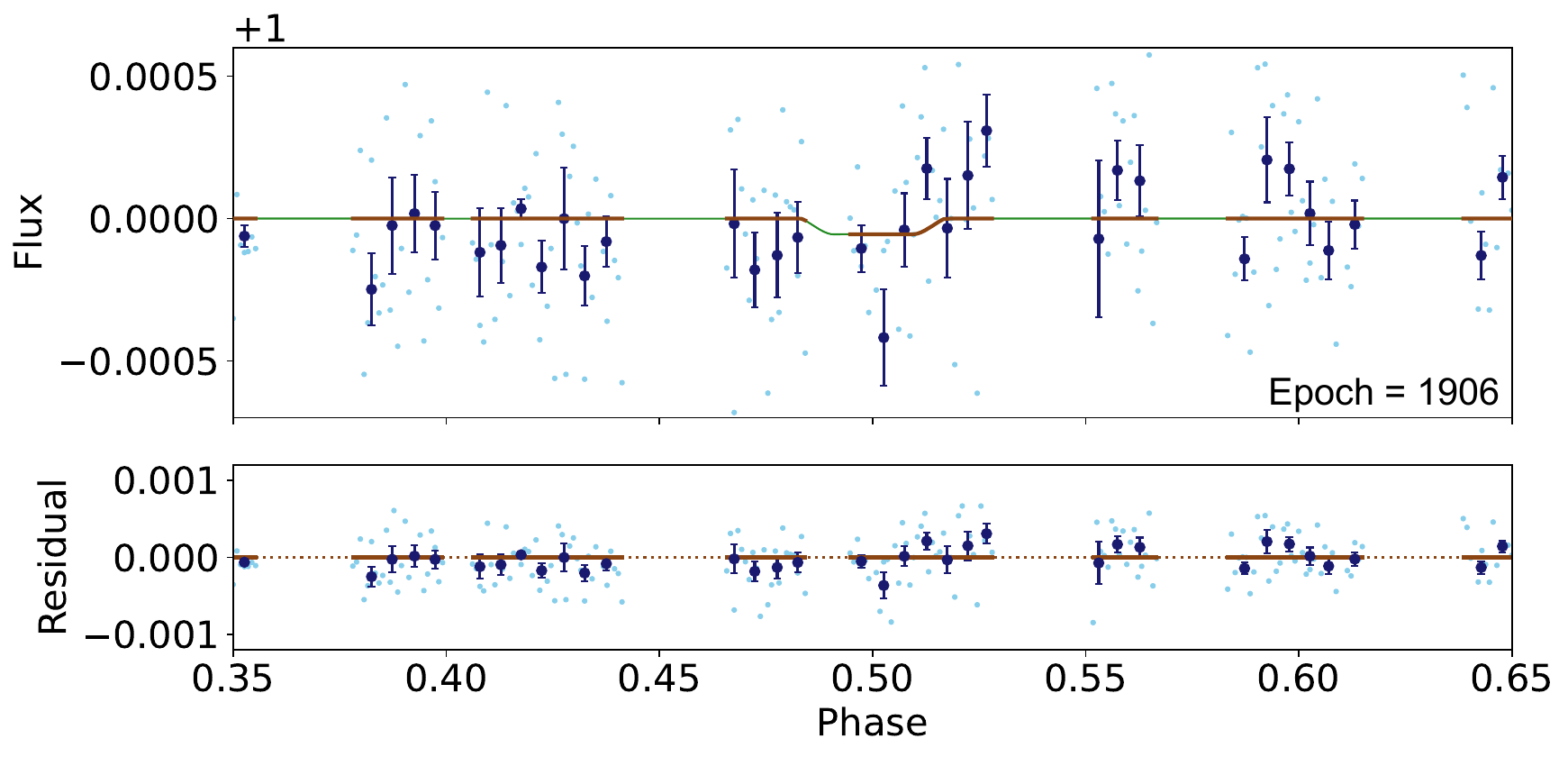} \\
    \end{tabular}
	\caption{(continued, part 2 of 2).}
	\label{fig:fig44}
\end{figure*}

\begin{figure*}
	\centering
	\includegraphics[width=\linewidth]{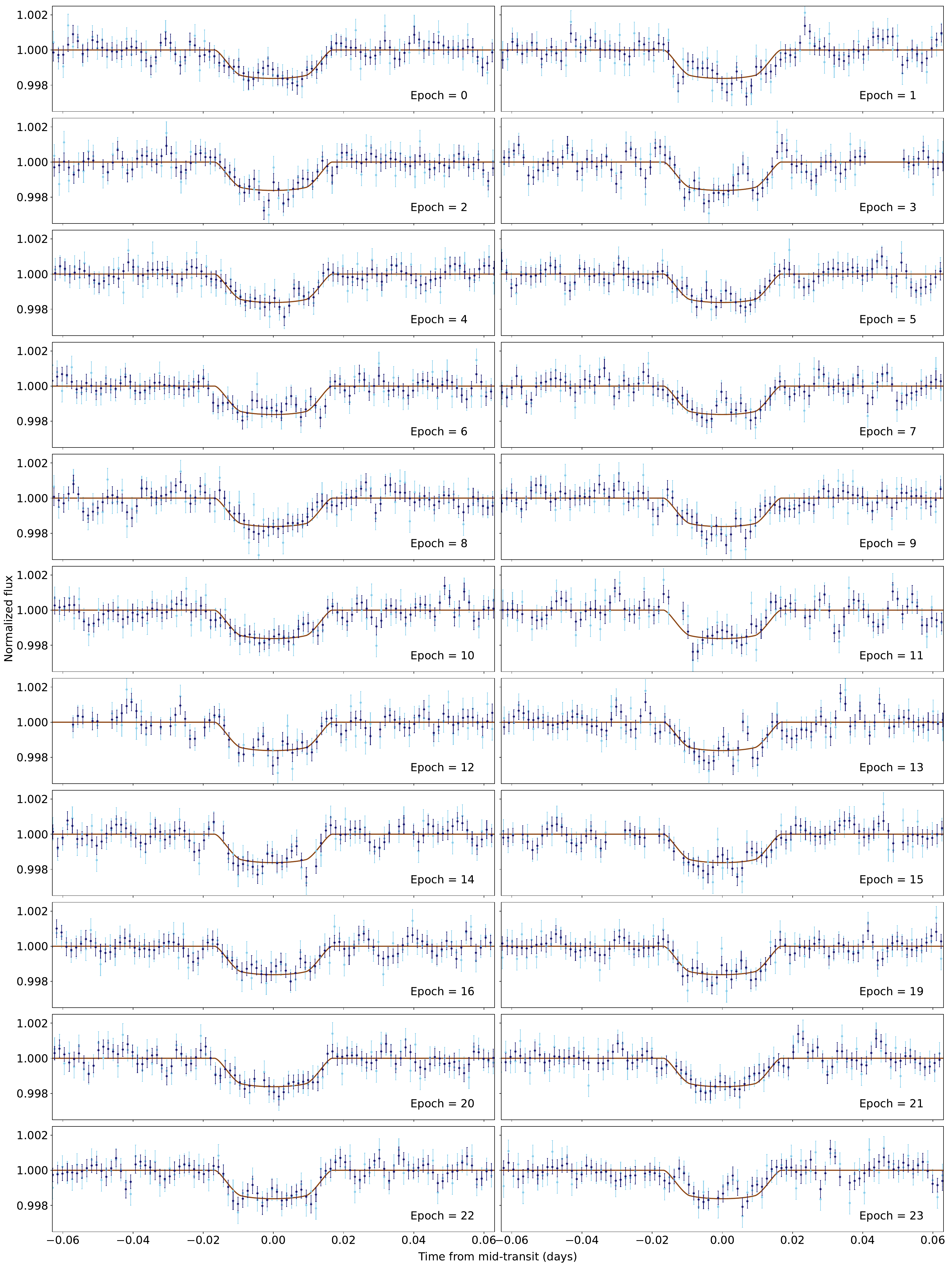}
	\caption{The normalised transit lightcurves before (light blue) and after (blue) wavelet denoising, along with the best-fit transit models (brown) (part 1 of 4).}
	\label{fig:fig51}
\end{figure*}

\addtocounter{figure}{-1}

\begin{figure*}
	\centering
	\includegraphics[width=\linewidth]{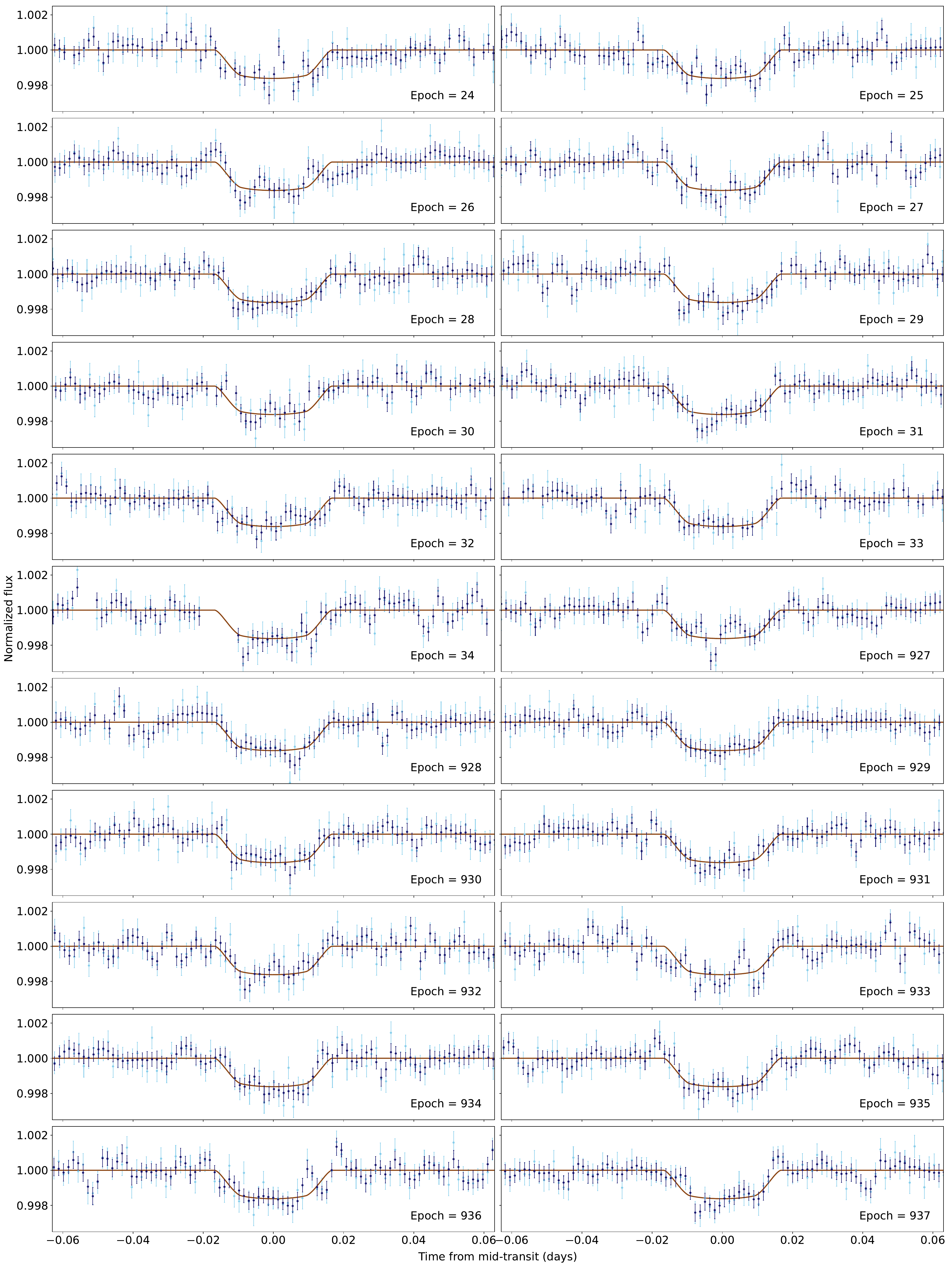}
	\caption{(continued, part 2 of 4).}
	\label{fig:fig52}
\end{figure*}

\addtocounter{figure}{-1}

\begin{figure*}
	\centering
	\includegraphics[width=\linewidth]{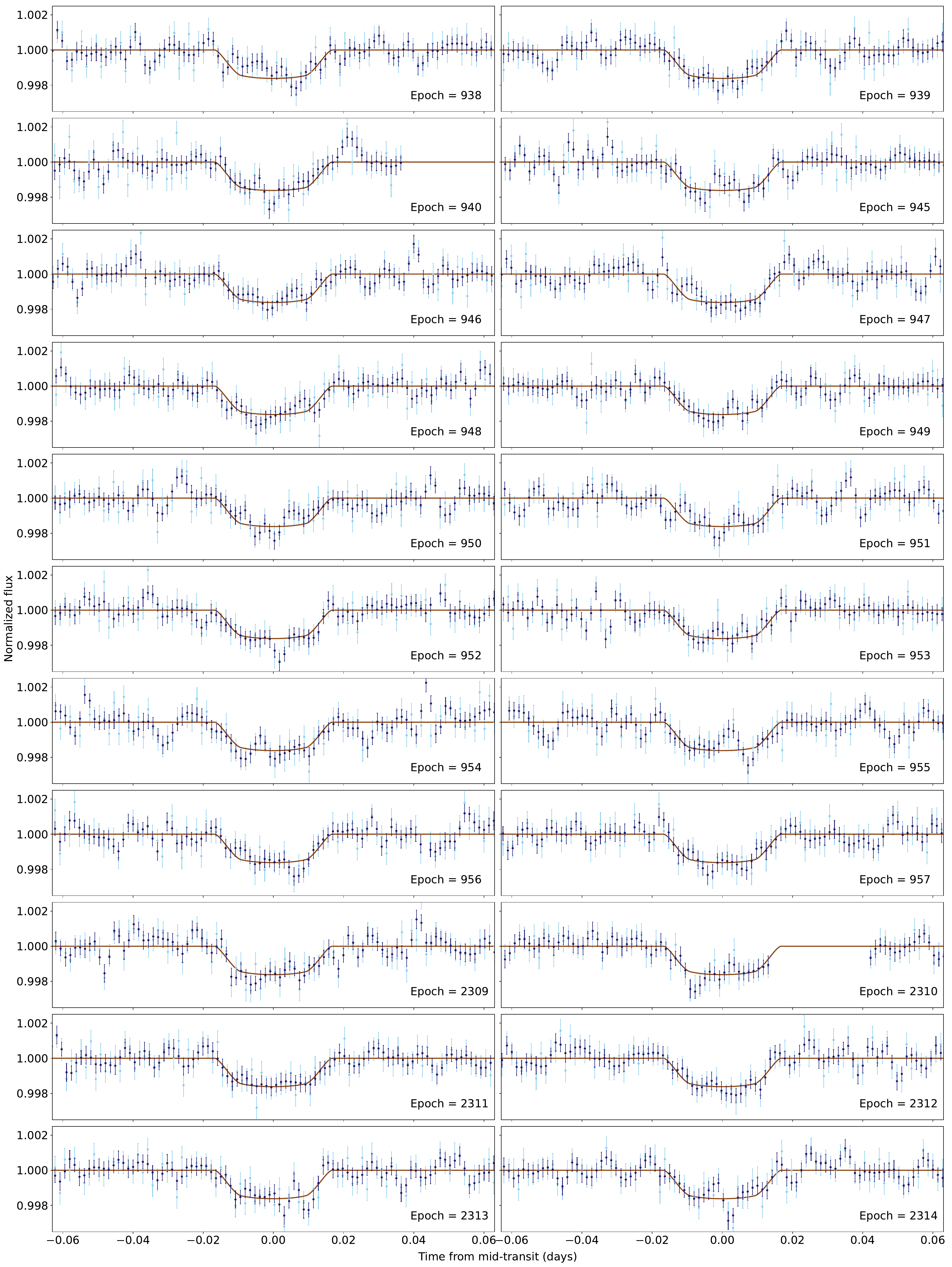}
	\caption{(continued, part 3 of 4).}
	\label{fig:fig53}
\end{figure*}

\addtocounter{figure}{-1}

\begin{figure*}
	\centering
	\includegraphics[width=\linewidth]{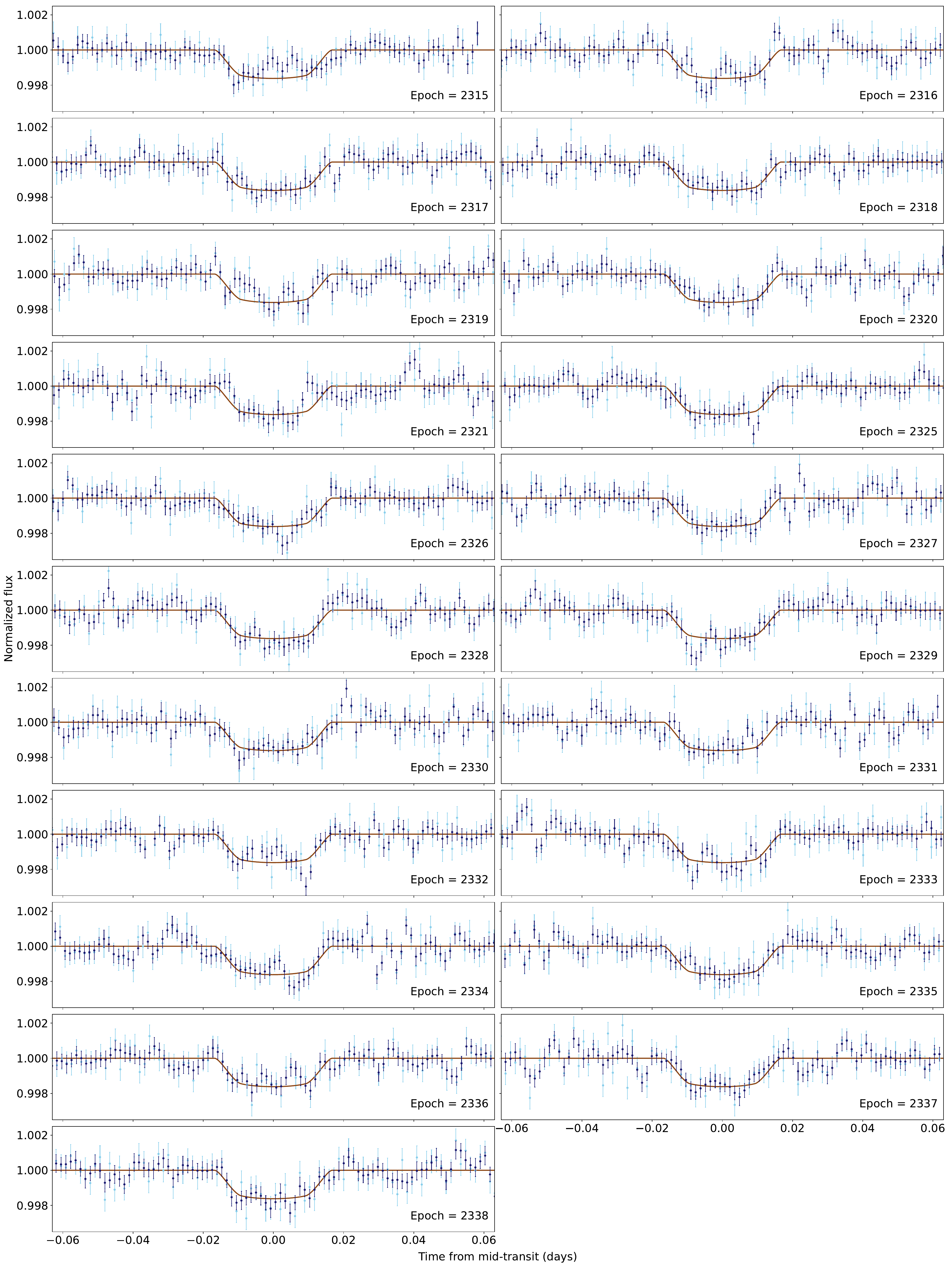}
	\caption{(continued, part 4 of 4).}
	\label{fig:fig54}
\end{figure*}

\end{appendix}

\end{document}